\documentclass[11pt]{aastex63}
\usepackage{graphicx}

\received{}
\revised{}
\accepted{}
\submitjournal{ApJ}
\shorttitle{Magnetic Field in SrcI}
\shortauthors{Hirota et al.}

\begin{document}
\title{Magnetic Field Structure of Orion Source~I}

\correspondingauthor{Tomoya Hirota}
\email{tomoya.hirota@nao.ac.jp}

\author{Tomoya Hirota}
\affiliation{Mizusawa VLBI Observatory, National Astronomical Observatory of Japan, Osawa 2-21-1, Mitaka, Tokyo 181-8588, Japan}
\affiliation{Department of Astronomical Sciences, SOKENDAI (The Graduate University for Advanced Studies), Osawa 2-21-1, Mitaka, Tokyo 181-8588, Japan}

\author{Richard L. Plambeck}
\affiliation{Radio Astronomy Laboratory, University of California, Berkeley, CA 94720, USA}

\author{Melvyn C. H. Wright}
\affiliation{Radio Astronomy Laboratory, University of California, Berkeley, CA 94720, USA}

\author{Masahiro N. Machida}
\affiliation{Department of Earth and Planetary Sciences, Faculty of Sciences, Kyushu University, Motooka 744, Nishi-ku, Fukuoka, Fukuoka 819-0395, Japan}

\author{Yuko Matsushita}
\affiliation{Department of Earth and Planetary Sciences, Faculty of Sciences, Kyushu University, Motooka 744, Nishi-ku, Fukuoka, Fukuoka 819-0395, Japan}

\author{Kazuhito Motogi}
\affiliation{Graduate School of Science and Engineering, Yamaguchi University, Yoshida 1677-1, Yamaguchi, Yamaguchi 753-8511, Japan}

\author{Mi Kyoung Kim}
\affiliation{Mizusawa VLBI Observatory, National Astronomical Observatory of Japan, Hoshigaoka 2-12, Mizusawa, Oshu, Iwate 023-0861, Japan}
\affiliation{Department of Child Studies, Faculty of Home Economics, Otsuma women's university, 12 Sanban-cho, Chiyoda-ku, Tokyo 102-8357, Japan}

\author{Ross A. Burns}
\affiliation{Mizusawa VLBI Observatory, National Astronomical Observatory of Japan, Osawa 2-21-1, Mitaka, Tokyo 181-8588, Japan}
\affiliation{Joint Institute for VLBI in Europe, Postbus 2, 7990 AA, Dwingeloo, The Netherlands}

\author{Mareki Honma}
\affiliation{Mizusawa VLBI Observatory, National Astronomical Observatory of Japan, Hoshigaoka 2-12, Mizusawa, Oshu, Iwate 023-0861, Japan}
\affiliation{Department of Astronomical Sciences, SOKENDAI (The Graduate University for Advanced Studies), Hoshigaoka2-12, Mizusawa-ku, Oshu, Iwate 023-0861, Japan}
\affiliation{Department of Astronomy, Graduate School of Science, The University of Tokyo, Hongo 7-3-1, Bunkyo-ku, Tokyo 113-0033, Japan}

\begin{abstract}
We observed polarization of the SiO rotational transitions from Orion Source~I (SrcI) to probe the magnetic field in bipolar outflows from this high mass protostar.  
Both 43~GHz $J$=1-0 and 86~GHz $J$=2-1 lines were mapped with $\sim$20~AU resolution, using the Very Large Array (VLA) and Atacama Large Millimeter/Submillimeter Array (ALMA), respectively.  
The $^{28}$SiO transitions in the ground vibrational state are a mixture of thermal and maser emission.
Comparison of the polarization position angles in the $J$=1-0 and $J$=2-1 transitions allows us to set an upper limit on possible Faraday rotation of $10^{4}$ radians~m$^{-2}$, which would twist the $J$=2-1 position angles typically by less than 10~degrees. 
The smooth, systematic polarization structure in the outflow lobes suggests a well ordered magnetic field on scales of a few hundred AU.
The uniformity of the polarization suggests a field strength of $\sim$30~milli-Gauss. 
It is strong enough to shape the bipolar outflow and possibly lead to sub-Keplerian rotation of gas at the base of the outflow.
The strikingly high fractional linear polarizations of 80-90\% in the $^{28}$SiO $v$=0 masers require anisotropic pumping. 
We measured circular polarizations of 60\% toward the strongest maser feature in the $v$=0 $J$=1-0 peak. 
Anisotropic resonant scattering (ARS) is likely to be responsible for this circular polarization.  
We also present maps of the $^{29}$SiO $v$=0 $J$=2-1 maser and several other SiO transitions at higher vibrational levels and isotopologues. 
\end{abstract}


\section{Introduction}
\label{sec1}

The Kleinmann-Low Nebula \citep[KL;][]{Kleinmann1967} in Orion is the nearest \citep[418~pc;][]{Kim2008,Menten2007} region of high-mass star formation.
Proper motion measurements with the NSF's Karl G. Jansky Very Large Array (VLA) indicate that the two most massive objects in this region, the Becklin-Neugebauer Object (BN) and radio source I (SrcI), are recoiling from one another at $\sim 40$~km~s$^{-1}$. 
Approximately 500 years ago, the projected separation of the two stars was less than 100 AU on the plane of the sky \citep{Rodriguez2017}.
An extensive system of bullets, bow shocks, and fingers, visible in lines of H$_2$, Fe{\sc{ii}}, and CO, also is centered on Orion-KL, and is estimated to have an age of 500-1000 years \citep{Bally2017}.  
Thus, the currently favored paradigm is that SrcI and BN were ejected from a  dynamical decay of a multiple system approximately 500 years ago, and that the ejection of the stars unbound the surrounding gas and circumstellar disks, creating the finger system \citep{Rodriguez2005, Gomez2008, Zapata2009, Bally2011, Goddi2011b, Zapata2011a, Bally2017, Rodriguez2017}.

Although this dynamical decay model has many attractive features, it has difficulty explaining a couple of observational results: (1) SrcI is moving at about half the speed of BN, which is known to be a 10$M_{\odot}$ B star \citep{Gezari1998, Rodriguez2005}.  Thus, by conservation of momentum, SrcI is expected to have a mass of roughly 20$M_{\odot}$ \citep{Bally2011, Goddi2011b}.
Although recent high angular resolution observations of the rotation curve of H$_2$O and salt lines imply a central mass of 15$M_{\odot}$ \citep{Ginsburg2018}, the mass inferred from observations of many other molecules, including 0.5~milli-arcsecond (mas) resolution observations of SiO masers, tends to be lower, typically in the range 5-8$M_{\odot}$ 
\citep{Kim2008,Matthews2010,Plambeck2016,Hirota2017,Kim2019}.
(2) SrcI launches an 18~km~s$^{-1}$ outflow into the surrounding cloud \citep{Plambeck2009, Greenhill2013, Hirota2017}.  
The outflow appears to propagate straight outward along a northeast-southwest (NE-SW) axis, perpendicular to SrcI's proper motion.  
Given that SrcI moving at 12~km~s$^{-1}$, it is unclear why the outflow lobes are not swept back by the ram pressure of the surrounding medium into a C-shape. 

These difficulties may be resolved if a magnetic field plays an important role in controlling the gas dynamics near SrcI.
The magnetic support from such a field could affect the rotation curve resulting in underestimate of the enclosed mass, and could shape an outflow straight outward into the surrounding medium. 
The curved trajectories of SiO masers at the base of the outflow \citep{Matthews2010,Issaoun2017} provide tantalizing evidence that magnetic fields control the gas kinematics near SrcI. 
Rotation of the outflow inferred from Si$^{18}$O observations \citep{Hirota2017} also hints that the outflow is launched by a magneto-centrifugal disk wind. 
Thus, it is important to investigate the relationship between outflow dynamics and magnetic field structure. 

SrcI is associated with a cluster of SiO masers, including the $^{29}$SiO and $^{30}$SiO isotopologues, in various vibrational levels \citep{Menten1995, Chandler1995, Wright1995, Baudry1998, Goddi2009a, Goddi2009b, Niederhofer2012, Greenhill2013, Kim2019}.  Even in the ground vibrational state, $v$=0, the $J$=1-0 and $J$=2-1 lines are mixture of maser and thermal emission \citep{Chandler1995, Wright1995, Baudry1998,Goddi2009a, Goddi2009b}. 
Polarization of these masers provides a means of probing the magnetic field near SrcI.  
Using the 45-m radio telescope at Nobeyama Radio Observatory (NRO) with a spatial resolution of 40\arcsec, 
\citet{Tsuboi1996} discovered that the 43~GHz $v$=0 $J$=1-0 masers were strongly polarized. 
Subsequently, \citet{Plambeck2003} mapped the linear polarization of the 86~GHz $v$=0 $J$=2-1 masers with the Berkeley-Illinois-Maryland Association (BIMA) Array with approximately 0.5\arcsec\ angular resolution.  

The BIMA data revealed that linear polarization in the $v$=0 $J$=2-1 line was as high as 50\% in some velocity ranges \citep{Plambeck2003}. 
With 0.5\arcsec\ resolution, the polarization position angle was found to be nearly uniform across the source, but was tilted by roughly 30~degrees with respect to the axis of the outflow.\footnote{In \citet{Plambeck2003}, the SiO emission was assumed to originate from a ``flared disk'' centered on SrcI, but now it is recognized as an outflow from the star.}  Surprisingly, the position angle of the $J$=2-1 linear polarization differed by 70~degrees from that of the $J$=1-0 line published by \citet{Tsuboi1996}.  
\citet{Plambeck2003} speculated that Faraday rotation by plasma in the foreground Orion H{\sc{ii}} region might plausibly explain this angular offset; a rotation measure of $3.3\times10^4$ radians m$^{-2}$ brought the $J$=1-0 and $J$=2-1 position angles into concordance, and also put the polarization direction along the axis of the outflow.
However, these observations had insufficient resolution to probe the magnetic field structure on the $\sim$100~AU scale of the outflow launching region. 

The theory of maser polarization was worked out by \citet{Goldreich1973} and has been revisited by \citet{Watson2009}, \citet{Perez-Sanchez2013}, \citet{Lankhaar2019}, and others.
Polarization occurs because selection rules cause the maser gain to differ for polarizations parallel and perpendicular to the magnetic field. 
The net  polarization may be either parallel or perpendicular to the field, depending on the angle between the field and the line of sight, $\theta$. 
In the simplest case where the excitation mechanism (collisions or radiation) pumps the magnetic sublevels equally, this transition from parallel to perpendicular occurs around 55~degrees, which is known as the van Vleck angle.
Key requirements are that the stimulated emission rate $R$ and the net decay rate of the maser transition state $\Gamma$ (i.e. including collisions and spontaneous emission) must be much less than the Zeeman splitting $g\Omega/2\pi$. 

Although the above basic theories of maser linear polarization predict the fractional polarization of $\sim$30\% at maximum \citep[e.g.][]{Perez-Sanchez2013}, a number of maser sources show higher linear polarization, including SrcI \citep{Tsuboi1996}. 
This can be explained by the anisotropic maser pumping \citep[e.g. ][]{Nedoluha1990, Watson2009, Lankhaar2019}, in which the radiation field for maser pumping is anisotropic owing to the presence of a central star. 
Furthermore, maser sources sometimes show significant levels of circular polarization. 
The Zeeman splitting of the maser line in an interstellar/circumstellar magnetic field is a possible origin. 
However, there are alternative non-Zeeman scenarios such as conversion from the linear to circular polarization as the maser radiation propagates through a region with a varying magnetic field \citep{Wiebe1998} and anisotropic resonant scattering \citep[ARS; ][]{Houde2013, Houde2014, Chamma2018}. 

In this paper, we present the results of Atacama Large Millimeter/Submillimeter Array (ALMA) and VLA observations that measure the linear polarization of both the 86~GHz $^{28}$SiO $v$=0 $J$=2-1 and 43~GHz $^{28}$SiO $v$=0 $J$=1-0 emission from SrcI with 0.05\arcsec\ (20~AU) angular resolution, to probe the magnetic field morphology in its bipolar outflow lobes. 
Based on our new results, we discuss possible origins of the above maser polarization mechanism. 
We also present the result of polarization measurement for the 86~GHz $^{29}$SiO $v$=0 $J$=2-1 emission at the base of the outflow.  

\section{Observations and Data Analysis}
\label{sec2}

\subsection{ALMA}
\label{sec2-1}

Observations with ALMA were carried out on 2017 October 12 and 17 in the full polarization mode consisting of  2 execution blocks (EBs) for each day. 
The array was in the C43-10 configuration with 51 antennas, providing the longest baseline length of 16~km available with ALMA. 
We used the ALMA band 3 receiver to detect 86~GHz SiO lines and their isotopologues listed in Table \ref{tab-line}. 
For the $^{28}$SiO $v$=0 $J$=2-1 and $v$=1 $J$=2-1 lines, we set up spectral windows with total bandwidths of 58.6~MHz and spectral resolutions of 122~kHz. 
For the other transitions, $^{28}$SiO $v$=2 $J$=2-1, $^{29}$SiO $v$=0 $J$=2-1, $^{29}$SiO $v$=1 $J$=2-1, and $^{30}$SiO $v$=0 $J$=2-1, we employed a coarser resolution of 244~kHz owing to the limited data rate. 

In addition, two broad spectral windows were configured at 96-99~GHz for continuum emission with total bandwidths of 1875~MHz each. 
The tracking center position of SrcI was RA(J2000)=$05^{\rm h}35^{\rm m}14\fs5129$ and Dec(J2000)=$-05\degr22\arcmin30\farcs576$. 
The total time of observations was 3.5~hours for both days, and total on-source time was 2.5~hours. 

Calibration used the pipeline script provided by East-Asian ALMA Regional Center (EA-ARC) through the CASA (Common Astronomy Software Applications) package. 
The absolute flux densities were calibrated using J0423-0120 for the first EBs in both days. 
The measured flux densities of the polarization calibrator, J0522-3627, in the first EBs were used as a reference flux for the second EBs. 
The measured flux densities of J0522-3627 were 4.96~Jy and 5.34~Jy at the first and second epochs, respectively. 
These values were different by 5\% from the cataloged value of 5.18$\pm$0.31~Jy and 5.63$\pm$0.20~Jy probably because of its variability, but are within the flux accuracy of ALMA band 3.
The bandpass calibration was made using J0423-0120. 
Residual phase calibration was made using the secondary calibrator J0541-0211. 
Another compact radio source J0539-0356 was included as a check source to verify the quality of calibration. 

The polarization calibrations were done by observing J0522-3627 in a standard manner, which showed fractional polarization of 2.8\% (those of Q and U are -1.5\% and -2.4\%, respectively). 
According to the ALMA Proposer's Guide\footnote{https://almascience.nrao.edu/documents-and-tools/cycle5/alma-proposers-guide}, the expected minimum detectable degree of  linear polarization is 1\% and 3\% for compact and extended sources, respectively. 

We note that circular polarization measurements were not assured in these ALMA cycle 5 observations.  
However, given the high circular polarization fraction in the $^{28}$SiO $v$=0 $J$=2-1 line, we will discuss possible detection of the circular polarization as described below. 
Here we summarize results of circular polarization measurements of calibrators in our ALMA data. 
For the polarization calibrator J0522-3627, we could not see any signature of the Stokes V leakage in the full polarization images. 
On the other hand, there exists possible Stokes V leakage in the images of the bandpass/flux calibrator J0423-0120 and the phase calibrator J0541-0211. 
These Stokes V images show point-like components coincide with the Stokes I emission. 
The flux densities of the Stokes I and V images of J0423-0120 are 1.31~Jy and -3~mJy, respectively, with an rms noise level of the Stokes V image of 0.02~mJy~beam$^{-1}$. 
Thus, the Stokes V/I ratio is 0.2\%. 
For J0541-0211, the Stokes V/I ratio is 0.09\% with Stokes I and V flux densities of 380~mJy and 0.3~mJy, respectively, and an rms noise level of the Stokes V image of 0.03~mJy~beam$^{-1}$. 
We could see no Stokes V emission peak in the check source J0539-0356 probably due to the low Stokes I intensity of 11~mJy with the the Stokes V/I ratio of $<$1\% (rms noise level of 0.01~mJy~beam$^{-1}$). 
It should be noted that the calibrators are point sources at the center of the beam and the leakage may be higher off-axis.  
The more recent ALMA Cycle 7 Proposer's Guide\footnote{https://almascience.nrao.edu/documents-and-tools/cycle7/alma-proposers-guide} indicates that the minimum detectable degree of circular polarization is generally 1.8\% of the peak flux within 1/10 of the primary beam, but can be a factor of 2 higher under unfavorable circumstances.
Although the emission region of SrcI ($\sim$2\arcsec) is smaller than 1/10 of the primary beam size of 70\arcsec, the accuracy in the extended SiO circular polarization emission could have higher  uncertainties. 

Using the resultant calibrated visibility data, synthesis images were made using CASA and Miriad software.
We did not apply self-calibration to the ALMA data because it provided no significant improvement (e.g. in case of a strong $^{28}$SiO $v$=1 $J$=2-1 maser feature, the flux density increased only $\sim$4\% ). 
For the continuum emission we used only line-free channels to make mfs (multi-frequency synthesis) images. 
Full-Stokes velocity channel maps were first made with the common velocity resolution of 0.5~km~s$^{-1}$ for $^{28}$SiO $v$=0 and $v$=1 transitions and 1.0~km~s$^{-1}$ for the other lines.  
In addition, channel maps with the coarser resolution of 2~km~s$^{-1}$ were also made for display purposes (see Appendix \ref{secA-map}).
The synthesized beam size was 49~mas$\times$47~mas with a position angle of -35~degrees for the 96~GHz continuum image. 
In the case of spectral lines, typical beam sizes are slightly larger, $\sim$55~mas with an almost circular beam shape.
The brightness temperature, $T_{B}$, can be obtained from the flux density, $F$, by $T_{B}$(K)=6.7$\times 10^{4} F$(Jy) with the observed frequency at 86~GHz and the beam size of 50~mas.
The shortest baseline length of 41~m provides a maximum recoverable scale of $\sim$1.0\arcsec\ at band 3.

Because linear polarization intensity is calculated by $P=\sqrt{Q^{2}+U^{2}}$, it is biased by the noise in the Stokes Q and U images. 
Therefore, the  linearly polarized intensity images were debiased by subtracting the rms noise over the emission-free channels/regions in the images; $P=\sqrt{Q^{2}+U^{2}-\sigma^{2}}$. 
Line-averaged polarization angles were calculated from Stokes Q and U images with velocity widths of 30~km~s$^{-1}$ centered at the systemic velocity of 5~km~s$^{-1}$. 
This is because averaging polarization angles over all channels can lead to wrong values owing to the $\pm$180~degree ambiguity in the angles (e.g., an average of 0$\degr$ and 180$\degr$ should be close to 0$\degr$ or 180$\degr$ but simple averaging yields $\sim90\degr$).  
Finally, fractional linear polarizations were calculated from the 30~km~s$^{-1}$-resolution P and Stokes I images.

\subsection{VLA}
\label{sec2-2}

Observations with VLA were made on 2018~March~6 with the most extended A-configuration in the full-polarization observing mode. 
The number of antennas was 27 and the baseline lengths ranged from 800~m to 36.6~km. 
The observed frequency was Q-band (43~GHz) and the target lines are listed in Table \ref{tab-line}. 
We included higher vibrationally excitation lines of $^{28}$SiO at $v$=1, 2, 3, and 4, but the $v$=3 and $v$=4 lines were not detected unlike the higher rotational level \citep{Kim2019}.
For these spectral lines, we assigned 7 spectral windows with 16~MHz bandwidth and 50~kHz resolution. 
In addition, we assigned 16 wideband spectral windows with 128~MHz bandwidth to observe continuum emission. 
The tracking center position of SrcI is the same as the ALMA observations. 
The observations lasted 5~hours to cover a sufficiently wide range of parallactic angle. The net on-source time was 2.5~hours.
We also had another observing session on 2018 June 9. 
However, we did not include these data because of the lack of a common amplitude and polarization calibrator, 3C286, which could introduce unexpected calibration uncertainties. 

The data were initially calibrated by the pipeline script provided by NRAO, but only for Stokes I of continuum emission. 
Thus, we conducted polarization calibration for both continuum and spectral line data using the CASA package. 
The absolute flux density and bandpass were calibrated using 3C286 for which we assumed a flux density of 1.53~Jy at 43.60~GHz. 
Phases were calibrated by observing the secondary calibrator J0541-0541. 
The polarization leakage and polarization angle were calibrated by observing 3C84 and 3C286, respectively.  
A fractional polarization of 13.2\% and polarization angle of 36~degrees were employed for 3C286 at the frequency of 43.50~GHz \citep{Perley2013}.

The accuracy of the circular polarization calibration is estimated to be 0.2\% from the leakage of Stokes V with respect to the Stokes I for the polarization calibrator 3C286 and the secondary calibrator J0541-0541, assuming that they have no circular polarization. 
We note that there is a significant systematic error showing a pair of positive and negative circularly polarized components in the calibrator images at a level of 0.2\% of the Stokes I, probably due to the beam squint of the VLA antennas. 
Nevertheless, we measure SiO circular polarization levels much higher than the above values without any sign of such a systematic pattern, so we expect these errors to have no significant effect on our analysis. 

The calibrated visibility data were used to make synthesis images in the same manner employed for the ALMA data.
Self-calibration was made using a single channel of the strong and compact $^{28}$SiO $v$=1 $J$=1-0 maser line. 
As a result, the observed flux density of the continuum emission increased by 20\%. 
The resultant synthesized beam size was 53~mas$\times$44~mas with a position angle of 30~degrees for the 43~GHz continuum emission.
The spectral line maps have a similar beam size with slightly different position angles.  Flux density can be converted to brightness temperature using the equation $T_{B}$(K)=2.7$\times 10^{5} F$(Jy) with the observed frequency at 43~GHz and a beam size of 50~mas.
The shortest baseline length in the VLA A-configuration, 800~m, provides a maximum recoverable scale of $\sim$1\arcsec \  at Q-band. 

\section{Results}
\label{sec3}

We observed in total 13 SiO $J$=1-0 and $J$=2-1 transitions as listed in Table \ref{tab-line}.  
All were detected except the two highest excitation lines, $^{28}$SiO $v$=3 $J$=1-0 and $v$=4 $J$=1-0. 
As listed in Table \ref{tab-line}, most of the observed SiO lines are dominated by maser emission, as their peak brightness temperatures exceed 10$^{3}$~K except for the $^{29}$SiO $v$=1 $J$=2-1 line. 
In particular, peak brightness temperatures are extremely high ($>10^{6}$~K) for the masers in the $^{28}$SiO $v$=0,1,2 $J$=1-0, $v$=1 $J$=2-1 and $^{29}$SiO $v$=0 $J$=2-1 lines.

We detected linearly polarized emission in 6 lines: $^{28}$SiO $v$=0,1,2 $J$=1-0, $^{28}$SiO $v$=0,1 $J$=2-1, and $^{29}$SiO $v$=0 $J$=1-0. 
All of them show brightness temperatures at the peak channels of $>$10$^{4}$~K and high linear polarization fraction of a few \% to 90\% at maximum. 
Spectra, moment maps, and channel maps for all the SiO transitions that we detected (Table \ref{tab-line}) are shown in Appendices \ref{secA-sp} and \ref{secA-map}.  
These maps show linear polarization intensities and angles with 2~km~s$^{-1}$ velocity resolution, adequate to see most of the individual maser features.
In this paper, we will focus mostly on the ordered polarization structures in the bipolar outflow lobes seen in the $v$=0 masers. 
Discussion of the vibrationally excited masers will be presented in forthcoming papers. 

\subsection{Continuum}
\label{sec3-1}

Figure \ref{fig-moment} shows images of the continuum emission from SrcI at 43 GHz and 96 GHz generated from the wideband spectral windows, superposed on the moment maps of $^{28}$SiO $v$=0 $J$=1-0 and $J$=2-1 lines, respectively. 
Note that the $^{28}$SiO $v$=0 $J$=1-0 maps are plotted with a smaller box than the $J$=2-1 maps because the $J$=1-0 emission region is smaller. 

The continuum emission traces a nearly edge-on circumstellar disk \citep[e.g.][]{Reid2007, Goddi2011b, Plambeck2013, Plambeck2016, Hirota2016b, Ginsburg2018, Ginsburg2019}.  Positions, sizes, peak brightnesses, and flux densities derived from Gaussian fits to these images are summarized in Table~\ref{tab-cont}.
The integrated flux densities of SrcI at 43~GHz and 96~GHz, 10~mJy and 58~mJy, respectively, are consistent with the previous results \citep[e.g.][]{Goddi2011b, Ginsburg2018}. 
The brightness temperatures at the peak positions are 740~K and 770~K for 43~GHz and 96~GHz, respectively.  
Extended dust emission from the Orion Hot Core is almost completely resolved out by our interferometric observations.

We did not detect polarization in the continuum emission, with 5$\sigma$ upper limits of 0.07~mJy~beam$^{-1}$ at 43~GHz and 0.06~mJy~beam$^{-1}$ at 96~GHz.  
These values correspond to fractional polarizations of $<$3\% and $<$0.4\% at 43~GHz and 96~GHz, respectively. 
Because the expected minimum detectable degree of polarization with ALMA is 1\% and 3\% for compact and extended sources, as mentioned above, the actual upper limit of the 96~GHz continuum is 1\%. 
The low polarization fraction of the continuum emission in SrcI is expected, since it is thought to be dominated by optically thick dust at frequencies of 43~GHz and higher \citep{Plambeck2016}.  
Furthermore, the gas densities on the disk surface are estimated to be of order 10$^{8}$-10$^{11}$~cm$^{-3}$ in order to pump the vibrationally excited SiO masers \citep{Goddi2009a}. 
Similar density values are also suggested by the continuum emission of SrcI \citep{Plambeck2016, Hirota2016b}. 
At such high densities, collisions are likely to randomize the dust grain orientations, leading to unpolarized emission.

\subsection{$^{28}$SiO $v$=0 lines}
\label{sec3-2}

\subsubsection{Velocity structure}
Integrated intensity images of the $^{28}$SiO $v$=0 lines are shown in Figure~\ref{fig-moment}.
The velocity range is determined based on spectra (see Figure \ref{fig-spall}) in Appendix \ref{secA-sp}. 
The lines show bipolar structure along the NE-SW direction perpendicular to the edge-on continuum disk.  
The $J$=1-0 emission is less extended than $J$=2-1.
Bright emission lobes are distributed away from the disk midplane by $\sim$0.2\arcsec\ ($\sim$80~au). 
Both the NE and SW lobes are extended to 0.5\arcsec \ and 1.0\arcsec \ from the disk midplane in the $J$=1-0 and 2-1 lines, respectively.  
These lobes can be traced farther from the disk in lower resolution maps \citep{Chandler1995, Plambeck2009, Goddi2009a, Greenhill2013}.
The NE lobe appears closer to SrcI in both $J$=1-0 and $J$=2-1.
This may be the effect of foreground opacity in the $^{28}$SiO $v$=0 lines and the inclination of the outflow axis, with the NE lobe tilted toward the observer. 

The moment~0 maps show compact peaks, probably maser clumps, that are spatially unresolved even at 0.05\arcsec\ angular resolution. 
The $J$=1-0 line map has more clumpy structures than the $J$=2-1 line in spite of comparable spatial resolutions. 
This result suggests that maser emission is more dominant in the $J$=1-0 line than the $J$=2-1 line. 
In 0.5~km~s$^{-1}$ channel maps the maximum brightness temperatures of the $J$=1-0 and $J$=2-1 lines are 2$\times$10$^{6}$~K (6.6~Jy~beam$^{-1}$ at -5~km~s$^{-1}$) and 5$\times$10$^{4}$~K (0.95~Jy~beam$^{-1}$ at 16~km~s$^{-1}$), respectively, clearly suggesting maser action. 

The radial velocities of the $^{28}$SiO lines seen in the moment~1 maps do not show systematic structures such as spatially symmetric blue- and red-shifted lobes or velocity gradients along the outflow axis. 
However, there is weaker collimated emission in the integrated intensity maps of the blue- and red-shifted components of the $J$=2-1 line as shown in Figure \ref{fig-28SiO21_bluered}, where one can see a velocity gradient along the northwest-southeast direction. 
The column emanates from the continuum source, and the direction of the velocity gradient is parallel to that of the rotating disk as traced by thermal molecular lines \citep{Hirota2014, Plambeck2016, Hirota2017,Ginsburg2018, Ginsburg2019} and vibrationally excited $^{28}$SiO masers \citep{Kim2008, Matthews2010,Issaoun2017,Kim2019}. 
The length of the rotating column is almost comparable to the rotating outflow traced by the 484~GHz Si$^{18}$O $J$=12-11 line \citep{Hirota2017} as shown in Figure \ref{fig-28SiO21_bluered}.  Thus, the $^{28}$SiO $v$=0 $J$=2-1 line also traces the rotation of the outflow driven by the disk wind as revealed by the Si$^{18}$O line.  

\subsubsection{Linear polarization}
Figure \ref{fig-sp} shows the spectra of the $^{28}$SiO $v$=0 $J$=1-0 and $^{28}$SiO $v$=0 $J$=2-1 lines integrated across the emission region.  
In these spectra, the fractional linear polarization of the $J$=1-0 line approaches 100\%, in particular at the $-6$~km~s$^{-1}$ component. The high fractional polarization of this feature is consistent with previous single-dish measurements made with the NRO~45~m telescope \citep{Tsuboi1996}.

\begin{figure*}[th]
\begin{center}
\includegraphics[width=7cm]{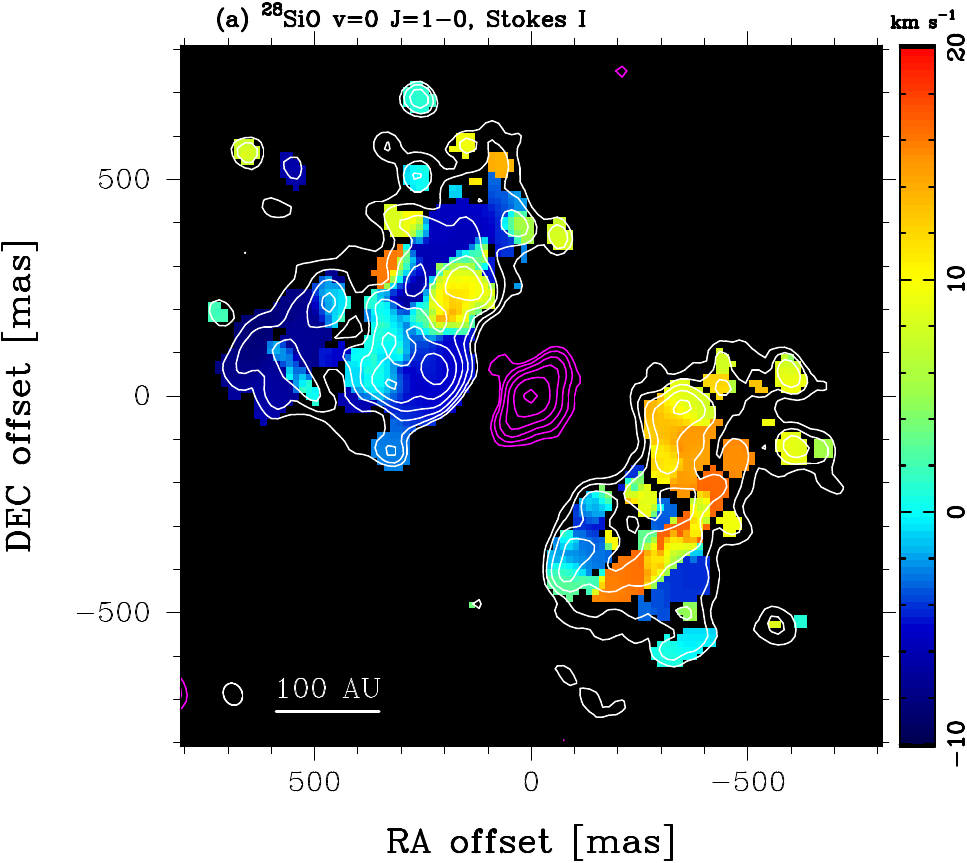}
\hspace{5mm}
\includegraphics[width=7cm]{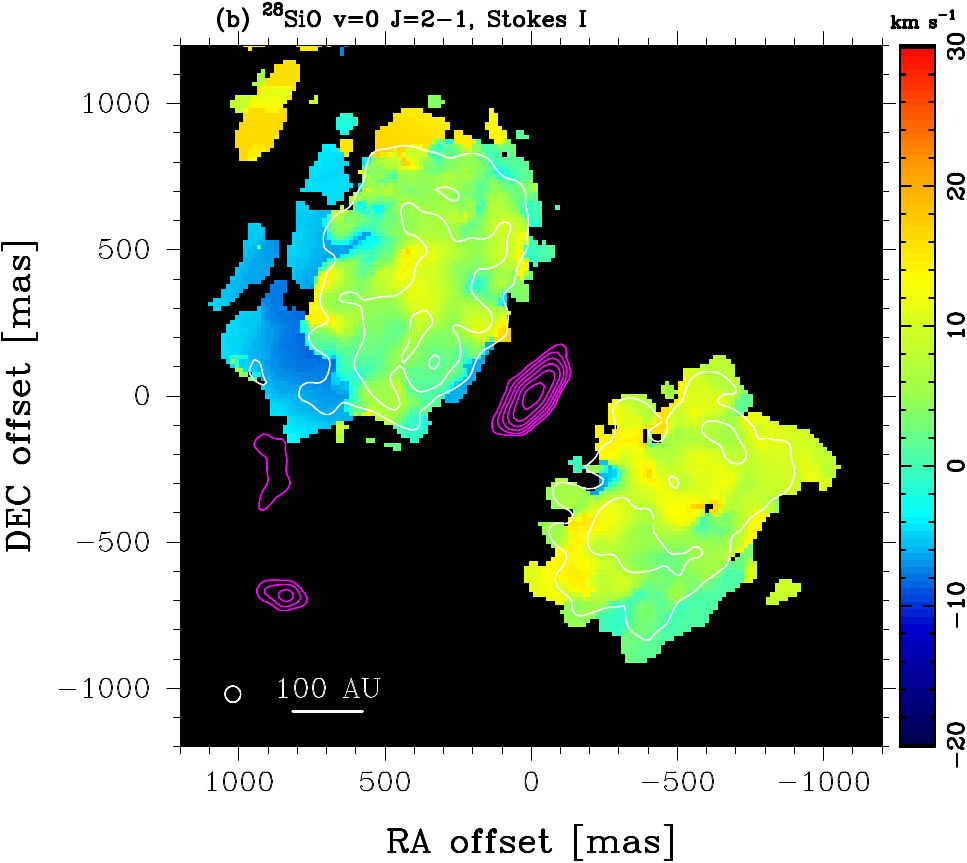}
\caption{Moment maps of (a) $^{28}$SiO $v$=0 $J$=1-0 and (b) $^{28}$SiO $v$=0 $J$=2-1 lines.
Moment maps are produced by using the velocity range from -10 to 20~km~s$^{-1}$ and from -20 to 30~km~s$^{-1}$ for the $J$=1-0 and 2-1 lines, respectively. 
Note that the $^{28}$SiO $v$=0 $J$=1-0 map (left panel) is plotted with a smaller box as it has smaller size of distribution. 
White contours show moment 0 and color scalesindicate moment 1 of the $^{28}$SiO lines. 
The contour  levels are 4, 8, 16, 32, 64, 128, and 256 times the rms noise levels of 62~mJy~beam$^{-1}$~km~s$^{-1}$ for $^{28}$SiO $v$=0 $J$=1-0 and  182~mJy~beam$^{-1}$~km~s$^{-1}$ for $^{28}$SiO $v$=0 $J$=2-1. 
Magenta contours show continuum emission at 43~GHz (left) and 96~GHz (right), with the contour levels of 4, 8, 16, 32, 64, 128 and 256 times the rms noise levels of 0.017~mJy~beam$^{-1}$ and 0.070~mJy~beam$^{-1}$, respectively. }
\label{fig-moment}
\end{center}
\end{figure*}

\begin{figure}[ht]
\begin{center}
\includegraphics[height=7cm]{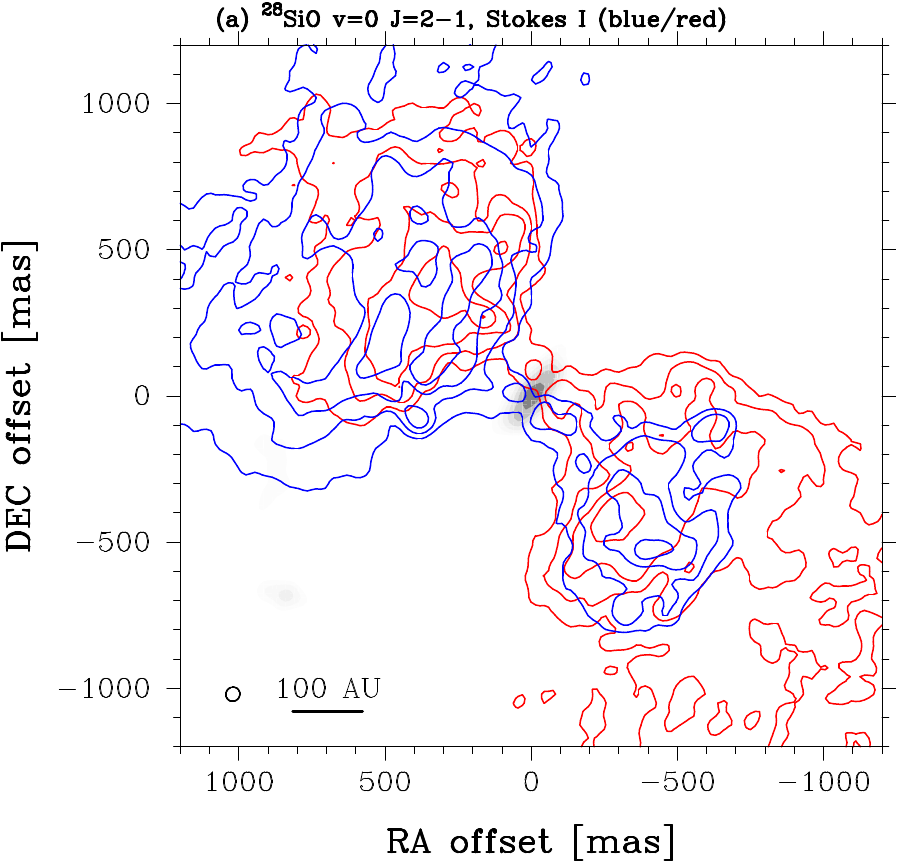}
\hspace{5mm}
\includegraphics[height=7cm]{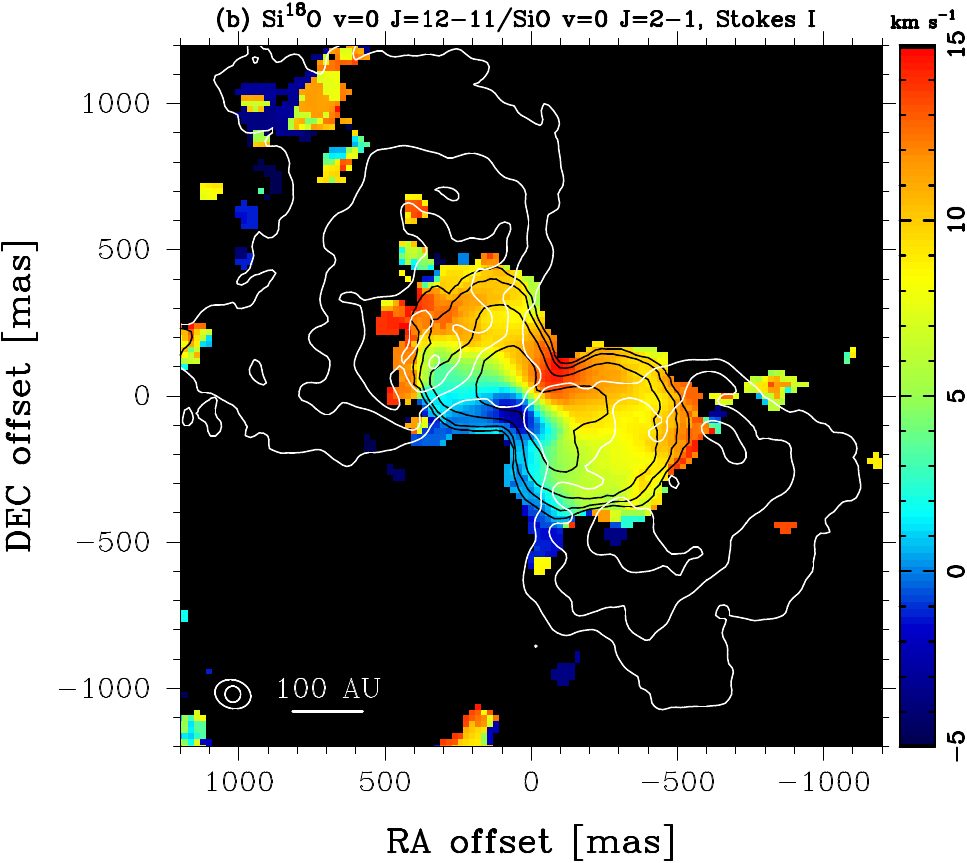}
\caption{
(a) The $^{28}$SiO $v$=0 $J$=2-1 line showing red- and blue-shifted emission extending from SrcI into the outer lobes.
Red contours shows the red-shifted components integrated from 10 to 30~km~s$^{-1}$ with the contour levels of 4, 8, 16, 32, and 64 times the rms noise level of 1.07~mJy~beam$^{-1}$~km~s$^{-1}$.
Blue contours shows the blue-shifted components integrated from -20 to 0~km~s$^{-1}$ with the contour levels of 4, 8, 16, and 32 times the rms noise level of 0.83~mJy~beam$^{-1}$~km~s$^{-1}$.
Gray scale map shows the 96~GHz continuum emission with the level of 4, 8, 16, 32, 64, and 128 times the rms noise level of 0.070~mJy~beam$^{-1}$.
(b) Moment 0 and 1 maps of the Stokes I of Si$^{18}$O $J$=12-11 line at 484.056~GHz \citep{Hirota2017}, with the black contour levels of 2, 4, 8, and 16 times the rms noise level of 0.56~Jy~beam$^{-1}$~km~s$^{-1}$.
White contours show the moment 0 image of the Stokes I of the $^{28}$SiO $v$=0 $J$=2-1 line, with the contour levels of 2, 4, 8, and 16 times 182~mJy~beam$^{-1}$~km~s$^{-1}$. 
}
\label{fig-28SiO21_bluered}   
\end{center}
\end{figure}

\begin{figure*}[th]
\begin{center}
\includegraphics[width=15.0cm]{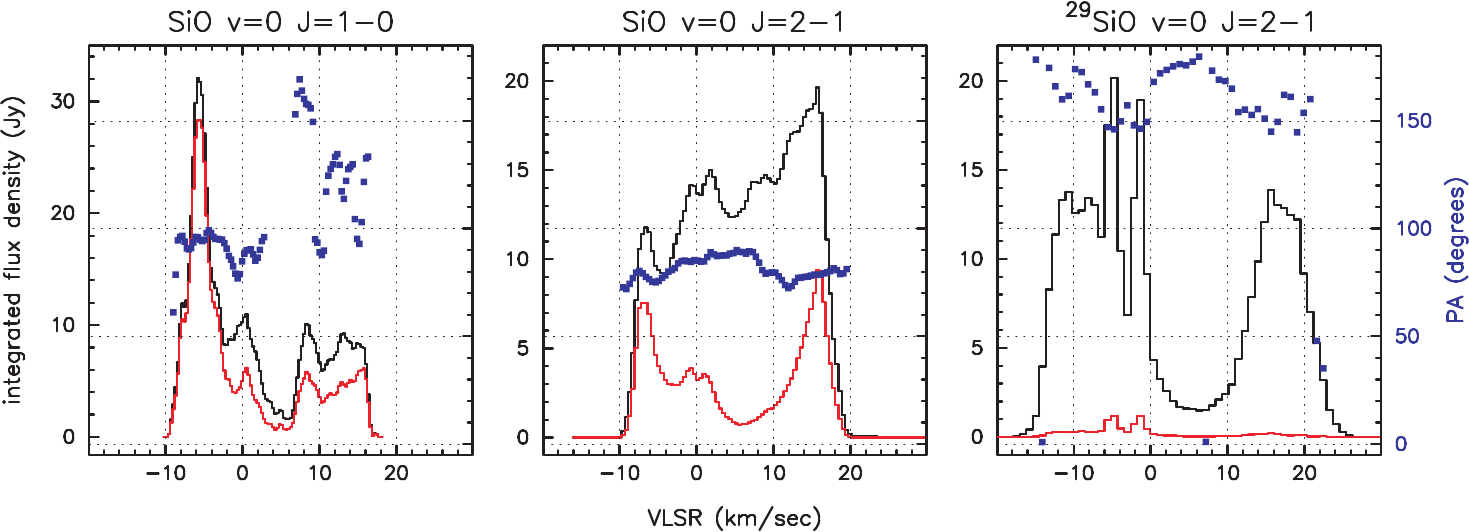}
\hspace{5mm}
\caption{Source-averaged spectra of total and linearly polarized flux density (black and red histograms, respectively), and electric vector position angle (blue squares) for the three SiO $v$=0 lines discussed in this paper.  The flux densities were integrated over boxes centered on SrcI that covered the emission regions; the box was 0.4\arcsec$\times$0.4\arcsec\ for $^{29}$SiO $J$=2-1, 2\arcsec$\times$2\arcsec\ for $^{28}$SiO $J$=1-0, and 4\arcsec$\times$4\arcsec\ for $^{28}$SiO $J$=2-1.  
The linearly polarized flux densities and position angles were obtained from vector averages of Stokes Q and U.   
Pixels with Stokes I less than a dynamic range cutoff (0.03~Jy~beam$^{-1}$ for $^{28}$SiO $J$=2-1, 0.05~Jy~beam$^{-1}$ for $^{29}$SiO $J$=2-1, and 0.2~Jy~beam$^{-1}$ for $^{28}$SiO $J$=1-0) were masked before creating the vector averages.
In the spectra, the signal-to-noise ratio of each linear polarization emission is larger than 10 for most of the channels which corresponds to the error in the polarization angle of less than $\sim$3~degrees. 
It can be hardly seen in the plot and hence, the error bar is not shown. 
}
\label{fig-sp}
\end{center}
\end{figure*}

\begin{figure*}[th]
\begin{center}
\includegraphics[width=7cm]{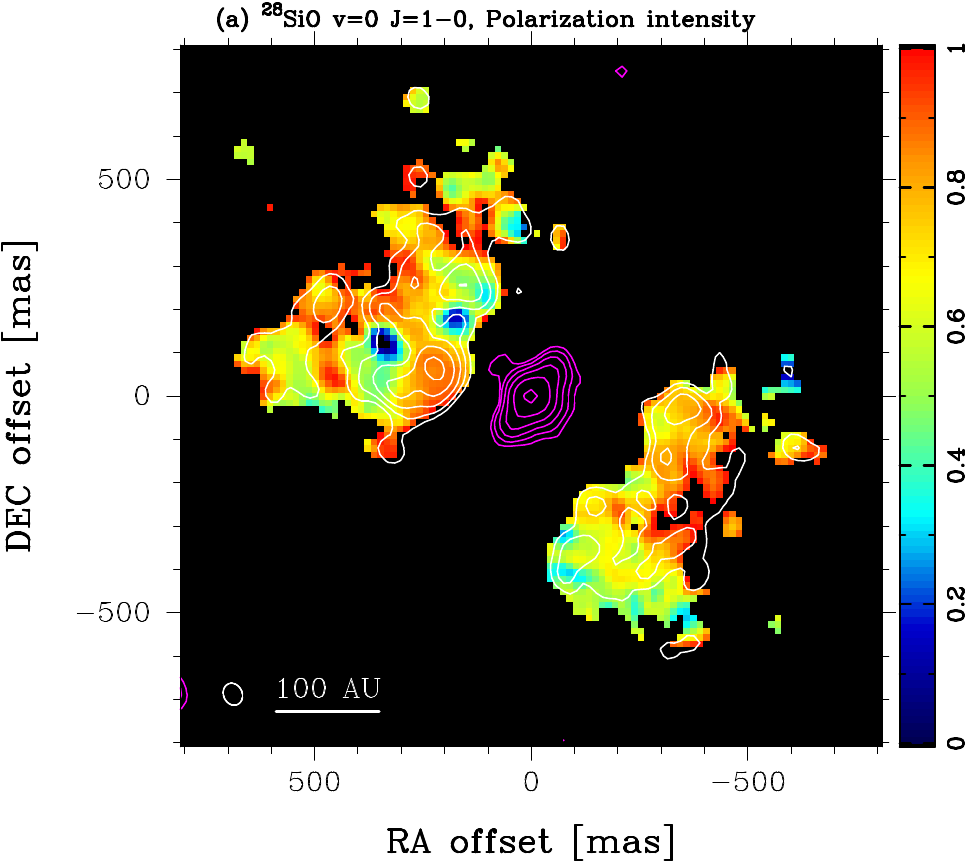}
\hspace{5mm}
\includegraphics[width=7cm]{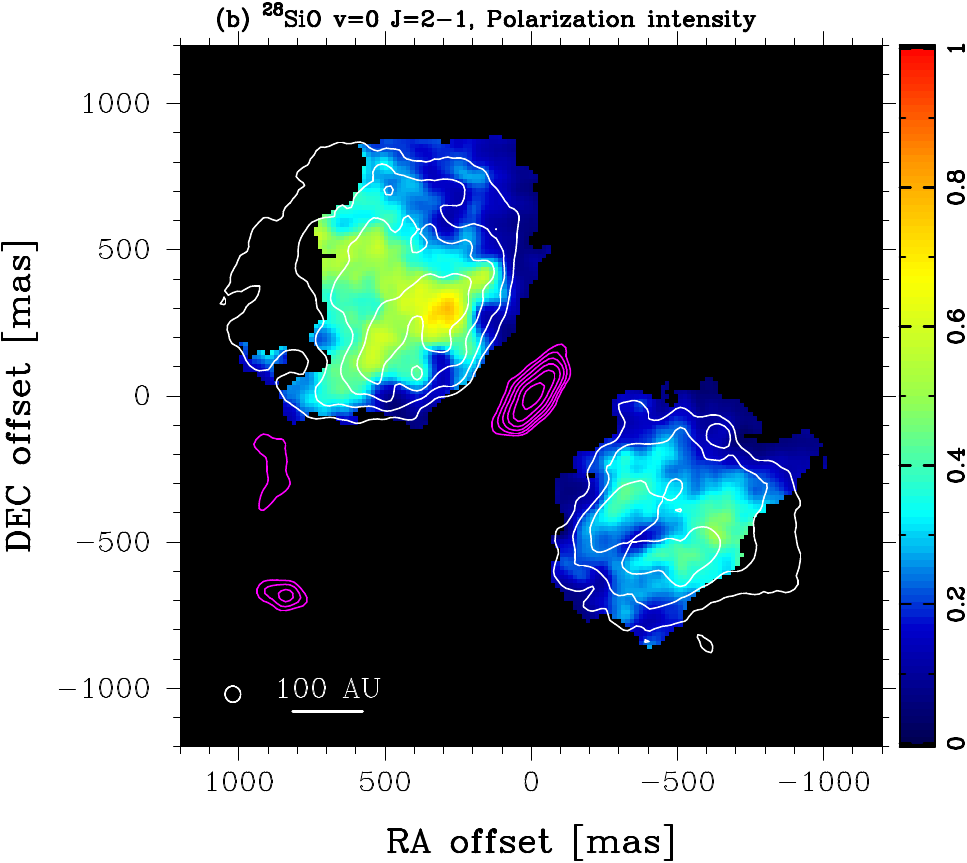}
\caption{Moment~0 of linear polarization intensity maps (white contour) and channel-averaged linear polarization fraction (color) maps of (a) $^{28}$SiO $v$=0 $J$=1-0 and (b) $^{28}$SiO $v$=0 $J$=2-1.
Moment maps are produced by using the velocity range from -10 to 20~km~s$^{-1}$ and from -20 to 30~km~s$^{-1}$ for the $J$=1-0 and 2-1 lines, respectively. 
Note that the $^{28}$SiO $v$=0 $J$=1-0 map (left panel) is plotted with a smaller box. 
The contour  levels are 4, 8, 16, 32, and 64 times the rms noise levels of 136~mJy~beam$^{-1}$~km~s$^{-1}$ for $^{28}$SiO $v$=0 $J$=1-0 and 27~mJy~beam$^{-1}$~km~s$^{-1}$ for $^{28}$SiO $v$=0 $J$=2-1. 
Magenta contours show the continuum emission as presented in Figure \ref{fig-moment}. }
\label{fig-ratio}
\end{center}
\end{figure*}

Figure \ref{fig-ratio} shows the moment 0 maps of the linear polarization intensities of the $v$=0 $J$=1-0 and $v$=0 $J$=2-1 lines of $^{28}$SiO. 
Again, the velocity range of the map is defined based on spectra (see Figure \ref{fig-spall} in Appendix \ref{secA-sp}). 
The rotating column shown in Figure~\ref{fig-28SiO21_bluered} is not significantly polarized and so does not appear in these images.
Polarized emission appears abruptly where the column expands into the extended lobes. 
We note that the rms noise level of the moment 0 map of the $^{28}$SiO $v$=0 $J$=2-1 line is higher than that of the $J$=1-0 line (Figure \ref{fig-moment}) while the linear polarization intensity map of the $^{28}$SiO $v$=0 $J$=2-1 line shows lower rms noise level than that of the $J$=1-0 line (Figure \ref{fig-ratio}). 
This is because the extended thermal emission component is more dominant in the Stokes I of the $^{28}$SiO $v$=0 $J$=2-1 line than that of the $J$=1-0 line, which produces the stronger sidelobe and hence, the higher rms level. 
In contrast, the polarization intensity map of the $^{28}$SiO $v$=0 $J$=2-1 line is free from such extended thermal emission. 
Thus, possible effects of sidelobes would be reduced, similar to both the Stokes I and polarization intensity maps of the $^{28}$SiO $v$=0 $J$=1-0 line. 

On average, the fractional linear polarizations of the $^{28}$SiO $v$=0 $J$=1-0 and 2-1 lines are $\sim$50-70\% and $\sim$20-50\%, respectively. 
The maximum fractional linear polarizations in individual channel maps (Appendix \ref{secA-map}) are as high as 90\% and 80\% for the $J$=1-0 and $J$=2-1 lines, respectively. 
We caution, however, that fractional linear polarizations derived from our aperture synthesis images may be overestimated if unpolarized thermal emission originates from a larger region than the polarized maser spots.  
Because of the lack of short baselines in the high resolution VLA and ALMA configurations, a greater fraction of the thermal emission will then be resolved out by  observations.

Theoretical models \citep{Watson2009, Perez-Sanchez2013, Lankhaar2019} predict that the fractional linear polarization of the $v$=0 SiO masers should decrease as the rotational quantum number (the angular momentum), $J$, increases.  Our observational results are consistent with this prediction.

\begin{figure*}[th]
\begin{center}
\includegraphics[width=7cm]{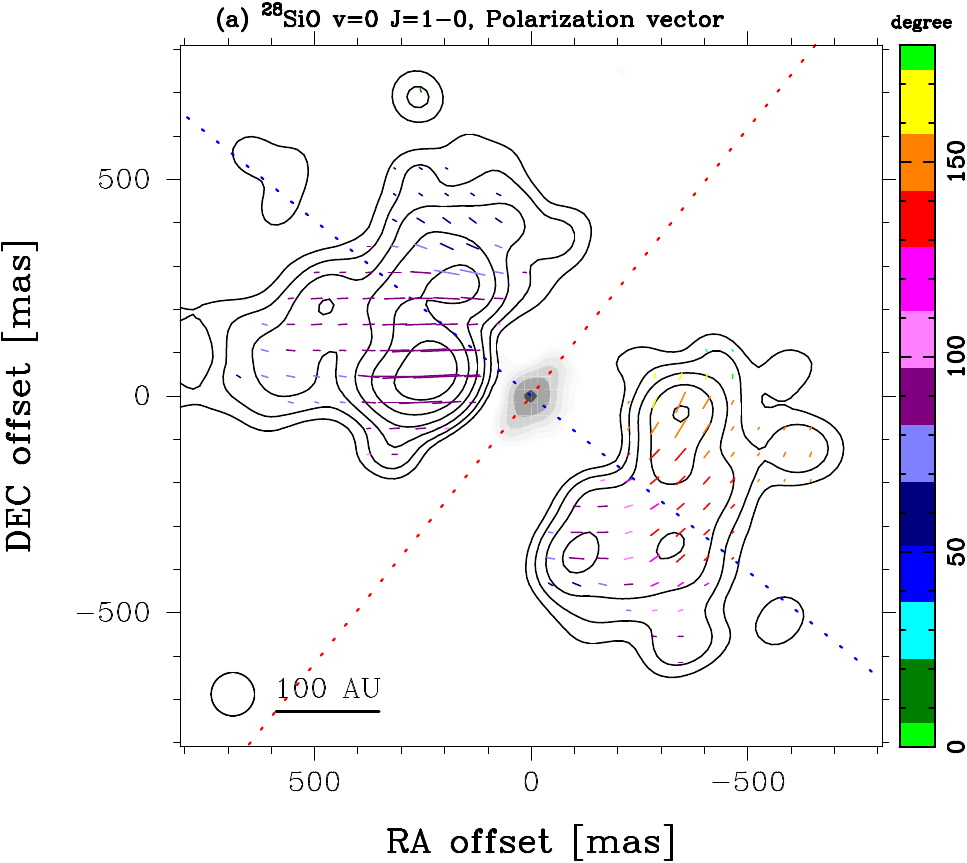}
\hspace{5mm}
\includegraphics[width=7cm]{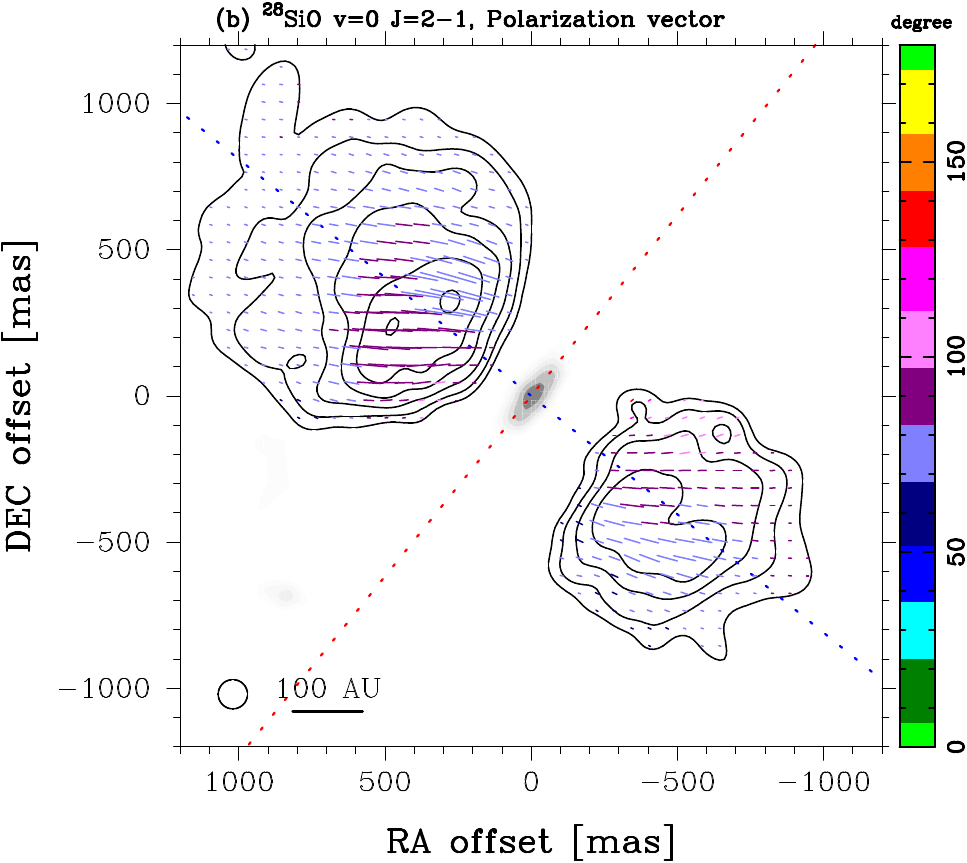}
\caption{Polarization vector of (a) $^{28}$SiO $v$=0 $J$=1-0 and (b) $^{28}$SiO $v$=0 $J$=2-1 lines averaged over the velocity range from -10 to 20~km~s$^{-1}$ superposed on moment 0 of linear polarization intensity. 
The maps are deconvolved with the 0.1\arcsec \ circular beam. 
Note that the $^{28}$SiO $v$=0 $J$=1-0 map (left panel) is plotted with a smaller box. 
The contour  levels are 4, 8, 16, 32, 64, and 128 times the rms noise levels of 108~mJy~beam$^{-1}$~km~s$^{-1}$ for $^{28}$SiO $v$=0 $J$=1-0 and 26~mJy~beam$^{-1}$~km~s$^{-1}$ for $^{28}$SiO $v$=0 $J$=2-1. 
Color-coded lines show position angle of the polarization vectors as indicated in the vertical bar at the right of the panel. 
The length of the polarization vector is proportional to the linear polarization intensity. 
The error in the polarization angle is smaller than 7~degrees for the linear polarization intensity higher than 4$\sigma$. 
The blue and red dashed lines indicate the position angle of the outflow axis (51~degrees) and disk midplane (141~degrees), respectively \citep{Plambeck2016}. 
Gray scale maps show continuum emission at 43~GHz (left) and 96~GHz (right). 
}
\label{fig-polmap}
\end{center}
\end{figure*}

Figure \ref{fig-polmap} plots the velocity-averaged linear polarization vectors, color coded to show polarization angle, with lengths proportional to the linearly polarized intensities. 
The linear polarization distributions are averaged over the velocity interval from $-10$~km~s$^{-1}$ to 20~km~s$^{-1}$.
We used a fixed velocity range for these averaged (single-channel) polarization angle/ratio images to compare the same velocity ranges for both $J$=1-0 and 2-1 lines.
These ranges are slightly narrower than those used for moment maps (Figures \ref{fig-moment} and \ref{fig-ratio}). 
To optimize the comparison of the $J$=1-0 and $J$=2-1 transitions, both images were convolved to 0.1\arcsec\ resolution.
This attenuates small scale structures sampled by the longest uv-spacings and more clearly reveals the overall polarization structure. 
The error in the polarization angle PA can be calculated by $\Delta{\rm (PA)}=\Delta P/2P$ in unit of radian where $P$ and $\Delta P$ are the linear polarization intensity and its rms noise level, respectively.  
For the polarization vector maps in Figure \ref{fig-polmap}, we plot the vectors with the signal-to-noise ratio of 4 or higher, and hence, the errors in the polarization angle, $\Delta{\rm (PA)}$ are smaller than 7~degrees. 

One can see systematic polarization patterns in both $J$=1-0 and $J$=2-1 lines but the overall trends differ. 
For the $J$=1-0 transition, the NE and SW lobes show different direction of the linear polarization. 
The averaged polarization angles in the NE lobe are $\sim$30-120~degrees (from light blue to magenta); polarization vectors more distant from the outflow axis are aligned with this axis while those in the central part of the lobe are rotated by $\sim$45~degrees from the axis.
In the SW lobe, on the other hand, the averaged polarization angle ranges from 30~degrees in the south to 170~degrees in the north (from light blue to yellow), gradually changing from parallel to perpendicular with respect to the polarization in the NE lobe. 
The mean polarization angle and its standard deviation are 81~degrees and 18~degrees, respectively, in the NE lobe, and 119~degrees and 34~degrees, respectively, in the SW lobe. 

In contrast, the $J$=2-1 transition shows a more ordered pattern of the polarization angles of 60-110~degrees and 50-130~degrees in the NE and SW lobes, respectively.  
The mean and standard deviation of polarization angles are 76~degrees and 7~degrees, respectively, in the NE lobe, and 81~degrees and 14~degrees, respectively, in the SW lobe.
In the central part of the NE lobe just south of the outflow axis, the polarization angles are offset by $\sim$45~degrees from those of the other parts, which are almost parallel to the outflow axis; this property is in good agreement with the $v$=0 $J$=1-0 line. 
In the SW lobe, there is a gradual change in the polarization angles toward the northern edge of the lobe where the vectors are inclined by $\sim$45~degrees from those of the outflow axis, similar to the center of the NE lobe. 
Except for these zones, the mean and standard deviation of the polarization angles in both NE and SW lobe are very similar. 

The 0.1\arcsec\ resolution $J$=2-1  linear polarization image presented in Figure~\ref{fig-polmap} is consistent with the 0.5\arcsec\ polarization image of the same transition obtained with the BIMA array in 2001 \citep[][Figure 6a]{Plambeck2003}. 
The Stokes I and linearly polarized intensities of the $^{28}$SiO $v$=0 $J$=2-1 transition that we observed for this line (Figure \ref{fig-sp}) also are comparable to those obtained with BIMA \citep[][Figure 5]{Plambeck2003}.

\subsubsection{Circular polarization}

Figures \ref{fig-vmap} shows the Stokes V maps integrated from -10 to 20~km~s$^{-1}$.  
Boxes in these figures indicate the positions of maser peaks for which Stokes I, linearly polarized intensity, and Stokes V spectra are plotted in Figure \ref{fig-IPV}. 
The circular polarization varies rapidly as a function of velocity and position.
Typical fractional circular polarizations at the brightest spectral features are 10-40\% for the $J$=1-0 transition and 1-10\% for the $J$=2-1 transition.  
While ALMA did not guarantee the accuracy of circular polarization calibration in Cycle 5 (ALMA Proposer's Guide$^{9}$), the observed magnitude of Stokes V is so large that it seems unlikely to be explained only by an instrumental effect.  

Although the recent model calculations by \citet{Lankhaar2019} predicted the highest circular polarization is associated with high linear polarization, the highest fractional circular polarizations do not necessarily coincide with the strongest maser peaks or with the positions of the highest fractional linear polarization (see Figure \ref{fig-vmap} and channel maps in Appendix \ref{secA-map}).  

\subsection{$^{29}$SiO $v$=0 $J$=2-1 line}
\label{sec3-3}

Figure \ref{fig-sp} also shows the spectrum of the $^{29}$SiO $v$=0 $J$=2-1 line.  
We note that the fractional linear polarization of this line is much lower than that of the $^{28}$SiO line.
The peak brightness temperature of 1.3$\times$10$^{6}$~K at the LSR velocity of $-1$~km~s$^{-1}$ indicates strong maser emission, as first suggested by \citet{Baudry1998}. 
Figure \ref{fig-29sio} shows the moment maps and polarization vector maps for this transition. 
The $^{29}$SiO emission region has a size of $<$0.5\arcsec, significantly smaller than those of the $^{28}$SiO $v$=0 lines. 
As shown in Figure~\ref{fig-29sio}, there is a clear velocity gradient in the northwest-southeast direction along the disk midplane due to rotation. 
The polarization vectors integrated over all the velocity components are nearly parallel with the midplane of the continuum disk toward its eastern and western margins, but suggest a radial distribution in the northern and southern regions. 
The polarization structure is more clearly seen by plotting the blue- and redshifted components separately, as in Figure~\ref{fig-29sio-bluered2}.
More detailed structures are seen in the channel map in Appendix \ref{secA-map}. 
We could not detect significant circular polarization of the $^{29}$SiO $v$=0 $J$=2-1 line. 
The Stokes V map with 30~km~s$^{-1}$-resolution shows peaks coincident with the Stokes I peaks at levels of 0.2-0.6\% of the Stokes I intensities. 
These levels are consistent with the expected instrumental residuals. 

\section{Discussion}
\label{sec4}

In this section, we discuss our analysis of the data in terms of the magnetic field structure traced by the polarized SiO maser emissions.

\subsection{Physical conditions for polarized masers}
\label{sec4-1}

Theories of maser polarization \citep{Goldreich1973, Watson2009, Perez-Sanchez2013, Lankhaar2019} predict that the maser radiation will be polarized parallel or perpendicular to the magnetic field direction in the plane of the sky as long as $g\Omega/2\pi \gg R$, where $g\Omega/2\pi$ is the Zeeman splitting in unit of Hertz, and $R$ is the stimulated emission rate per second, given by \citep{Goldsmith1972}
\begin{equation}
R =  B_{J+1, J}U = \frac{8 \pi^{3} \mu^{2}}{3 h^2}\frac{(J+1)}{(2J+3)} \frac{2kT_{B}}{\lambda^2} \frac{d\Omega_{b}}{c}.
\label{eq-r}
\end{equation}
Here $B_{J+1, J}$ is the Einstein B-coefficient for the $J+1 \rightarrow J$ transition, $U$ is the radiation energy density, $\mu$ is the dipole moment, $T_{B}$ is the brightness temperature, $\lambda$ is the wavelength, and $d\Omega_{b}$ is the beaming angle of the maser transition. 
There is no way of measuring the beaming angle $d\Omega_{b}$; probably it is in the range 0.01-1 steradian.  One may, however, obtain a conservative upper limit for $R$ by assuming $d\Omega_{b} = 4\pi$ steradians (isotropic radiation). 
Using this upper limit, \citet{Plambeck2003} verified that $g\Omega/2\pi > R$ for BIMA observations of the $^{28}$SiO $v$=0 $J$=2-1 masers ($\lambda$=0.35~cm), assuming a magnetic field strength of $B=1$~milli-Gauss (mG), $g\Omega/2 \pi=0.2$~Hz for 1~mG, $\mu=3.1$~Debye \citep{Raymonda1970}, and $T_{B}$=2000~K.

We estimated the stimulated emission rate $R$ for the polarized masers listed in Table \ref{tab-line} using the same parameters employed by \citet{Plambeck2003}, but with revised brightness temperatures given by the new higher resolution data.  
Equation (\ref{eq-r}) can be expressed as 
\begin{equation}
R  \sim  1.4 \left(\frac{T_{B}}{10^{6}~\rm{[K]}}\right) {\rm{sec}}^{-1} {\rm{\ for \ }} J=1-0
\end{equation}
and
\begin{equation}
R  \sim  7 \left(\frac{T_{B}}{10^{6}~\rm{[K]}}\right) {\rm{sec}}^{-1} {\rm{\ for \ }} J=2-1.
\end{equation}
The magnitude of the Zeeman splitting, $g \Omega/2\pi B = -230$~Hz~G$^{-1}$, can be derived from the relationship $\mu_{N}gBm = \hbar g\Omega m/2$, where $g$ is the Land\'e g-factor of the SiO line, $-0.15$ \citep{Davis1974}, $\mu_{N}$ is the nuclear magneton, and $m$ is the quantum number of the magnetic substate. 
The condition $g\Omega/2\pi>R$ is satisfied with the magnetic field strengths of $B>$13~mG ($^{28}$SiO $v$=0 $J$=1-0), 1.5~mG ($^{28}$SiO $v$=0 $J$=2-1), and 40~mG ($^{29}$SiO $v$=0 $J$=2-1) for the maximum brightness temperatures of the Stokes I as summarized in Table \ref{tab-line}. 
It is unlikely that the brightest maser spots are radiating isotropically, however, so the inequality probably is satisfied for magnetic field strengths $\gtrsim 1$~mG.

For larger stimulated emission rates, $R \gtrsim\, g\Omega/2\pi$, the maser polarization no longer is guaranteed to be parallel or perpendicular to the magnetic field \citep{Nedoluha1990,Lankhaar2019}.
Thus, the $v$=1 and $v$=2 masers, which are 1 to 2 orders of magnitude brighter than the $v$=0 masers, may not be good probes of the magnetic field direction.  \citet{Plambeck2003} found that the polarization position angles of the SiO $v$=1 $J$=2-1 masers near SrcI were time variable, hence considered them an unreliable tracer of the magnetic field morphology.  
However, these 0.5\arcsec\ resolution observations did not resolve individual maser spots, so it is possible that the position angle changes were caused by variations in the relative brightness of individual masers within the synthesized beam.  In this paper we consider only the $v$=0 maser polarizations, which are likely to satisfy the more stringent $g\Omega/2\pi>R$ condition. 

The net decay rates of the $^{28}$SiO $v$=0 $J$=1-0 and 2-1 masers are determined by either the collisional excitation/deexcitation rate, $n$(H$_{2}$)$R_{ij}$ or the spontaneous emission rate, $A_{ij}$, where $n$(H$_{2}$), $R_{ij}$, and $A_{ij}$ are the hydrogen molecule density, collisional rate coefficient, and Einstein A-coefficient, respectively. 
Radiative transfer calculations for the $J$=1-0 $^{28}$SiO masers by \citet{Goddi2009a} suggest that the $v$=0 transition is inverted at densities $n$(H$_{2}$)$<$10$^7$ cm$^{-3}$. 
The masers are quenched when the collision rate becomes comparable to the stimulated emission rate. 
This should set a firm upper limit on gas density, and hence, we adopt a density of $10^{6}$ cm$^{-3}$ in the $^{28}$SiO $v$=0 emission region. 
From the plots in \citet{Palov2006}, we estimate that the total collisional rates from the $v$=0 $J$=1 and $J$=2 levels into all other rotational levels are of order 10$^{-9}$~cm$^{3}$~s$^{-1}$, which gives a collisional loss rate $\Gamma$ of order 10$^{-3}$~s$^{-1}$.
The Einstein-A coefficients, 3.0$\times10^{-6}$~s$^{-1}$ and 2.9$\times10^{-5}$~s$^{-1}$ for the $^{28}$SiO $v$=0 $J$=1-0 and 2-1 lines, respectively \citep{Barton2013}, are smaller than the collisional rate, and hence, the dominant source of the decay rate is the collisional excitation/deexcitation. 
The derived decay rate of $\Gamma$ is smaller than the Zeeman splitting $g\Omega/2\pi$. 
It is significantly smaller than $R$, so the masers are in the saturated regime.

\subsection{$^{28}$SiO $v$=0 lines}
\label{sec4-2}

We detected high linear polarization in both the $J$=1-0 and $J$=2-1 transitions which provides us with the data to map the polarization distribution across the extended lobes of the $^{28}$SiO outflow from SrcI.  
The observed fractional linear polarizations of $\sim$50-70\% and $\sim$20-50\% (Figure \ref{fig-ratio}) are not easily explained unless the masers are anisotropically pumped \citep{Watson2009}.
Nevertheless, as long as $g\Omega/2\pi > R$ the overall pattern of the polarization vector maps is expected to be parallel or perpendicular to the magnetic field, and hence, useful as a probe of the field direction. 

\subsubsection{Comparison with previous polarization measurements}
\label{sec4-2-1}

The polarized emission of the 43~GHz $v$=0 $J$=1-0 masers was first detected by \citet{Tsuboi1996} with the NRO 45~m radio telescope at a spatial resolution of 40\arcsec. 
Figure~2 of \citet{Tsuboi1996} shows a series of spectra at polarizer position angles of 0-157~degrees. 
The peak at -7~km~s$^{-1}$ is strongest for a polarizer position angle of 67.5~degrees, and almost disappears at 157.5~degrees as noted in \citet{Tsuboi1996}. 
However, the position angles shown in their Figure~4 are mostly in the range 140-180~degrees.  The polarization angles in our integrated spectrum (Figure~\ref{fig-sp}) are $\sim$90~degrees, consistent with their Figure 2, and not with Figure 4. 
This is because \citet{Tsuboi1996} did not use a conventional definition of the polarization angle consistently (Tsuboi, M., private communication).

\subsubsection{Faraday rotation}
 \label{sec4-2-2}

Figure \ref{fig-polmap} shows that the polarization angles of both the $^{28}$SiO $v$=0 $J$=1-0 and $J$=2-1 lines vary smoothly across the source, but are rotated with respect to each other in some regions.  
As already discussed, the error in the polarization angle is proportional to the signal-to-noise ratio of the linear polarization intensity, $\Delta{\rm (PA)}=\Delta P/2P$, and is smaller than 7~degrees for vectors plotted in Figure \ref{fig-polmap}. 
It allows us to discuss systematic linear polarization structures for both lines. 
The difference in the $J$=1-0 and $J$=2-1 position angles is plotted in Figure \ref{fig-padiffrev}.  
The differences are slightly increasing toward the northwestern and southeastern sides of the NE lobe and the southwestern edge of the SW lobe. 

It is possible that the two SiO transitions originate in different volumes of gas with different magnetic field orientations.  
In Figure \ref{fig-sp}, the polarization angle of the $^{28}$SiO $v$=0 $J$=2-1 line shows a smooth profile, while that of $J$=1-0 shows a big jump from $\sim$90~degrees to $\sim$170~degrees around 5~km~s$^{-1}$. 
The $J$=1-0 line also shows highly scattered polarization angles from 10 to 20~km~s$^{-1}$, which is much larger than the observed uncertainty of $\sim$3~degrees for the integrated spectra (i.e. signal-to-noise ratios greater than 10). 

Such a difference would be caused by a stronger maser emission in the $J$=1-0 line as seen in the different spatial structures between $J$=1-0 and $J$=2-1 lines; the former line maps dominate compact/clumpy structures compared with those of latter ones. 
In particular, maser lines are sensitive to the physical conditions and geometry of the emission regions which may result in striking difference in the spatial distributions and hence, the profiles of the integrated spectra.
Nevertheless, similarity in the polarization angles in the NE lobe averaged over the entire velocity range of emission strongly supports the scenario that both $J$=1-0 and $J$=2-1 lines trace overall outflow structures in the same volume of gas. 

If both the $J$=1-0 and 2-1 lines originate from the same volume of gas, the different polarization structures could be explained by the propagation effects. 
Here, we consider the possibility that Faraday rotation causes the position angle discrepancies. 
Faraday rotation occurs when linearly polarized radiation travels through ionized gas with a magnetic field component along the line of sight.   
The plane of polarization is rotated by
\begin{equation}
\theta(\lambda) = \lambda^{2}~RM,
\label{eq-zeeman}
\end{equation}
where $\theta$ is rotation of the polarization angle in radians and $\lambda$ is wevelength in meters. 
The rotation measure $RM$ is given by 
\begin{eqnarray}
RM & = & \frac{e^{3}}{2 \pi m_{e}^{2} c^{4}} \int n_{e}B_{\parallel} dl \nonumber \\
  & = & 4.0\times10^{3} \int \left( \frac{n_{e}}{10^{3} \ \mbox{cm}^{-3}}\right) \left( \frac{B_{\parallel}}{\mbox{10 mG}}\right) d\left( \frac{l}{\mbox{100 au}}\right)  [\mbox{rad~m}^{-2}]. 
\end{eqnarray}
Here $n_{e}$ is the electron density in cm$^{-3}$ and $B_{\parallel}$ is the component of the magnetic field along the line-of-sight in mG. 
For the $^{28}$SiO $v$=0 $J$=1-0 and 2-1 lines with wavelengths of 0.71~cm and 0.35~cm, respectively, a polarization angle difference of 1~degree corresponds to $RM$ of 460~rad~m$^{-2}$. 

The maximum position angle difference of $\sim$60~degrees in the SW lobe corresponds to a rotation measure of 3$\times10^{4}$~rad~m$^{-2}$. 
On the other hand, the NE lobe shows smaller and mostly negative differences ranging from $\sim-20$~degrees to 10~degrees. 
The corresponding $RM$ is from $-1\times10^{4}$~rad~m$^{-2}$ to $5\times10^{3}$~rad~m$^{-2}$.
The result suggests that there could be Faraday rotation within the outflow and/or in the foreground field. 

To explain an apparent 60~degrees systematic difference in the position angles of the $J$=2-1 and $J$=1-0 transitions seen in older BIMA and NRO data, \citet{Plambeck2003} estimated an $RM$ of $3.3\times10^{4}$~rad~m$^{-2}$, and suggested that the Faraday rotation originated in the foreground H{\sc{ii}} region.  
As noted in section \ref{sec4-2-1}, however, the $J$=1-0 position angles derived from our new VLA observations differ substantially from the previously published NRO results, so it is no longer necessary to account for such a large position angle discrepancy.  
It also seems unlikely that the foreground H{\sc{ii}} region has a magnetic field reversal that coincides precisely with the small gap between the NE and SW outflow lobes.

While the foreground H{\sc{ii}} region may contribute to a smooth component of the Faraday rotation, the more plausible explanation for the structure in Faraday rotation is that it originates from ionized gas associated with the outflow, possibly from a radio jet that is marginally seen in the continuum map of SrcI \citep{Reid2007} or from molecular gas that is partially ionized by shocks. 
If the radio jet is responsible for the Faraday rotation, the ionization rate would be higher close to the outflow axis, whereas the observed RM is higher toward the edges of the SiO outflow lobes. Thus, the latter interpretation is more likely. 

If we assume that Faraday rotation occurs in a layer that is 1000~AU with a magnetic field strength of 10~mG, then the electron density in this layer would need to be 100~cm$^{-3}$ in order to introduce a polarization angle difference of 10~degrees between the $J$=1-0 and $J$=2-1 lines. 
If the total gas density is 10$^{6}$~cm$^{-3}$, as discussed in the previous section, this would correspond to a fractional ionization of 10$^{-4}$. 
The estimated electron density and fractional ionization are much higher than chemical model calculations for the shocked molecular gas \citep[e.g. $n_{e} \sim 10^{-3}$~cm$^{-3}$; ][]{Caselli1997}.
A deeper column of ambient gas may be responsible for the jump in the $J$=1-0  polarization angle  around 5~km~s$^{-1}$. 

If we assume that Faraday rotation is responsible for the difference in the $J$=1-0 and 2-1 position angles, then we can solve for the rotation measure at each position where both lines are detected, and compute the intrinsic $^{28}$SiO $v$=0 polarization angles.  
These are presented in Figure \ref{fig-pola0}. 
The polarization vectors in the SW lobe are almost parallel to the outflow axis while those in the NE lobe still are significantly inclined to this axis, by as much as 45~degrees in the eastern side.  
Note that the rotation corrections for the $J$=2-1 line are typically only 5 to 10~degrees and 20~degrees at maximum in the SW lobe, so in directions where the $J$=1-0 line is not detected it is reasonable to assume that the intrinsic angles are equal to the uncorrected the $J$=2-1 angles.

We rule out the possibility of much larger Faraday rotations that could produce the 1-0 and 2-1 position angle agreement by chance (e.g., $\sim$45\degr\ at 86~GHz and $\sim$4$\times$45\degr\ = 180\degr\ at 43 GHz) because these would require much higher electron column densities and/or magnetic fields.  
\citet{Rao1998} also estimated an upper limit on the RM toward the Orion Hot Core of 3$\times$10$^4$~rad~m$^{-2}$ based on the close agreement of dust polarization vectors at 3.3 and 1.3~mm wavelength.
Finally, the well ordered $^{29}$SiO $v$=0 polarization vector map with radial pattern close to the disk as discussed later (Figures \ref{fig-29sio} and \ref{fig-29sio-bluered2}) also suggests small rotations at 86~GHz.

\subsubsection{Foreground absorption or scattering?}
\label{sec4-2-3}

In principle, differential absorption by foreground magnetically aligned dust grains could rotate the observed maser polarizations, but we expect this to be unimportant because dust opacities at 43 and 86~GHz are small.  For masers in the ground vibrational state, absorption or scattering by SiO molecules in cold foreground gas is a bigger concern.
If these molecules are in a magnetic field, their absorption coefficients are likely to be polarization-sensitive.  Then, if the foreground magnetic field direction is twisted relative to the field in the outflow, the polarization of maser radiation propagating through this gas could be rotated. 

The high circular polarizations observed for the $v$=0 masers suggest that anisotropic resonant scattering \citep[ARS; ][]{Houde2013, Houde2014, Chamma2018} by foreground SiO could be significant, as discussed further in section~\ref{sec4-2-5}. 
Simple absorption will be important only if the collisional rate is comparable to or larger than the stimulated emission rate $R$, so that absorbed photons can be lost through collisional deexcitation.  Otherwise, a photon that is absorbed will quickly be reemitted in the same polarization state.  
The values we computed for $R$ in section~\ref{sec4-1} were orders of magnitude larger than the collisional rates, but these were upper limits that assumed isotropic maser radiation.  
If the masers are highly beamed, $R$ will be substantially smaller.  
For example, if the solid angle of the radiation is 0.01~steradians, $R\leq10^{-3}$~sec$^{-1}$, comparable to the collisional rate. 

If foreground absorption or scattering is significant, we expect to find greater discrepancies between the $J$=1-0 and 2-1 position angles at blueshifted velocities (V$_{\rm LSR} < 5$~km~s$^{-1}$), since absorption by foreground gas against the SrcI continuum is much more prominent at these velocities \citep{Plambeck2016}.  
We searched unsuccessfully for such a correlation.  
In addition, we found no obvious correlations with the projected distance to the SrcI disk midplane, or with the offset from the axis of the outflow, or with the fractional circular polarization.  
This, along with the generally good agreement of the $J$=1-0 and 2-1 position angles across most of the outflow, suggests that foreground effects do not significantly skew the measured polarization directions.

\subsubsection{Estimate of the magnetic field strength}
\label{sec4-2-4}

The large number of polarization vectors measured across the outflow lobes in the $^{28}$SiO $v$=0 $J$=2-1 line makes it feasible to estimate the magnetic field strength using the Davis-Chandresekhar-Fermi (DCF) method \citep{Davis1951,Chandrasekhar1953}. 
Normally the DCF method is applied to dust polarization observations.
The DCF method assumes that turbulent motions result in small-scale fluctuations of the magnetic field lines, depending on the field strength. 
The fluctuation of the magnetic field can be measured through that of the linear polarization direction of the dust continuum emission. 

If the same assumptions as used for the dust polarization are valid, we can apply the DCF method to the polarized SiO maser \citep[e.g.][]{Lee2018}. 
The outflow in SrcI becomes turbulent once it emerges from the rotating column near SrcI (e.g. Figure \ref{fig-28SiO21_bluered}). 
It is also clear from the channel maps (see FIgures B2 and B21 in Appendix \ref{secA-map}) that the velocity field in the lobes is pretty turbulent and not simple rotation. 
In this case, the large scale polarization structure could probe the overall magnetic field structure in the outflow lobes while the smaller scale fluctuation could reflect turbulence due to the interaction between the outflow and ambient gas in the extended lobes.
Thus, we apply the DCF method to the polarization structure of the SiO lines in the SrcI outflow. 

Following the formulation in \citet{Crutcher2004}, the plane of sky component of the magnetic field is given by
\begin{equation}
B_{pos} \sim 0.5 \sqrt{4\pi\rho}\ \frac{\delta{\rm V}}{\delta\phi_{\rm rad}} \sim 9.3 \frac{\sqrt{{n \rm(H}_2{\rm )}} \Delta{\rm V}}{\delta\phi_{\rm deg}}\ \mu{\rm G}
\label{eqn-DCF}
\end{equation}
where $\rho$ is the gas density, $\delta$V is the velocity dispersion along the line of sight, and $\delta\phi$ is the dispersion in the polarization position angles.  
The factor of 0.5 is the correction factor derived from numerical simulations by \citet{Ostriker2001}. 
On the right hand side, $\Delta$V = $(8\,{\rm ln} 2)^{1/2}\, \delta$V is the FWHM linewidth in units of km~s$^{-1}$, $n$(H$_2$) is the molecular hydrogen density in cm$^{-3}$, and $\delta\phi_{\rm deg}$ is the polarization angle dispersion in degrees.

Examination of the channel maps in Figures B4 and B23 in Appendix \ref{secA-map} shows that the polarization position angles are consistent from channel to channel, so we compute the polarization angle dispersion from a channel-averaged image. 
The dispersion of the polarization angle (without any correction for Faraday rotation) is 9.5~degrees.  The polarization angle variations are attributable in part to the large scale field geometry, which should be removed before using the DCF formula.  
\citet{Houde2009,Houde2011,Houde2016} have developed a formalism to remove the large scale structure using a 2-point correlation function; see \citet{Chuss2019} for an example. 
Here, however, we adopt the simpler strategy used by \citet{Pattle2017} in which a smoothed position angle map is subtracted from the high resolution map before the dispersion is calculated.

Accordingly, we convolved the SiO Stokes I, Q, and U images with an 0.5\arcsec \ FWHM Gaussian, generated a low resolution polarization angle map from these data, then subtracted this image from the 0.1\arcsec \ resolution image on a pixel-by-pixel basis.
The result is shown in Figure~\ref{fig-PAresid}, and a histogram of the residual polarization angle values is shown in Figure~\ref{fig-PAhisto}.  
Although the dispersion computed for all the data is 6.1~degrees, this is biased by outliers which most likely are due to incomplete removal of the large scale structure.  
In addition, the position angle uncertainty due to noise can be as large as $\Delta{\rm (PA)} = \sigma /(2 \sqrt{U^2+Q^2}) ~\sim 6$\degr\ with our choice of a 5$\sigma$ minimum cutoff in Q and U. Therefore, we adopt a somewhat smaller dispersion $\delta\phi$ = 4\degr, as illustrated by the red curve in Figure~\ref{fig-PAhisto}.  
If we analyze the data of the NE and SW lobes separately (see Figure \ref{fig-PAhisto}), we obtain the standard deviation of the polarization angle residuals of 4.8~degrees and 7.4~degrees, respectively. 
Obviously, the field estimates for the two lobes separately will differ by only about 50\%, and hence, we will employ the result for all the data. 

The velocity width of the $^{28}$SiO line is attributable to large scale outflow motions and rotation as well as turbulence, but the Stokes I images in Figures B2 and B21 in Appendix \ref{secA-map} suggest that these systematic motions make only a modest contribution to the $^{28}$SiO linewidths. 
Therefore, we adopt an FWHM linewidth $\Delta$V $\sim$ 10~km~s$^{-1}$.
The hydrogen density within the lobes is the final parameter needed to estimate the field strength. 
As discussed above, we adopt a density of $10^{6}$ cm$^{-3}$ in the $^{28}$SiO $v$=0 $J$=2-1 emission region in order to satisfy the maser pumping model \citep{Goddi2009a}. 

Inserting these estimates into equation~(\ref{eqn-DCF}) gives a plane of sky magnetic field strength of 23~mG.  
An estimate for the total field strength is $4B_{\rm los}/\pi$ \citep{Crutcher2004}, or $\sim$30~mG.
Assuming that the magnetic field strength  scales as the square root of the gas density, e.g.  $B \propto n^{0.49}$ \citep{Vlemmings2007} and $B \propto n^{0.65}$ \citep{Crutcher2012}, a magnetic field strength $\sim$30~mG is consistent with that estimated from OH maser observations \citep{Cohen2006} in a density $n$(H$_{2}$)$\sim10^7$~cm$^{-3}$, and with an estimate of 300~mG for the $v$=1 SiO masers close to SrcI \citep{Matthews2010} where density $n$(H$_{2}$)$\sim10^9$~cm$^{-3}$. 
We note that the above scaling laws come from density regimes of gravitationally contracting cores where the magnetic field is assumed to be well coupled with the gas. 
Although the derived field strength from the SiO data agrees with these relationships, they may not apply to the field in the outflow. 

A magnetic field of order 30~mG could explain why the bipolar outflow from SrcI is not swept back into two trailing arms by the ram pressure of the ambient molecular cloud as SrcI plows through the cloud at $\sim 12$~km~s$^{-1}$.  
The magnetic pressure associated with a 30~mG field is $B^2/8\pi \sim 4\times10^{-5}$~dynes~cm$^{-2}$, much larger than the ram pressure $\rho v^2 \sim 4 \times 10^{-6}$~dynes~cm$^{-2}$, assuming a molecular hydrogen density of 10$^{6}$~cm$^{-3}$ in the ambient gas.
The magnetic energy density balances the rotational energy density at 18~km~s$^{-1}$ and density 10$^{6}$~cm$^{-3}$. 
It offers support for the idea that sub-Keplerian rotation could account for underestimates of the central mass calculated from rotation curves of spectral lines close to the launching region of the outflow \citep[e.g.][]{Kim2008,Matthews2010,Plambeck2016,Hirota2017,Kim2019}.
Possibly the H$_2$O line used as a probe of the velocity field by \citet{Ginsburg2018} is confined more closely to the disk surface, and so is not confused by the outflow that has magnetic support.

\subsubsection{Circular Polarization}
\label{sec4-2-5}

\begin{figure}[thb]
\begin{center}
\includegraphics[width=7cm]{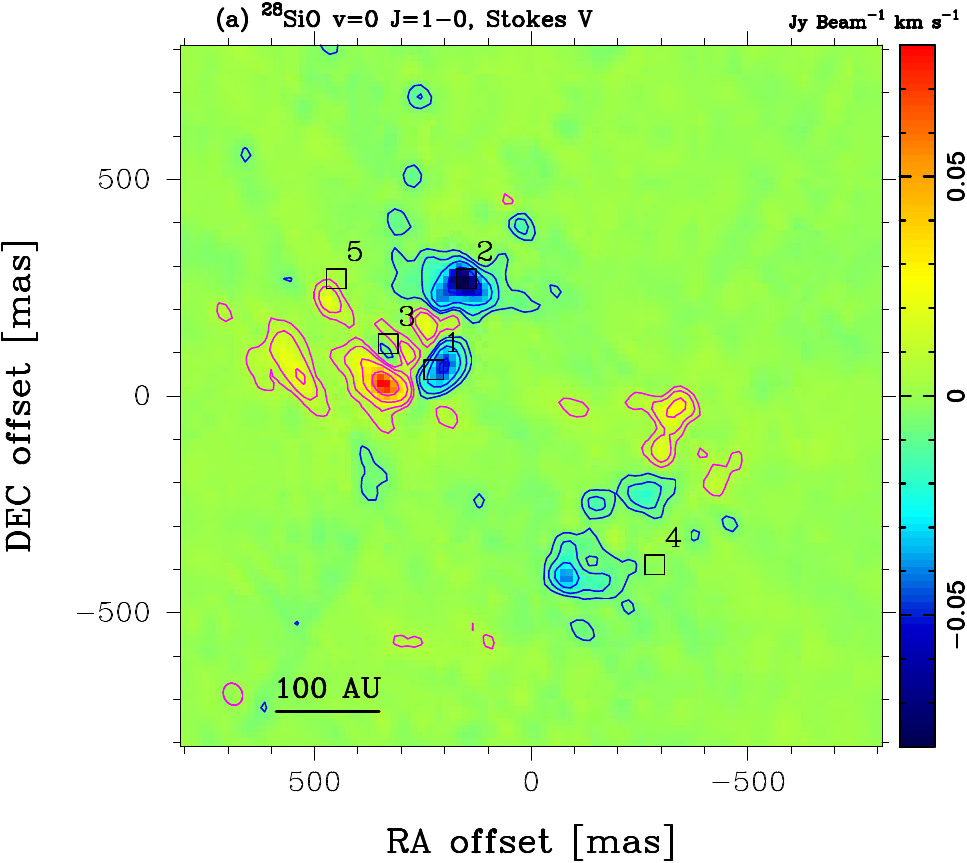}
\hspace{5mm}
\includegraphics[width=7cm]{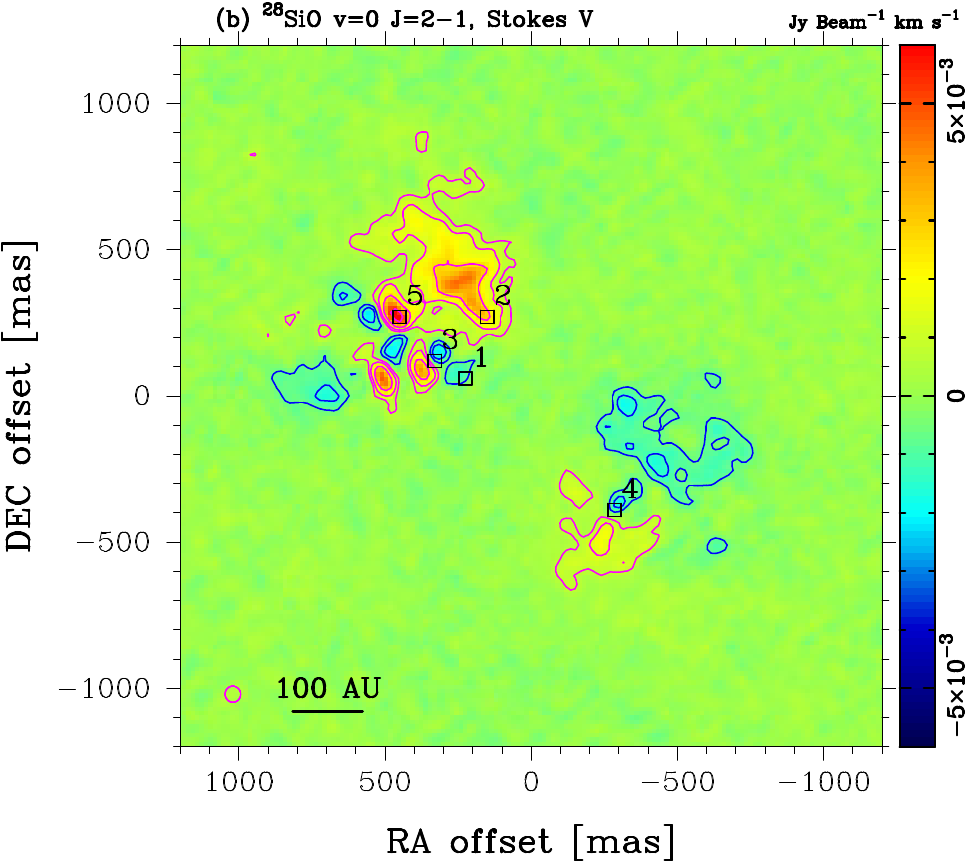}
\caption{Stokes V maps of (a) $^{28}$SiO $v$=0 $J$=1-0 and (b) $^{28}$SiO $v$=0 $J$=2-1 lines.  
Note that the $^{28}$SiO $v$=0 $J$=1-0 map (left panel) is plotted with a smaller box.
Stokes V maps are produced by averaging over the velocity range from -10 to 20~km~s$^{-1}$.
The contour  levels are $\pm$4, $\pm$8, $\pm$16, $\pm$32, and $\pm$64 times the rms noise levels of 1.44~mJy~beam$^{-1}$~km~s$^{-1}$ for $^{28}$SiO $v$=0 $J$=1-0 and 0.17~mJy~beam$^{-1}$~km~s$^{-1}$ for $^{28}$SiO $v$=0 $J$=2-1.
Magenta and blue solid lines show the positive and negative levels, respectively. 
Maser features used to extract spectra in Figure \ref{fig-IPV} are indicated with boxes. 
Peak 1; Stokes I and linear polarization intensity peak of $J$=1-0 in the NE lobe, 2; Stokes V negative minimum of $J$=1-0, 3; Stokes I peaks of $J$=2-1 in the NE lobe, 4; linear polarization intensity peaks of $J$=2-1 in the SW lobe, and 5; Stokes V positive maximum of $J$=2-1.}
\label{fig-vmap}
\end{center}
\end{figure}

\begin{figure}[bht]
\begin{center}
\includegraphics[width=6.5cm]{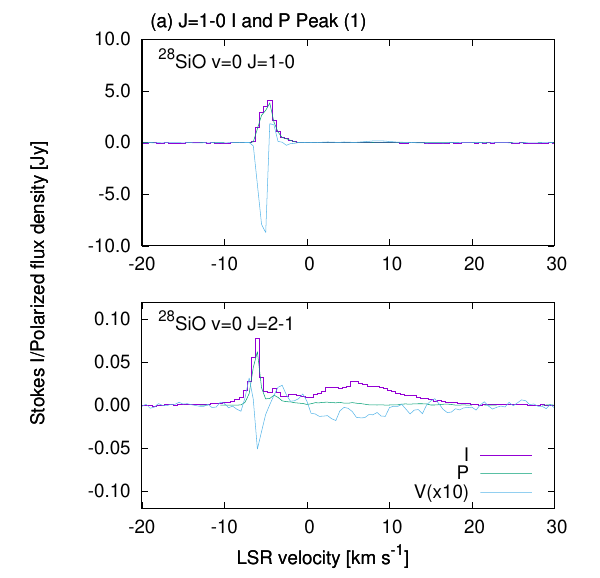}
\includegraphics[width=6.5cm]{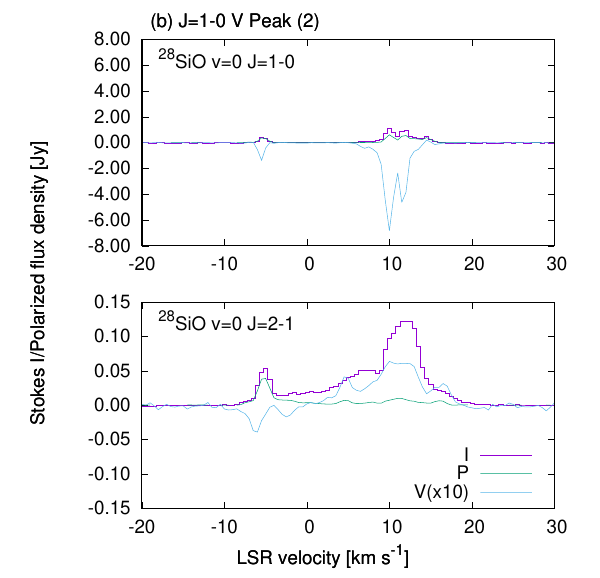} \\
\vspace{5mm}
\includegraphics[width=6.5cm]{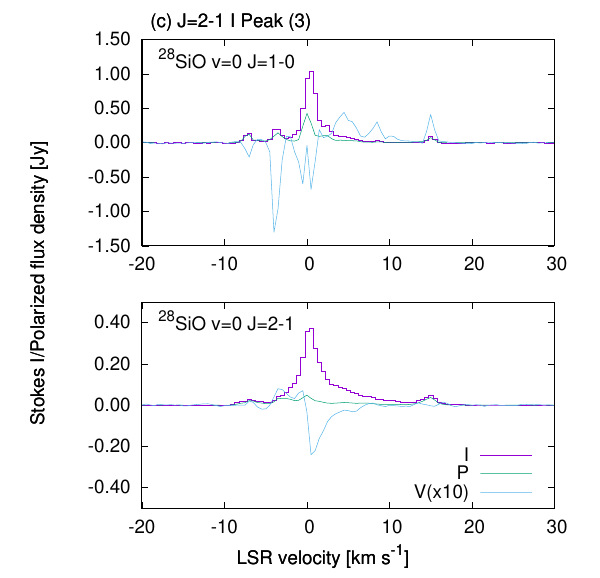}
\includegraphics[width=6.5cm]{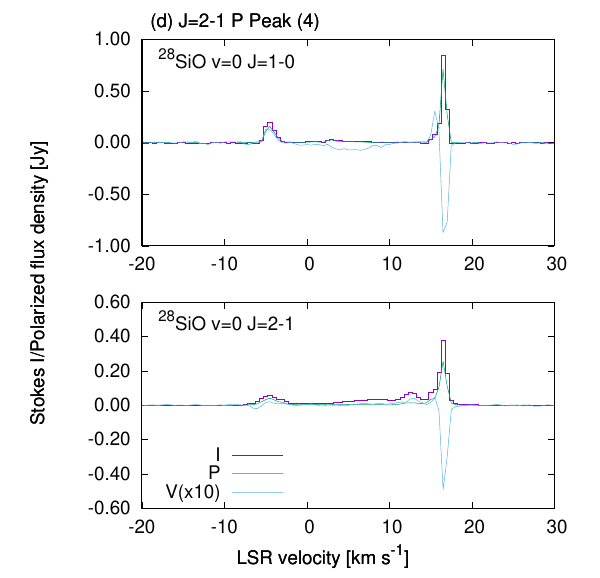}\\
\vspace{5mm}
\includegraphics[width=6.5cm]{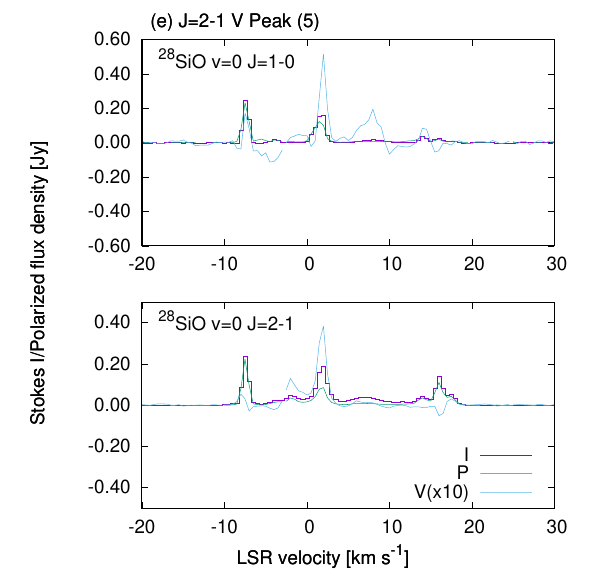} 
\vspace{10mm}
\caption{Stokes~I, linear polarization intensity (P), and Stokes~V spectra toward the bright peaks in the $^{28}$SiO $v$=0 $J$=1-0 (top) and 2-1 (bottom) emission in each panel.
All the spectra are averaged over 0.045\arcsec$\times$0.045\arcsec \ region as indicated in Figure \ref{fig-vmap}. 
}
\label{fig-IPV}
\end{center}
\end{figure}

\begin{figure*}[th]
\begin{center}
\includegraphics[width=7cm]{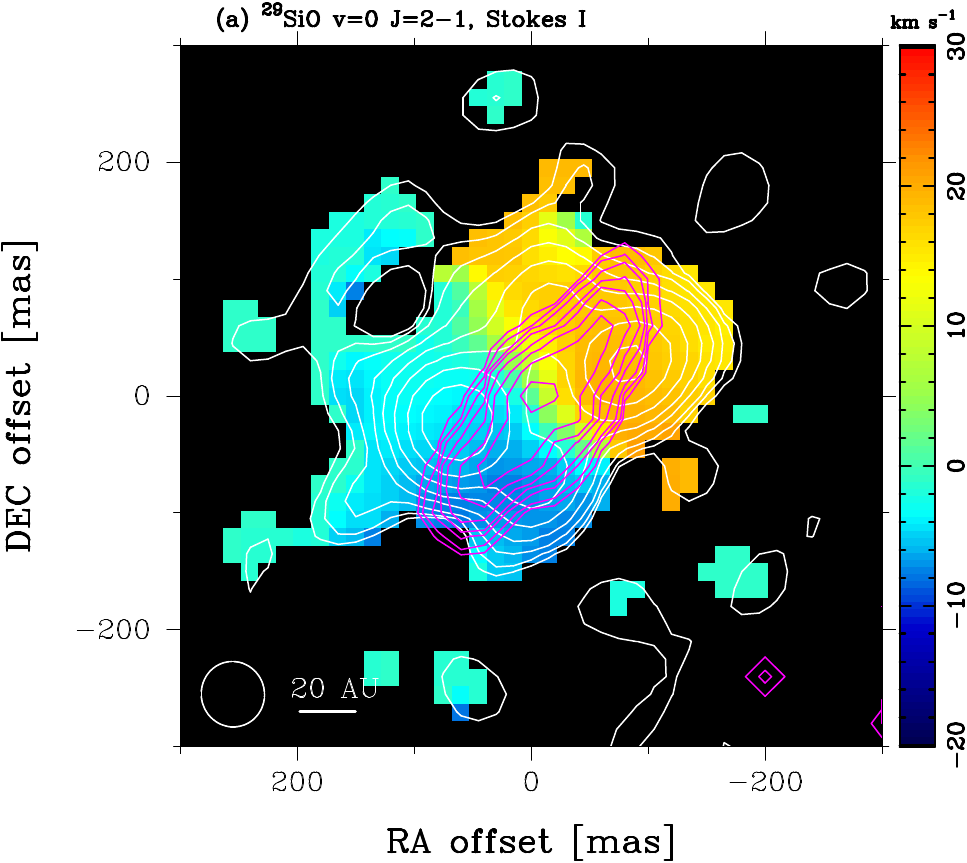}
\vspace{5mm}
\includegraphics[width=7cm]{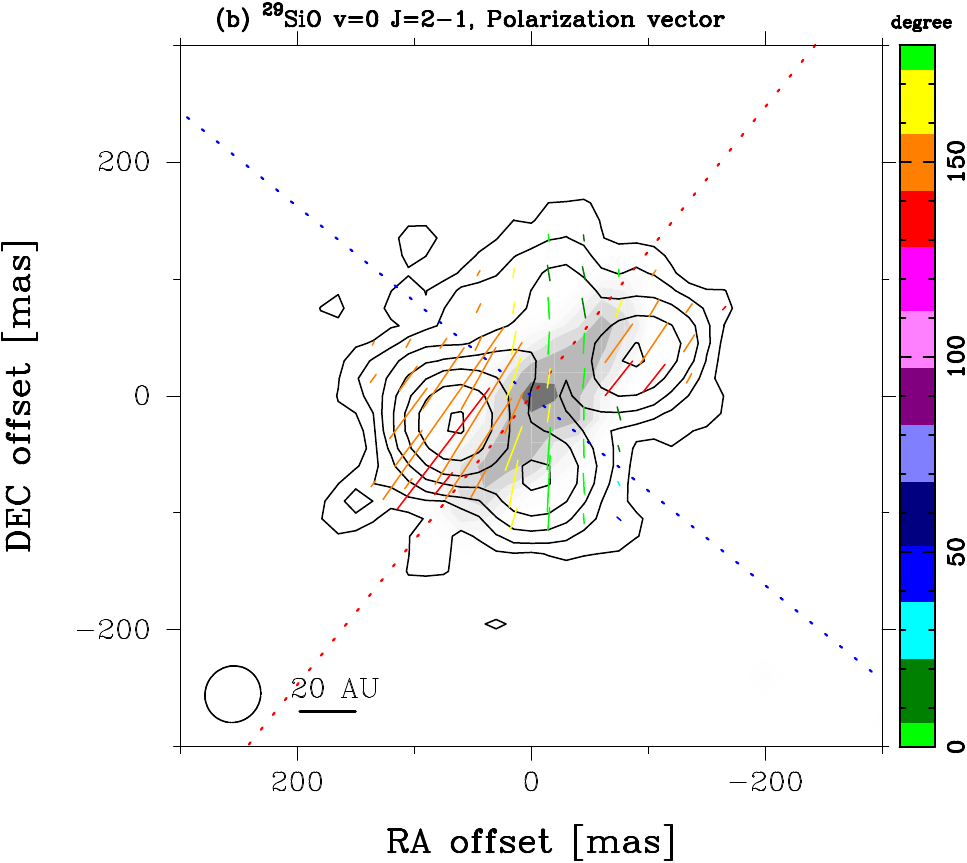}
\vspace{5mm}
\caption{Same as Figures \ref{fig-moment} and \ref{fig-polmap} but for the $^{29}$SiO $v$=0 $J$=2-1 line.
Moment maps are produced by using the velocity range from -20 to 30~km~s$^{-1}$. 
(a) White contours show moment 0 and color indicate moment 1 of the $^{29}$SiO $v$=0 $J$=2-1 line.
The contour  levels are 4, 8, 16, 32, 64, 128, 256, 512, and 1024 times the rms noise levels of 107~mJy~beam$^{-1}$~km~s$^{-1}$.
Magenta contours show the 99~GHz continuum emission at 30~mas resolution \citep{Wright2020}. 
The contour  levels are 4, 8, 16, 32, 64, 128, 256, and 512 times the rms noise levels of 0.011~mJy~beam$^{-1}$. 
(b) Polarization vector of the $^{29}$SiO $J$=2-1 $v$=0 line averaged over the velocity range from -10 to 20~km~s$^{-1}$ superposed on moment 0 of linear polarization intensity. 
The contour  levels are 4, 8, 16, 32, 64, 128, and 256 times the rms noise levels of 15.9~mJy~beam$^{-1}$~km~s$^{-1}$.
The error in the polarization angle is smaller than 7~degrees for the linear polarization intensity higher than 4$\sigma$. 
Gray scale shows the 99~GHz continuum emission at 30~mas resolution \citep{Wright2020}. 
}
\label{fig-29sio}
\end{center}
\end{figure*}

\begin{figure}[ht]
\begin{center}
\includegraphics[width=7cm]{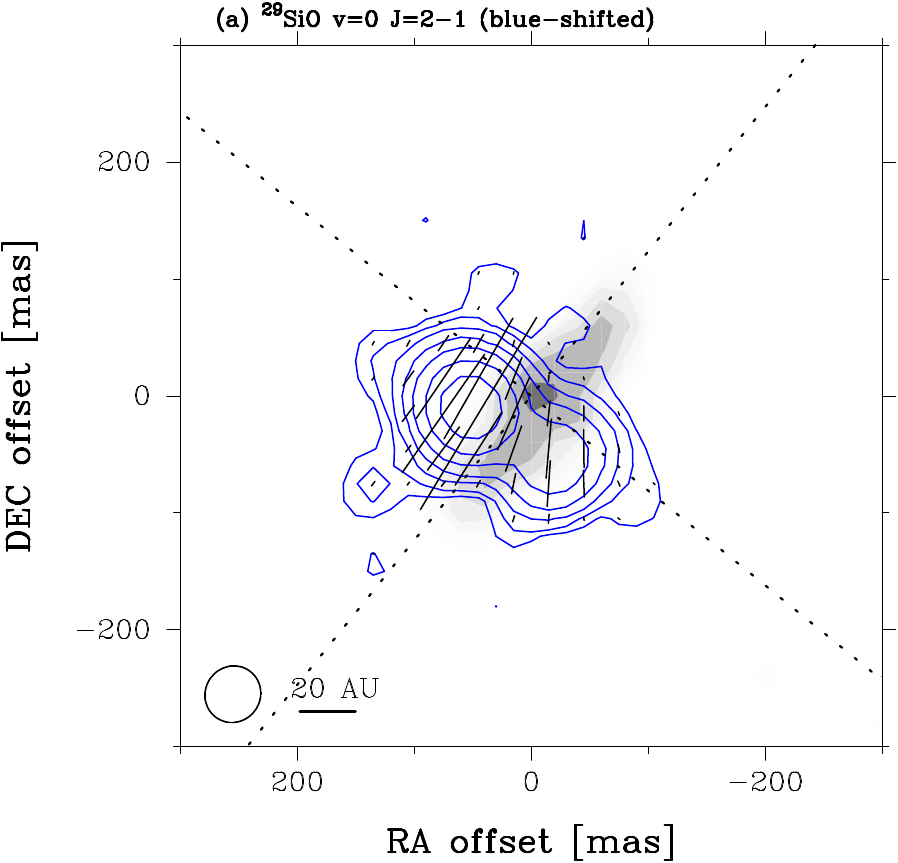}
\hspace{5mm}
\includegraphics[width=7cm]{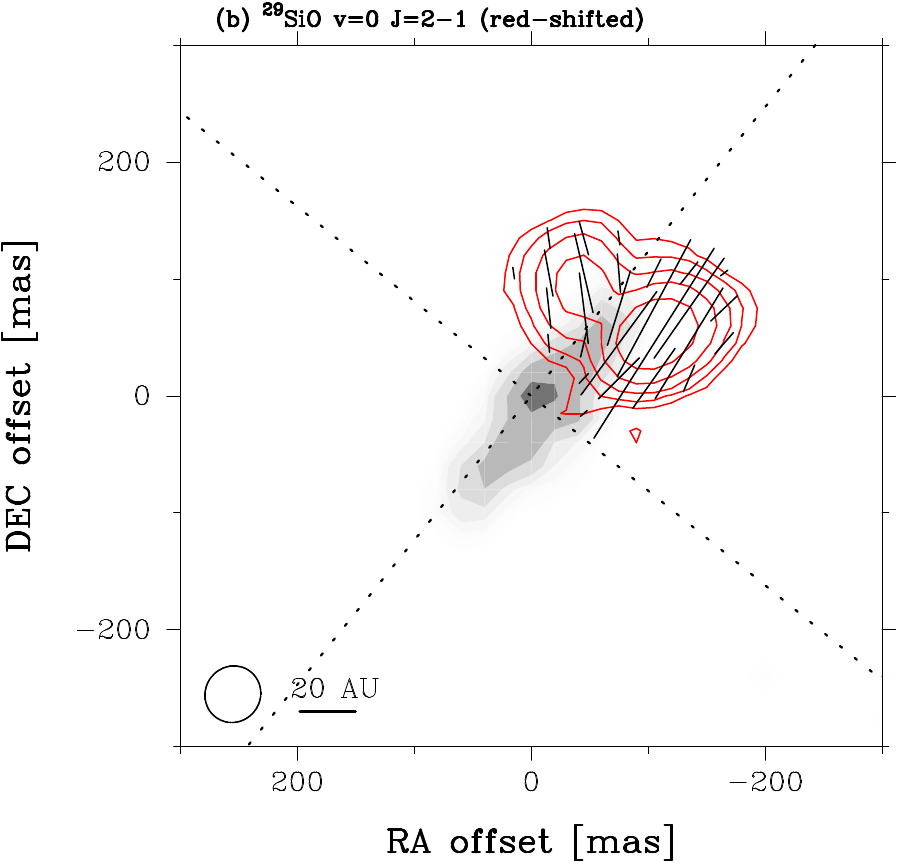}
\caption{
Linear polarization intensity map of $^{29}$SiO $v$=0 $J$=2-1 line superposed the 99~GHz continuum emission at 30~mas resolution \citep{Wright2020}.
(a) Blue-shifted and (b) red-shifted components of the $^{29}$SiO $v$=0 $J$=2-1 line integrated from -10 to 0~km~s$^{-1}$ and 10 to 20~km~s$^{-1}$, respectively. 
The contour  levels are 4, 8, 16, 32, 64, 128, and 256 times the rms noise levels of 0.43~mJy~beam$^{-1}$~km~s$^{-1}$ and 0.34~mJy~beam$^{-1}$~km~s$^{-1}$ for the blue- and red-shifted components, respectively. 
}
\label{fig-29sio-bluered2}
\end{center}
\end{figure}

\begin{figure*}
\begin{center}
\includegraphics[width=7cm]{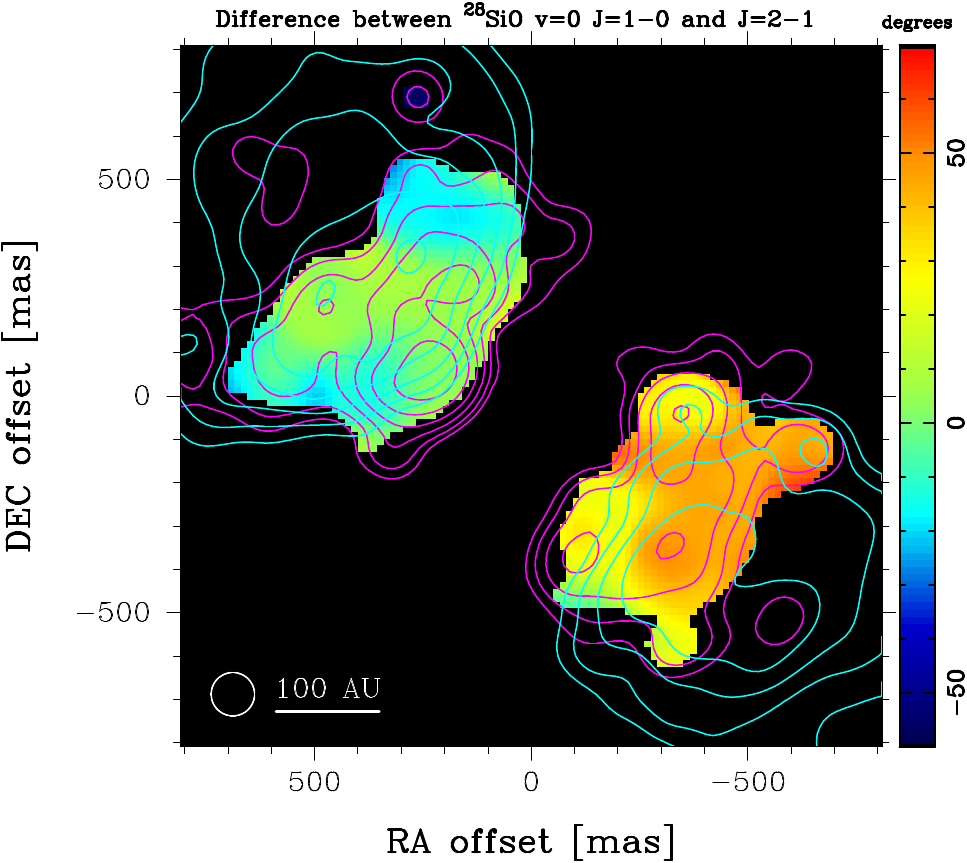}
\caption{Difference in average polarization angle (PA) between $^{28}$SiO $v$=0 $J$=1-0 and $J$=2-1 (see Figure \ref{fig-polmap}), PA($J$=1-0)$-$PA($J$=2-1).
The contour  shows the moment 0 of linear polarization intensities of $^{28}$SiO $v$=0 $J$=1-0 (magenta) and $J$=2-1 (cyan) lines as shown in Figure \ref{fig-polmap}.
All the maps are deconvolved with the 0.1\arcsec \ circular beam. 
The velocity range of the map is from -10 to 20~km~s$^{-1}$. 
The polarization angle rotation of 1~degree corresponds to $RM$ of 460~rad~m$^{-2}$.
}
\label{fig-padiffrev}
\end{center}
\end{figure*}

\begin{figure*}[th]
\begin{center}
\includegraphics[width=7cm]{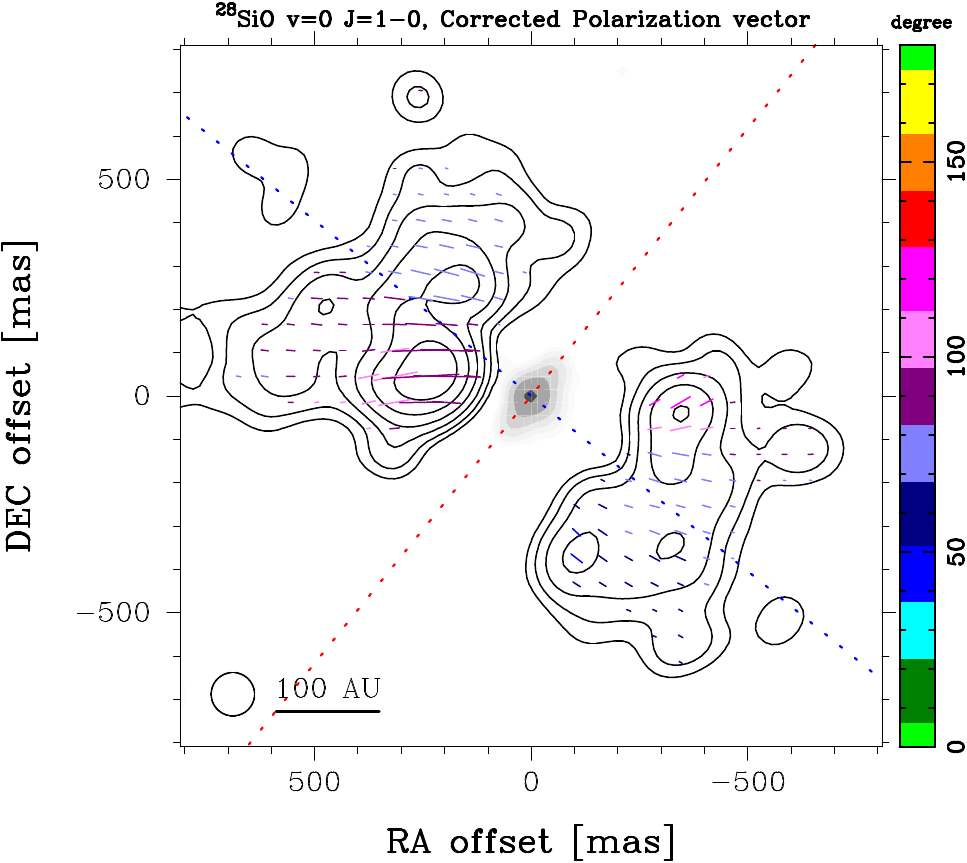}
\caption{Polarization vector maps corrected for Faraday rotation. 
The contours are the linear polarization intensity maps of the $^{28}$SiO $v$=0 $J$=1-0 and 2-1 lines, respectively. 
All the maps are deconvolved with the 0.1\arcsec \ circular beam. 
The velocity range of the map is from -10 to 20~km~s$^{-1}$. }
\label{fig-pola0}
\end{center}
\end{figure*}

\begin{figure}[ht]
\begin{center}
\includegraphics[width=7cm]{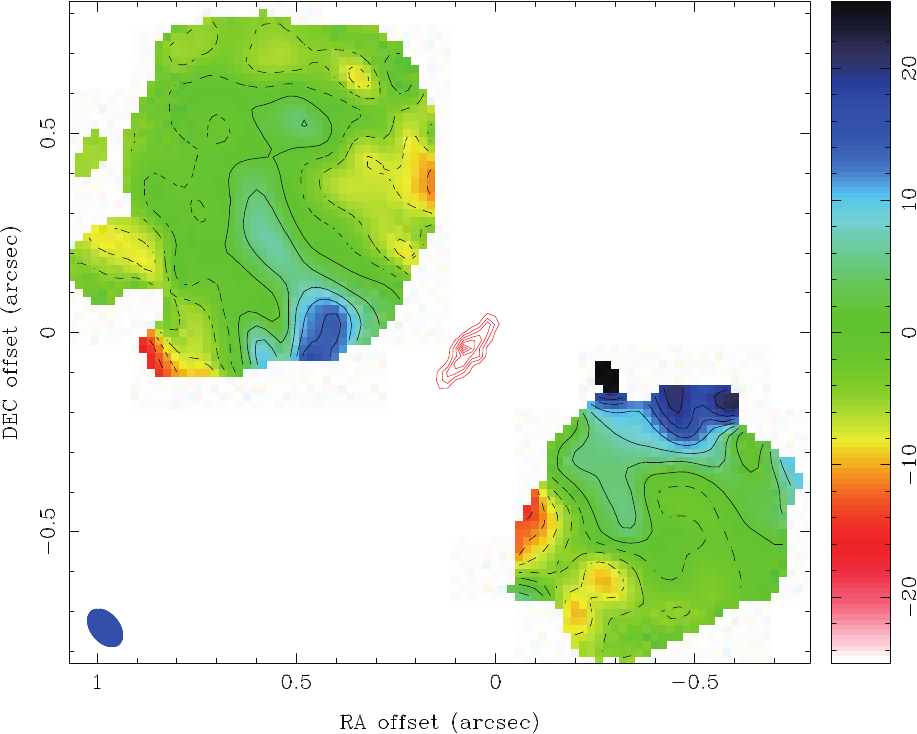}
\caption{$^{28}$SiO $v$=0 $v$=2-1 polarization angle residuals at 0.1\arcsec \ resolution (beam, lower left corner) after subtracting an 0.5\arcsec \ resolution smoothed polarization angle image in order to remove structure from the large scale magnetic field geometry.  Black contours range from -9\degr\ to $+13$\degr\ in 2\degr\ steps.  Red contours show the 96 GHz SrcI continuum image at 30~mas resolution \citep{Wright2020} in 1~mJy~beam$^{-1}$ steps beginning at 1~mJy~beam$^{-1}$. 
}
\label{fig-PAresid}
\end{center}
\end{figure}

\begin{figure}[ht]
\begin{center}
\includegraphics[width=10cm]{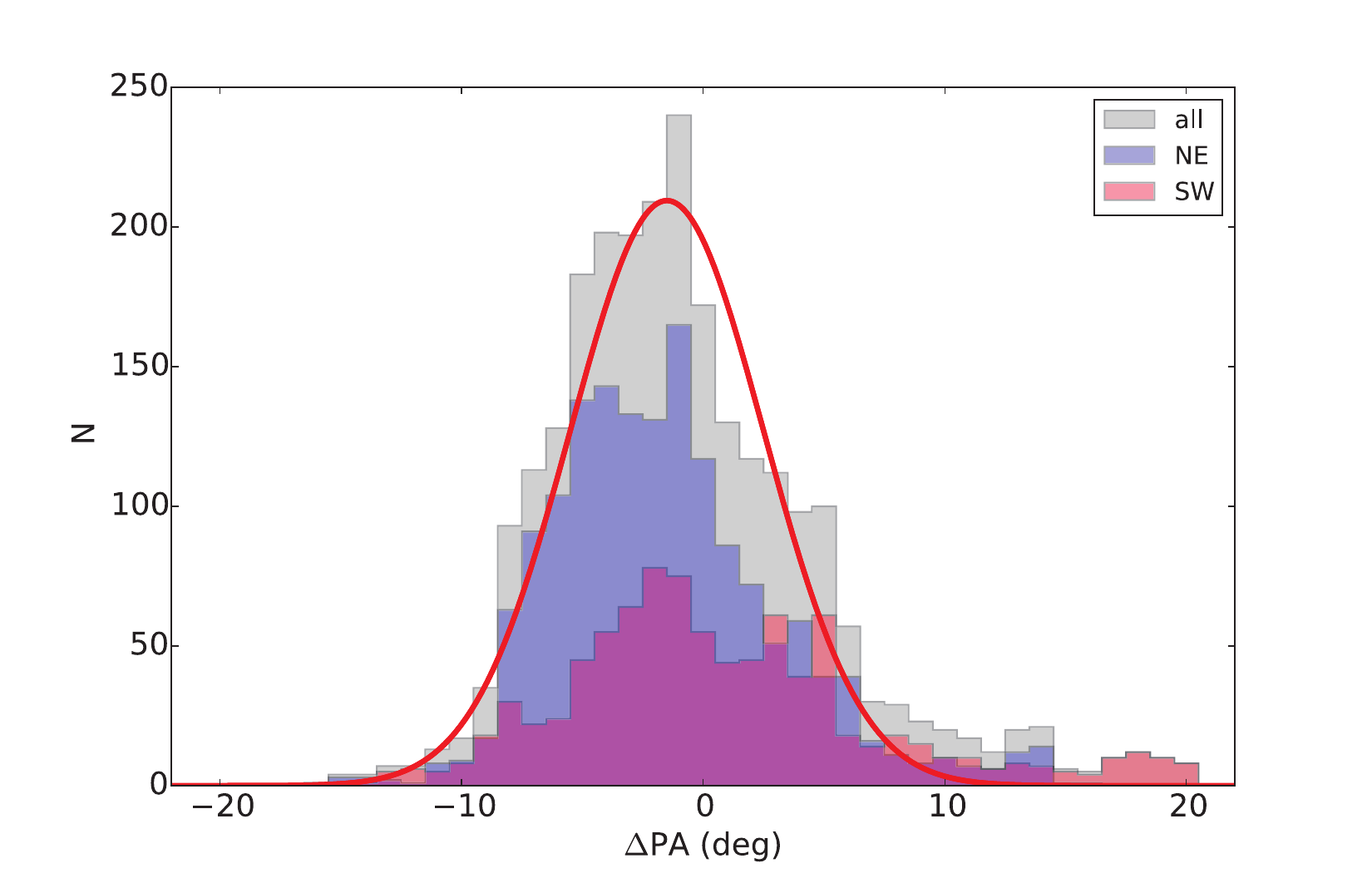}
\caption{Histogram of the position angle residuals in Figure~\ref{fig-PAresid}.  
Histograms are plotted for the SW lobe (red), the NE lobe (blue), and both lobes (gray).
The red curve shows a Gaussian with a dispersion of 4~degrees, used to estimate the magnetic field strength using the Davis-Chandrasekhar-Fermi method for all regions. }
\label{fig-PAhisto}
\end{center}
\end{figure}

As shown in Figure \ref{fig-vmap}, Stokes~V components are clearly detected in the $^{28}$SiO $v$=0 $J$=1-0. 
For the spectra presented in Figure~\ref{fig-IPV}, the fractional circular polarization is as high as 60\% for the $J$=1-0 line. 
These values are much higher than the calibration uncertainties as discussed in Section \ref{sec2-2}, and hence, we rule out the possibility that the circular polarization is an instrumental effect. 
In spite of the calibration uncertainties, the circular polarization signatures are also detected in the $^{28}$SiO $v$=0 $J$=2-1 line. 

A pair of positive and negative components in the Stokes V spectra could be attributed to the Zeeman effect, which shifts the frequency of the observed spectra with opposite signs of circular polarization as reported for the HI line \citep{Troland1982} and later for the SiO masers \citep{Barvainis1987, Vlemmings2017, Shinnaga2017}. 
Then the circularly polarized line profiles can be fitted by
\begin{equation}
V = b \frac{dI}{d\nu},
\label{eq-Zeeman}
\end{equation}
where $b=z B \cos \theta$ is proportional to the magnetic field strength along the line of sight, $B \cos \theta$ and $z$ is the Zeeman splitting coefficient, $g \Omega/2\pi$. 
Here we can ignore the leakage from the Stokes I because it is much smaller than the Stokes V components in our data. 

Assuming uniform magnetic field strengths for all the velocity components, we were able to fit some of the velocity components in the Stokes V spectra to the Zeeman pattern as expressed by equation (\ref{eq-Zeeman}).
The magnitude of the Zeeman splitting, $g \Omega/2\pi$=230~Hz~G$^{-1}$, corresponds to 1.6~m~s$^{-1}$~G$^{-1}$ and 0.8~m~s$^{-1}$~G$^{-1}$ at 43~GHz ($J$=1-0) and 86~GHz ($J$=2-1), respectively.  

\begin{figure}[bht]
\begin{center}
\includegraphics[width=6.5cm]{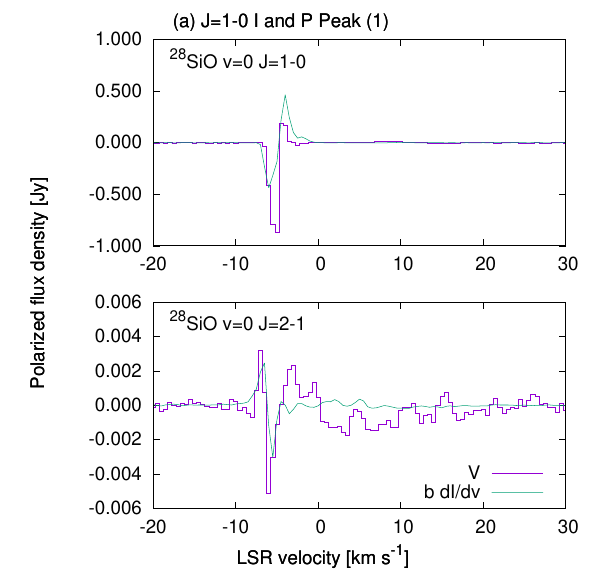}
\includegraphics[width=6.5cm]{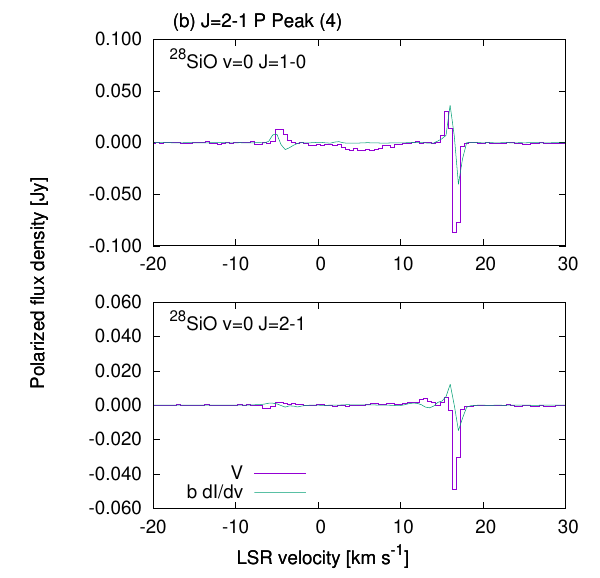} 
\vspace{10mm}
\caption{Stokes~V spectra and the best fit results of the derivative of Stokes~I (equation \ref{eq-zeeman}) toward peaks 1 and 4 indicated in Figure \ref{fig-vmap}. 
All the spectra are averaged over 0.045\arcsec$\times$0.045\arcsec \ region as indicated in Figure \ref{fig-vmap}. 
}
\label{fig-spstokes}
\end{center}
\end{figure}

We present the fitting results for the selected two positions showing the most clear signatures of possible Zeeman splitting pattern. 
Our fits give $b$=0.05 and $b=-0.16$ for the $J$=1-0 line as shown in Figure \ref{fig-spstokes}. 
These values would suggest magnetic field strengths of 5-16~G. 
If the Zeeman pattern of the $J$=2-1 lines is also real, similar magnitudes of the field strengths of 8-10~G ($b$=0.04 and 0.05) would be inferred. 
\citet{Elitzur1998} predicts that maser amplification tends to enhance the fractional circular polarization relative to the case of the thermal Zeeman effect.  
This could lead to overestimates of the magnetic field strength, perhaps by as much as an order of magnitude.

Such large magnetic fields are implausible.  
A 10~G field corresponds to a magnetic pressure $B^2/8\pi$ = 4~dynes~cm$^{-2}$, 6 orders of magnitude greater than the turbulent pressure $\rho\,\Delta {\rm{V}}^2 /3$ that one obtains assuming a molecular hydrogen number density $n \sim 10^{6}$~cm$^{-3}$ and a turbulent linewidth $\Delta \rm{V} ~\sim 10$~km s$^{-1}$.  
The inferred fields also are 2-3 orders of magnitude greater than fields estimated with the DCF method.

The Zeeman interpretation also is improbable if the opposite signs of the splittings in the $J$=1-0 and 2-1 lines at the same positions are real. 
For example, although the $J$=1-0 and $J$=2-1 spectra for Peak 1 shown in the left panels of Figure~\ref{fig-spstokes} both appear to show a clear Zeeman pattern, the magnetic fields inferred from the two lines have opposite signs.
Similar discrepancies can be seen in the Stokes V maps and spectra (Figures \ref{fig-vmap} and \ref{fig-spstokes}). 

We conclude that Zeeman splitting cannot explain the observed circular polarizations.  
Although the calibrations of Stokes V in the ALMA data are highly uncertain, the $^{28}$SiO $v$=0 $J$=2-1 lines could give additional support for our conclusions that the Zeeman interpretation is improbable. 
Further more accurate measurements of the Zeeman splittings in the $v$=0 $J$=2-1 masers would help interpretation. 

It is likely, instead, that the Stokes V spectra arise from a conversion of linear to circular polarization.  
This can occur if the linearly polarized radiation propagates through foreground gas with a different magnetic field direction\footnote{\citet{Nedoluha1994} also consider the case of intensity dependent circular polarization that occurs when the stimulated emission rate $R$ becomes comparable with the Zeeman splitting $g\Omega/2\pi$.  A change of magnetic field direction is not needed.  Since we estimate that $R \ll g\Omega/2\pi$ for the $v$=0 masers, we do not consider this case here.}.
Numerical simulations of this effect for $v$=1 SiO masers in a turbulent velocity field by \citet{Wiebe1998} produced fractional circular polarizations of up to $\sim$10\% for magnetic field strengths of $\sim$10~mG. 
Another possible mechanism producing circular polarization is a gradient of the velocity and/or magnetic field along the maser path \citep{Elitzur1992}. 

More recently, an alternative model of ARS was proposed to explain high degree of circular polarization detected in masers and thermal lines \citep{Houde2013, Houde2014, Chamma2018}.
ARS occurs when linearly polarized radiation from molecules in the background scatters off a population of these molecules in the foreground where the magnetic field direction is different.
Again, the effect is to convert some of the linearly polarized background radiation to circular polarization. 
\citet{Houde2014} notes that a complete conversion of linear to circular polarization is possible. 
Thus, ARS is likely for the $v$=0 SiO transitions toward Orion because there appears to be a large column density of foreground SiO in the cooler outer regions of the outflow.  
This is indicated by the absence of $^{28}$SiO $v$=0 emission along the plane of the disk suggesting foreground absorption, and by blueshifted absorption in many other molecular lines \citep{Plambeck2016}.  
If the ARS is responsible for producing circular polarization up to 20-60\%, it should become the dominant mechanism to rotate the linear polarization angles by redistributing the ratio between the Stokes Q and U. 
However, we found no correlation between the fractional circular polarization and the position angle discrepancies between the $J$=1-0 and 2-1 lines, as stated in the last paragraph of Section \ref{sec4-2-3}.  
It is unlikely that ARS would rotate the polarizations of the two transitions by the same amount. 
Thus, we expect the effects of ARS on the rotation of the linear polarization angles is small, or limited to small spatial scales and/or velocity ranges around the regions of strong circular polarization. 

\subsubsection{Does the SrcI outflow impact the Hot Core?}
\label{sec4-2-6}

SrcI is offset just 1\arcsec \ from the famous Orion Hot Core, a dense clump with many molecular lines \citep[e.g.][]{Wright2017, Wright2020}. 
The relationship between SrcI and the Hot Core is still uncertain; is their proximity merely a chance projection effect or are they physically/dynamically linked?
Observational evidence suggests that the Hot Core is heated externally, perhaps by shock waves generated by SrcI's outflow \citep{Goddi2011a, Zapata2011b}, but there does not seem to be a clear chemical signature of this interaction at the boundary between the outflow and Hot Core. 
It is also proposed that the explosive outflow in this region is the alternative candidate heating source of the Hot Core \citep{Zapata2011b}, although neither of these two possibilities could be ruled out. 

\begin{figure*}[th]
\begin{center}
\includegraphics[width=16cm]{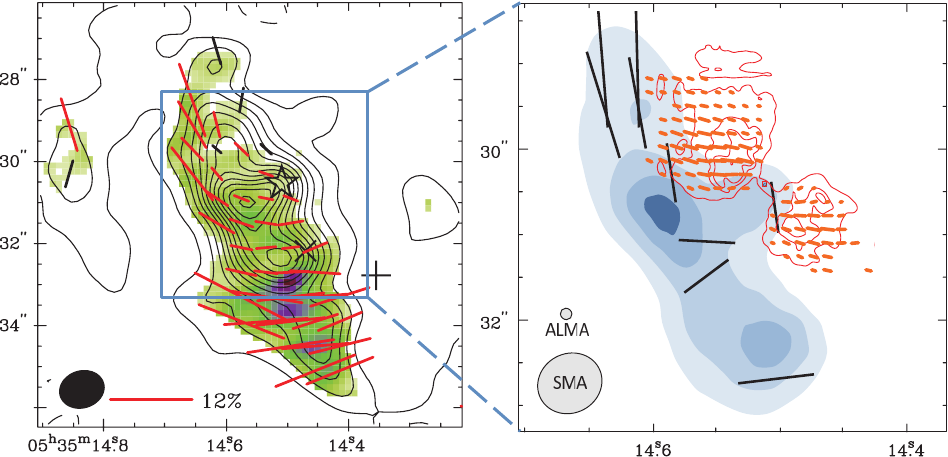}
\caption{(left) The 345~GHz dust polarization measured with the SMA with a 1.1\arcsec$\times$0.9\arcsec \ synthesized beam \citep{Tang2010}. 
The contour interval is 150~mJy~beam$^{-1}$. 
Line segments indicate the dust polarization. SrcI is denoted by the black star. 
(right) The linear polarization image of the $^{28}$SiO $v$=0 $J$=2-1 line overlaid on the highest resolution dust polarization image shown by \citet{Tang2010}. 
The synthesized beam size is 0.8\arcsec$\times$0.7\arcsec \ and the contour interval is 120~mJy~beam$^{-1}$. 
}
\label{fig-dustpol}
\end{center}
\end{figure*}

Another method to investigate the relationship between SrcI and the Hot Core is to study how the magnetic field of the outflow connects with that in the Hot Core gas, as measured by dust polarization.
Toward Orion-KL, 4\arcsec \ resolution polarization observations made with BIMA at 90 and 230 GHz revealed an abrupt kink in the polarization direction just SE of SrcI \citep{Rao1998}. 
More sensitive 1\arcsec \ resolution observations with the SMA at 345 GHz \citep{Tang2010}, shown in the left panel of Figure \ref{fig-dustpol}, indicate that this kink is part of a smooth, symmetric twist of the polarization vectors within the Hot Core. 

In the right panel of Figure \ref{fig-dustpol}, we overlay our SiO $J$=2-1 polarization vectors on the highest resolution image from \citet{Tang2010}. 
As discussed above, the polarization vectors of the $^{28}$SiO $v$=0 $J$=2-1 line should be good tracers of the magnetic field direction near SrcI, albeit with a 90~degree ambiguity. 
The dust polarization is remarkably strong and smooth, and not aligned with the SrcI outflow, suggesting a pre-existing well-organized magnetic field direction. 
There appears to be no correlation between the magnetic field directions inferred from the dust and the SiO (whether or not the SiO linear polarization is parallel or perpendicular to the field). 
Given that the field in the outflow is strong, one would expect to find some evidence for a kink in the dust polarization adjacent to the outflow, but no kink is apparent in the existing data.
If future observations find that there is really no connection between the two, and no obvious depolarization along the interface, this will suggest that the Hot Core is physically remote from SrcI, despite their proximity in the plane of the sky. 
It will also suggest that shocks or MHD waves launched by the outflow are not heating the Hot Core. 

\subsection{$^{29}$SiO $v$=0 $J$=2-1 line}
\label{sec4-3}

This transition maps the polarization close to SrcI. 
Figure \ref{fig-29sio} shows spatial and velocity structures of Stokes I (moment 0 and 1 maps) of the $^{29}$SiO $v$=0 $J$=2-1 line along with its linear polarization intensity and vectors.
The polarization vectors are mostly radial pointing toward SrcI or parallel to the disk midplane. 

The $^{29}$SiO $J$=2-1 $v$=0 line could be a good probe for magnetic field structure at the base of the disk wind. 
As discussed in Section \ref{sec4-1}, the condition $g\Omega/2\pi>R$ is satisfied for the $^{29}$SiO $v$=0 $J$=2-1 maser with the magnetic field strength of 40~mG. 
If the beaming factor of maser is about 10, $R$ becomes smaller by a factor of 10 and hence, comparable to the Zeeman frequency for a 4~mG field. 
Given that we find a field strength of $\sim$30~mG farther out in the lobes, it is likely that the $^{29}$SiO polarization does trace the field direction. 
We note that the Faraday rotation could cause uncertainties in the polarization angle of $\sim$10~degrees, when interpreting the magnetic field structure from the polarization results of the $^{29}$SiO $v$=0 $J$=2-1 line. 
However, if it is caused by the foreground effect, the polarization angle would show systematic smooth change similar to the $^{28}$SiO $v$=0 $J$=2-1 line. 
Furthermore, it is unlikely that the symmetric pattern of the polarization vectors is caused by the Faraday rotation coincidentally. 

\subsection{Three dimensional structure of magnetic field}
\label{sec4-4}

We now consider the question of the three dimensional magnetic field morphology in the bipolar outflow from SrcI.  Several previous works have used the Goldreich-Kylafis effect \citep{Goldreich1981} to study the magnetic field geometries in the outflows or jets from low mass stars \citep{Girart1999,Cortes2005,Ching2016,Lee2018}.  
As in the simplest cases of maser emission, the polarizations can be parallel or perpendicular to the projection of the field on the plane of the sky. 
If there is dust polarization along the same, or a very nearby, line of sight, it is possible to resolve this 90~degree ambiguity by comparing the spectral line polarization with the dust polarization, which invariably is polarized perpendicular to the field direction. 
Unfortunately, dust emission is almost completely resolved out in our data, and comparisons with lower resolution dust images, as described in Section~\ref{sec4-2-6}, revealed no obvious correlations between the SiO maser and dust polarization directions.

In the original model of maser polarization developed by \citet{Goldreich1973}, the relationship between the polarization and field directions depends only on the angle $\theta$ between the field and the maser propagation direction, which is assumed to be very close to the line of sight.  When $\theta<55$~degrees (i.e. van Vleck angle), the polarization vectors are parallel to the magnetic field vectors in the plane of sky; when $\theta>55$~degrees, then the polarization and magnetic field vectors are perpendicular to each other.  However, this simple relationship holds only if the masers are isotropically pumped.  
The extremely high fractional polarizations observed for the SiO masers in the SrcI outflow strongly suggest that these masers are anisotropically pumped, in which case the parallel to perpendicular transition angle varies depending on the excitation direction \citep[e.g. Figures A.10-A.18 in ][in case of the SiO $v$=1 masers]{Lankhaar2019}.  
This makes it very difficult to make definitive statements about the field morphology. 

According to the polarization images of the $v$=0 SiO masers, our observations reveal three distinct zones: 
(1) close to the disk with radial E-vectors in the $^{29}$SiO $v$=0 $J$=2-1 line; (2) the rotating column with well ordered rotation but without significant polarized emission, most evident in the $^{28}$SiO $v$=0 $J$=2-1 line; and (3) the outer NE-SW lobes with ordered linear polarization and substantial circular polarization in both the $^{28}$SiO $v$=0 $J$=1-0 and 2-1 lines.
No significant polarization was seen in the rotating column. 
This may be due to lack of strong polarization component in thermal emission or depolarization by the tight toroidal field.  
The former case would be more likely because the SiO $v$=0 masers are possibly quenched at higher density gas as traced by the Si$^{18}$O line (Figure \ref{fig-28SiO21_bluered}) and $^{29}$SiO $v$=0 maser \citep[see Figure \ref{fig-29sio}][]{Goddi2009a}. 
Magnetic field reconnection may play a role in reconfiguring the field topology from the rotating column to the more uniform field observed in the extended lobes.
\citet{Zweibel1989} suggests that reconnection must play a role because of the short time scale available as the outflow expands.  Magnetic energy is injected on small scales, where reconnection occurs, and allows the rotating field to conform to the large scale field seen in the extended lobes. 

\begin{figure*}[th]
\begin{center}
\includegraphics[width=4cm]{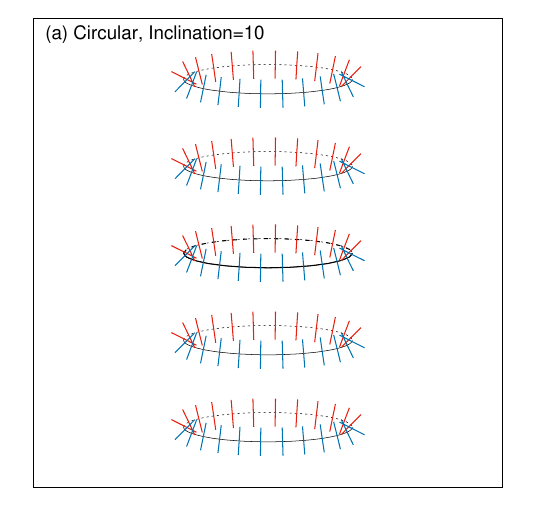}
\includegraphics[width=4cm]{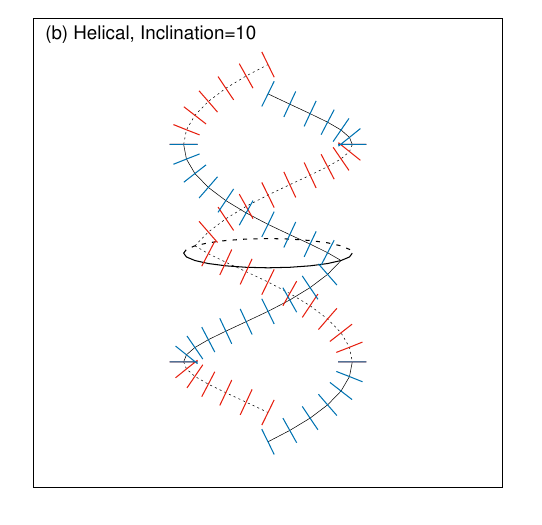}
\includegraphics[width=4cm]{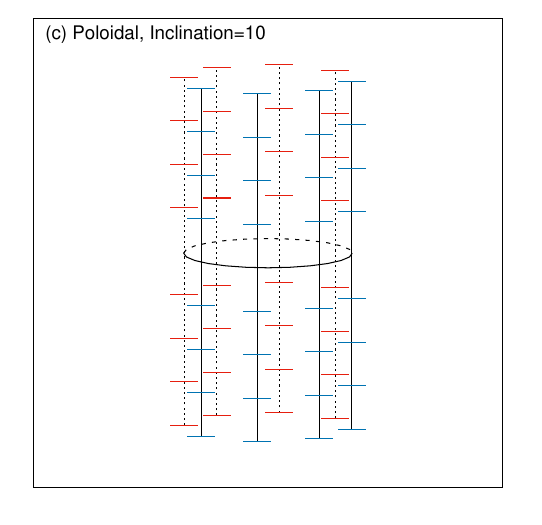}
\includegraphics[width=4cm]{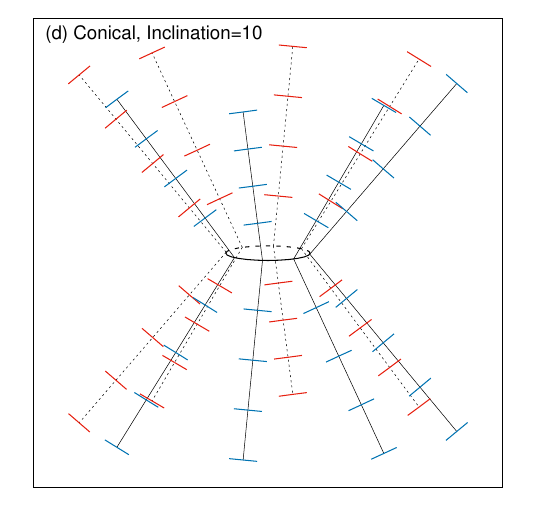}
\vspace{10mm}
\caption{Models of polarization structure projected on the sky plane ($x$-$z$ plane where $x$ and $z$ are horizontal and vertical axes, respectively). 
The observers are located on the $y$-axis ($y$=-$\infty$) with an inclination angle of 10~degrees. 
The black thin line shows the disk midplane. 
Red and blue bars indicate the polarization vectors projected on the sky plane. 
Dashed and solid lines indicate the magnetic field lines projected on the sky plane. 
Red/dashed and blue/solid lines are located at the front ($y<$0) and rear ($y>$0) side along the line-of-sight, respectively. 
From left to right: Circular field parallel to the disk midplane, helical field, poloidal field parallel with each other, and conical field.
}
\label{fig-model}
\end{center}
\end{figure*}

To interpret our observational results, we made simple models to illustrate magnetic field configuration revealed by the polarization structures. 
In this paper, we just consider two simple geometries for both toroidal and poloidal fields with disk inclination angles of 79-90~degrees \citep{Wright2020}, as shown in Figure \ref{fig-model}. 
These are circular and helical geometries for toroidal fields and parallel and conical geometries for poloidal fields. 

For both toroidal and poloidal field configuration, most of the magnetic field vectors are close to perpendicular to the line of sight, and hence, polarization vectors are more likely to be perpendicular to the magnetic field vectors on the plane of the sky. 
At first glance, there is no model that is relevant to our polarization map of the SiO lines under the assumption as described above. 
However, looking at panel (b) of Figure \ref{fig-model}, it appears that red (front side) vectors seem to match the observations for the NE lobe, and blue (back side) vectors match for the SW lobe. 
If this interpretation is valid, the SiO line would be optically thin and the emission from the rear side is dominant only in the SW lobe. 
These assumptions are still uncertain considering the fact that the optical depths of the SiO masers would be {\it{negative}} or the saturated masers are optically thick. 

The $^{29}$SiO polarization vectors close to SrcI appear to be radial or parallel to the disk midplane. 
These could be consistent with toroidal or helical magnetic fields at the base of the outflow if the polarization and magnetic field vectors are parallel.  A helical field could also be consistent with the $^{28}$SiO polarizations in the outflow lobes if the polarized emission favors either the front or rear side of the outflow lobes. However, a detailed simulation would be required to fully test these possibilities.

\section{Summary}
\label{sec5}

We have carried out full polarization observations of multiple SiO lines in the $J$=1-0 (43~GHz) and 2-1 (86~GHz) transitions using VLA Q-band and ALMA band~3, respectively, toward a high-mass protostar candidate Orion Source~I (SrcI) at resolutions of 50~mas. 
Results of this paper are listed below: 

\begin{enumerate}
\item In total 11 transitions are detected: $^{28}$SiO $v$=0 $J$=1-0 and 2-1, $^{28}$SiO $v$=1 $J$=1-0 and 2-1, $^{28}$SiO $v$=2 $J$=1-0 and 2-1, $^{29}$SiO $v$=0 $J$=1-0 and 2-1, $^{29}$SiO $v$=1 $J$=2-1, and $^{30}$SiO $v$=0 $J$=1-0 and 2-1. Higher vibrationally excited lines of $^{28}$SiO $J$=1-0 $v$=3 and $v$=4 are not detected.  
Most of the detected transitions have high peak brightness temperatures indicative of maser emission; some of them (e.g. $^{28}$SiO $v$=0) are a mixture of thermal and maser emission.
\item Maser emission from the $v$=0, ground vibrational state transitions of $^{28}$SiO $J$=1-0 and 2-1 are distributed up to $\sim$1\arcsec\ from SrcI, in both the NE and SW lobes of the bipolar outflow that emanates from its circumstellar disk.  
Distributions of the other lines are more compact with sizes of less than $\sim$0.5\arcsec, possibly emitted from the base of the outflow. 
\item Linearly polarized emission is detected in 6 transitions; $^{28}$SiO $v$=0 $J$=1-0 and 2-1, $^{28}$SiO $v$=1 $J$=1-0 and 2-1, $^{28}$SiO $v$=2 $J$=1-0, and $^{29}$SiO $v$=0 $J$=1-0. 
The continuum emission from SrcI does not show significant polarization at either 43~GHz or 96~GHz. 
\item Moderately strong $^{28}$SiO $v$=0 $J$=1-0 and 2-1 lines could trace well ordered magnetic field structures as suggested by their polarization structures. 
The linear polarization fractions are on average $\sim$50-70\% and $\sim$20-50\% for the $J$=1-0 and 2-1 lines, respectively, while the largest values are up to 80-90\%. Such a high fractional linear polarization can be explained by anisotropic pumping or possibly by masers that are highly saturated. 
\item Polarization vectors in the $^{28}$SiO $v$=0 $J$=1-0 and 2-1 lines are consistent with each other in the NE lobe while they show offsets as large as  $\sim$60~degrees in the SW lobe.
The difference can be explained by the Faraday rotation with the rotation measure with an order of $10^{4}$~rad~m$^{-2}$, or possibly by absorption or scattering by SiO molecules in foreground gas.
\item The large number of polarization vectors in the $^{28}$SiO $v$=0 $J$=2-1 maser lines allows us to estimate the field strength in the outflow lobe to be about 30~mG using the Davis-Chandresekhar-Fermi method. 
Magnetic support may then explain why the outflow lobes are not swept back by the ram pressure of the surrounding gas as SrcI moves through the surrounding medium at 12~km~s$^{-1}$.  Magnetic support might also lead to sub-Keplerian rotation velocities for gas at the base of the outflow, possibly explaining discrepancies in SrcI mass estimates based on molecular line rotation curves.
\item 
Strong circular polarization up to 60\% and 20\% is detected for the $^{28}$SiO $v$=1-0 and 2-1 lines, respectively, although the Stokes V measurements of the ALMA data ($J$=2-1 line) would contain large uncertainties. 
The extremely high circular polarization fractions are most likely due to ARS rather than the Zeeman splitting which requires extremely strong magnetic field strengths of $\sim$10~G to explain the observed spectra. 
\item The polarization vectors are radial pointing toward SrcI or parallel to the disk midplane. 
The $^{29}$SiO $v$=0 $J$=2-1 line could be a good probe for magnetic field structure at the base of the disk wind.
\end{enumerate}

\bigskip

This paper makes use of the following ALMA data: ADS/JAO.ALMA\#2017.1.00497.S.
ALMA is a partnership of ESO (representing its member states), NSF (USA) and NINS (Japan), together with NRC (Canada), NSC and ASIAA (Taiwan), and KASI (Republic of Korea), in cooperation with the Republic of Chile. The Joint ALMA Observatory is operated by ESO, AUI/NRAO and NAOJ. 
The National Radio Astronomy Observatory is a facility of the National Science Foundation operated under cooperative agreement by Associated Universities, Inc.
TH is financially supported by the MEXT/JSPS KAKENHI Grant Numbers 16K05293, 17K05398, and 18H05222. 
RAB is supported by East Asia Core Observatory Association (EACOA) as an EACOR fellowship. 
Data analysis were in part carried out on common use data analysis computer system at the Astronomy Data Center, ADC, of the National Astronomical Observatory of Japan. 

{\it Facilities:} \facility{ALMA,VLA}.

{\it Software:} CASA \citep{McMullin2007}; Miriad \citep{Sault1995}.

\clearpage

\begin{deluxetable}{cccrrrcc}
\tablewidth{0pt}
\tabletypesize{\scriptsize}
\tablecaption{Observed lines
\label{tab-line}}
\tablehead{
\colhead{Frequency\tablenotemark{a}}   & \colhead{}                 & \colhead{}           & \colhead{$E_{l}$} & \colhead{Stokes I} &  \colhead{Pol. intensity}  & \colhead{$R$} & \colhead{} \\
\colhead{(MHz)}                      & \colhead{Molecule} & \colhead{Transition} & \colhead{(K)}     & \colhead{$T_{\rm{peak}}$ (K)}   & \colhead{$T_{\rm{peak}}$ (K)}   & \colhead{(s$^{-1}$)} & \colhead{Note}
}
\startdata
VLA Q-band &              &               &          &                    & & \\
43423.8530  & $^{28}$SiO   & $v$=0, $J$=1-0    &    0     & 2$\times10^{6}$    & 1.7$\times10^{6}$ & \quad 3 & \\
43122.0747  & $^{28}$SiO   & $v$=1, $J$=1-0    & 1769     & 1.3$\times10^{8}$  & 1.1$\times10^{7}$ &   188   & \\
42820.5864  & $^{28}$SiO   & $v$=2, $J$=1-0    & 3521     & 8$\times10^{7}$    & 3$\times10^{6}$   &   114 & \\
42519.3826  & $^{28}$SiO   & $v$=3, $J$=1-0    & 5256     & $<3\times10^{2}$   & $<2\times10^{2}$  & \nodata & \\
42218.4543  & $^{28}$SiO   & $v$=4, $J$=1-0    & 6974     & $<3\times10^{2}$   & $<2\times10^{2}$  & \nodata & \\
42879.9465  & $^{29}$SiO   & $v$=0, $J$=1-0    &    0     & 5$\times10^{3}$    & $<2\times10^{2}$  & \nodata & $^{28}$Si/$^{29}$Si$\sim$26\tablenotemark{b} \\
42373.4250  & $^{30}$SiO   & $v$=0, $J$=1-0    &    0     & 8$\times10^{3}$    & $<2\times10^{2}$  & \nodata & $^{28}$Si/$^{30}$Si$\sim$44\tablenotemark{b} \\
\hline
ALMA Band~3 &              &               &          &                    & & \\
86846.9850  & $^{28}$SiO   & $v$=0, $J$=2-1    &    2     & 5$\times10^{4}$    & 5$\times10^{4}$   & \quad 0.4 & \\
86243.4277  & $^{28}$SiO   & $v$=1, $J$=2-1    & 1771     & 2$\times10^{7}$    & 3$\times10^{6}$   &   139    & \\
85640.4531  & $^{28}$SiO   & $v$=2, $J$=2-1    & 3523     & 1.2$\times10^{4}$  & $<2\times10^{1}$  & \nodata  & \\
85759.1940  & $^{29}$SiO   & $v$=0, $J$=2-1    &    2     & 1.3$\times10^{6}$  & 9$\times10^{4}$   & \quad 9 & $^{28}$Si/$^{29}$Si$\sim$26\tablenotemark{b}  \\
85166.9579  & $^{29}$SiO   & $v$=1, $J$=2-1    & 1760     & 6$\times10^{2}$    & $<2\times10^{1}$  & \nodata & $^{28}$Si/$^{29}$Si$\sim$26\tablenotemark{b}  \\
84746.1663  & $^{30}$SiO   & $v$=0, $J$=2-1    &    2     & 2$\times10^{3}$    & $<3\times10^{1}$  & \nodata & $^{28}$Si/$^{30}$Si$\sim$44\tablenotemark{b}  
\enddata
\tablenotetext{a}{\citet{Muller2005}}
\tablenotetext{b}{\citet{Tercero2011}}
\tablecomments{$R$ is the stimurated emission rate if the maser emission is isotropic (equation \ref{eq-r}). }
\tablecomments{Uppler limits indicate the rms noise level (1$\sigma$). }
\end{deluxetable}

\begin{deluxetable}{lcc}
\tablewidth{0pt}
\tabletypesize{\scriptsize}
\tablecaption{Continuum emission
\label{tab-cont}}
\tablehead{
\colhead{}                       & \colhead{VLA Q-band}                             & \colhead{ALMA Band~3}
}
\startdata
Frequency                        & 43~GHz                                            & 96~GHz      	       	       	       	       	       \\
Beam size and position anlge     & 0.053\arcsec$\times$0.044\arcsec, +30.1$^{\circ}$ & 0.049\arcsec$\times$0.047\arcsec, $-$35.3$^{\circ}$ \\
R.A. (J2000)                     & 05h35m14.51677s                                   & 05h35m14.51851s     	       	       	       	       \\
Decl. (J2000)                    & -05$^{\circ}$22\arcmin30\arcsec.6237              & -05$^{\circ}$22\arcmin30\arcsec.6280	       	       \\
Peak Intesity                    & 2.6$\pm$0.3~mJy~beam$^{-1}$                       & 13.5$\pm$1.5~mJy~beam$^{-1}$		       \\
Integrated Flux\tablenotemark{a} & 10$\pm$1~mJy                                   & 58$\pm$6~mJy  	       	 		       \\
Deconvolved size and position angle\tablenotemark{b} &
                                   0.099\arcsec$\times$0.057\arcsec, +140$^{\circ}$  & 0.151\arcsec$\times$0.041\arcsec, +142$^{\circ}$    \\
rms of Pol. Intensity            & 0.014~mJy~beam$^{-1}$                             & 0.012~mJy~beam$^{-1}$
\enddata
\tablenotetext{a}{The VLA image was integrated over a box; the source has a bright spot in the middle of a disk, and is not well-described by a Gaussian fit which gives a flux 7.9~mJy. Error estimates include 10\% systematic calibration uncertainty. }
\tablenotetext{b}{Fitting errors are smaller than 0.001\arcsec \  and 0.1$^{\circ}$. }
\end{deluxetable}

\clearpage

\clearpage

\appendix
\section{All spectra of observed SiO transitions}
\label{secA-sp}

In this appendix, Figure A1 shows all the spectra of Stokes I and linear polarizations of SiO transitions observed in the present study as listed in Table \ref{tab-line}. 

\renewcommand{\thefigure}{\thesection\arabic{figure}}
\setcounter{figure}{0}

\begin{figure}[th]
\begin{center}
\includegraphics[width=5.0cm]{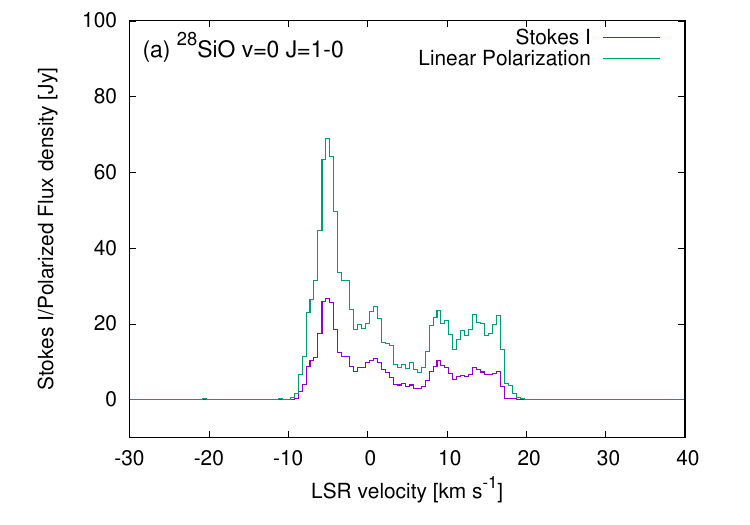}
\includegraphics[width=5.0cm]{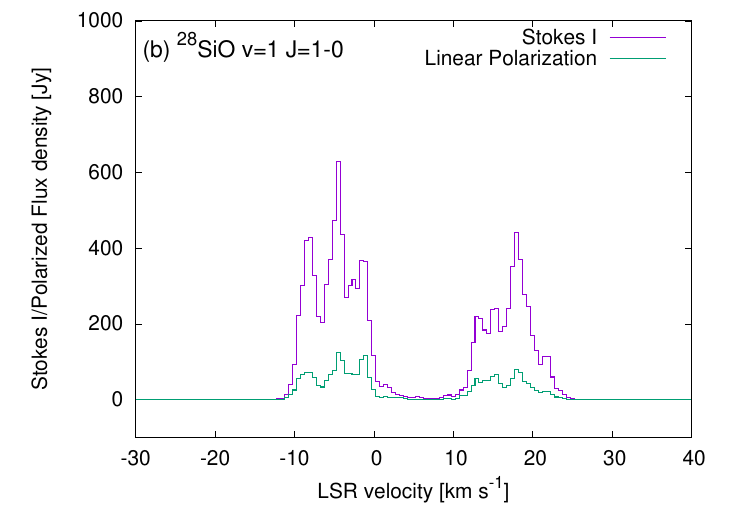}
\includegraphics[width=5.0cm]{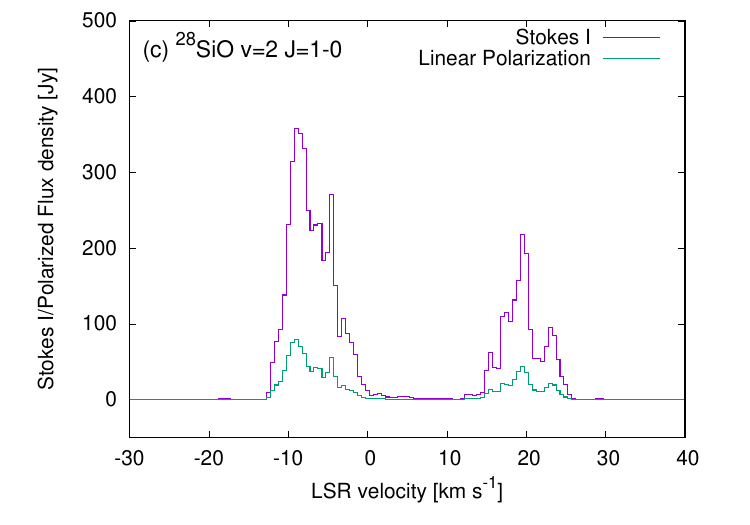} \\
\vspace{5mm}
\includegraphics[width=5.0cm]{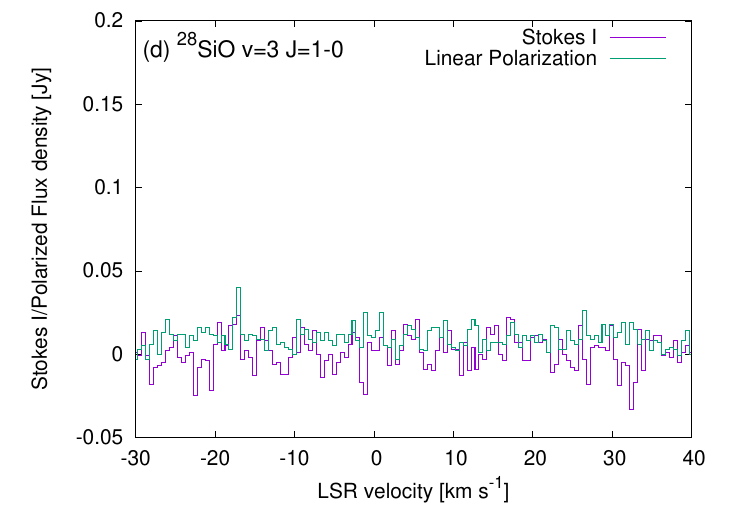}
\includegraphics[width=5.0cm]{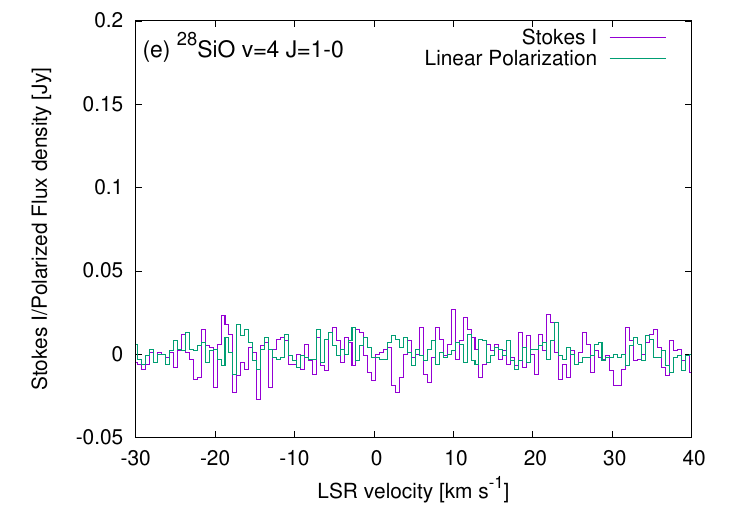} \\
\vspace{5mm}
\includegraphics[width=5.0cm]{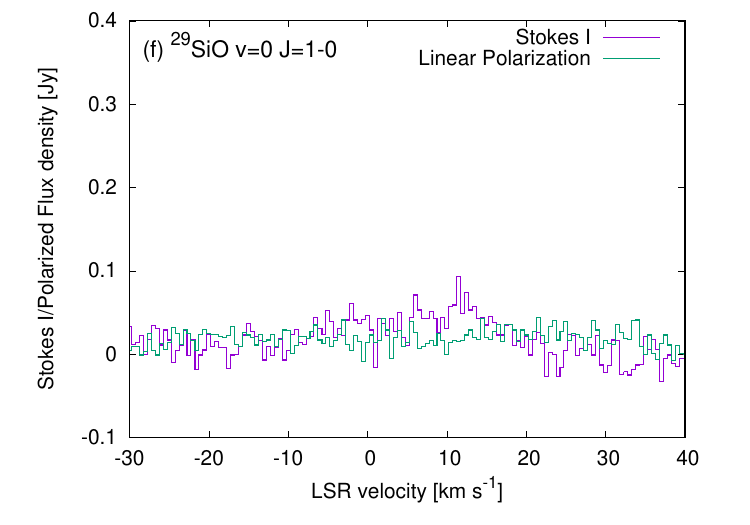}
\includegraphics[width=5.0cm]{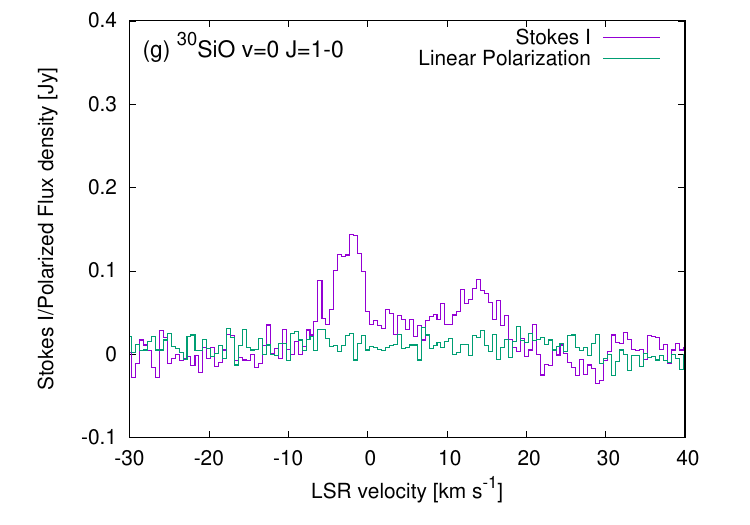} \\
\vspace{5mm}
\includegraphics[width=5.0cm]{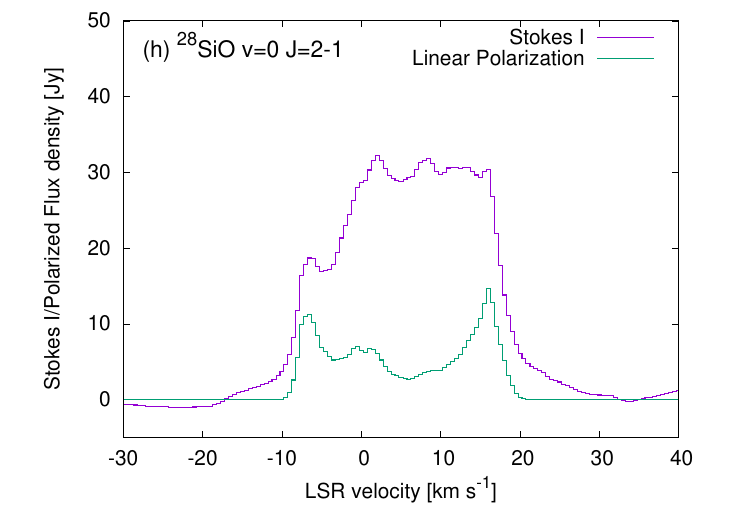}
\includegraphics[width=5.0cm]{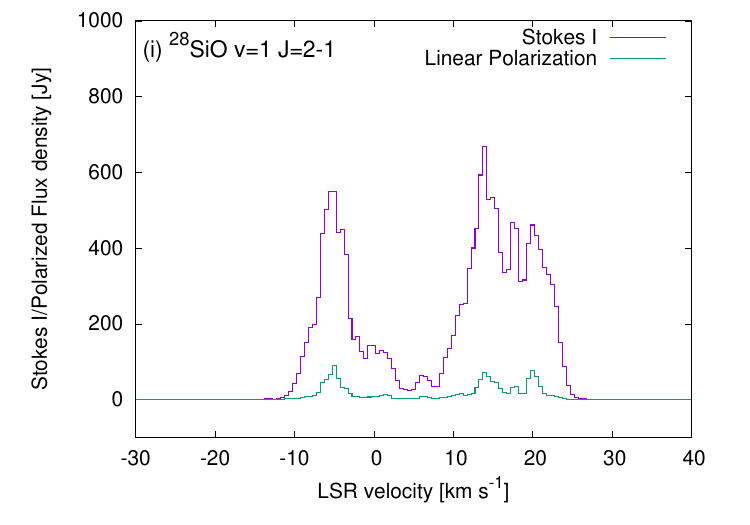}
\includegraphics[width=5.0cm]{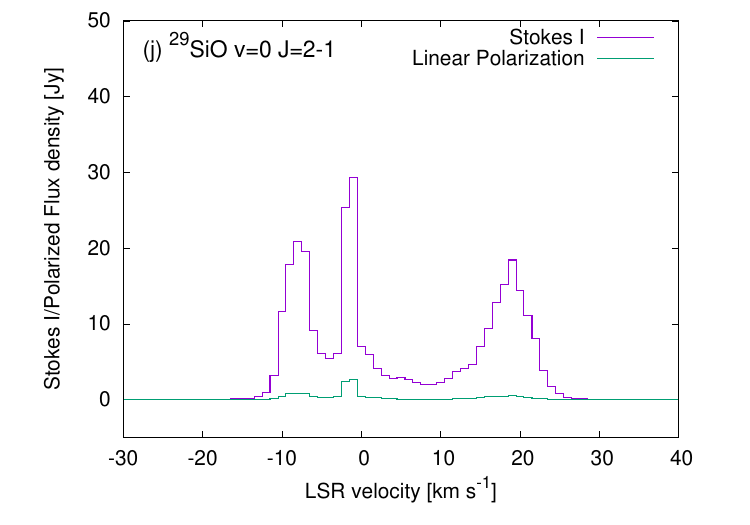} \\
\vspace{5mm}
\includegraphics[width=5.0cm]{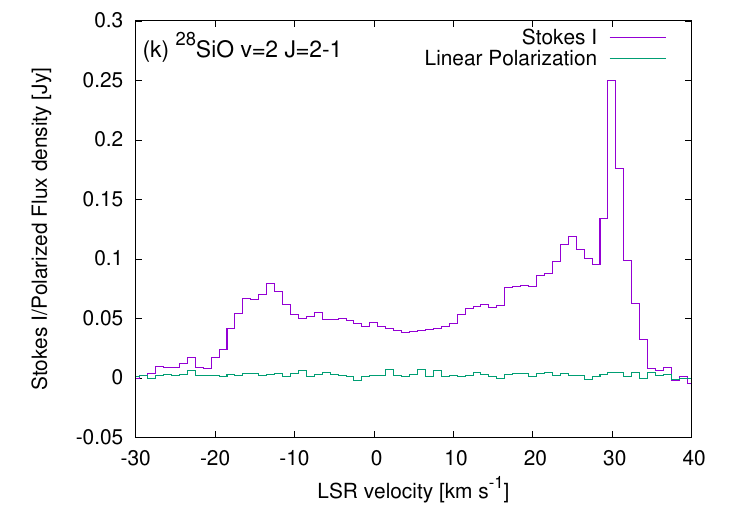}
\includegraphics[width=5.0cm]{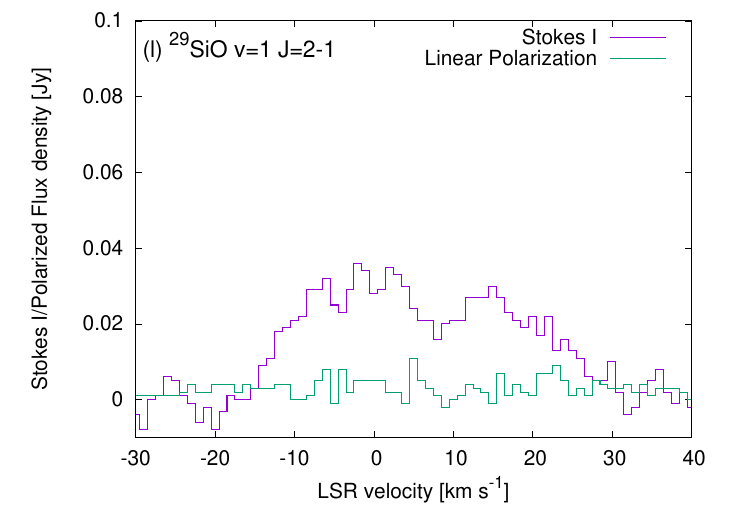}
\includegraphics[width=5.0cm]{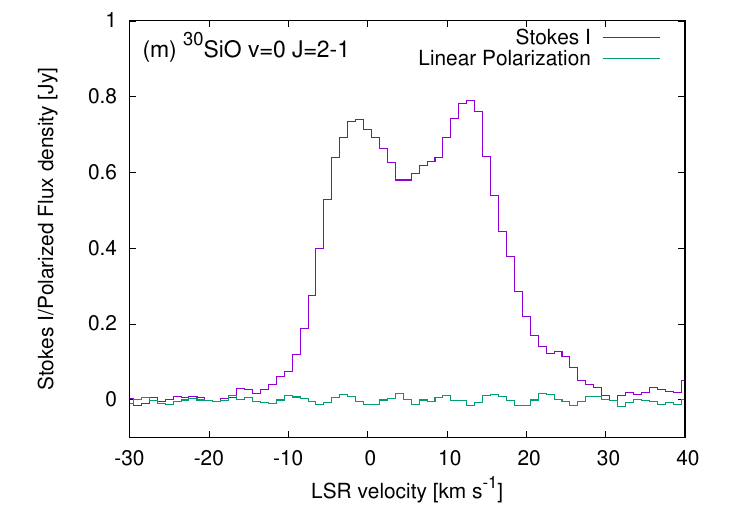}\\
\vspace{10mm}
\caption{Integrated spectra of all the SiO lines. 
All spectra are integrated over 4\arcsec$\times$4\arcsec \ for the $^{28}$SiO $v$=0 $J$=1-0 and 2-1 lines, 1\arcsec$\times$1\arcsec \ for the $^{28}$SiO $v$=1,2,3,4 $J$=1-0 and 2-1 and $^{29}$SiO $v$=1 $J$=2-1 lines, and 1.5\arcsec$\times$1.5\arcsec \ for the $^{29}$SiO and $^{30}$SiO $v$=0 $J$=1-0 and 2-1 lines. }
\label{fig-spall}
\end{center}
\end{figure}

\clearpage
\renewcommand{\thefigure}{\thesection\arabic{figure}}
\setcounter{figure}{0}

\section{All maps of observed SiO transitions}
\label{secA-map}

In this section, we show all the moment maps, polarization vector maps, and channel maps of SiO transitions observed in the present study as listed in Table \ref{tab-line}. 
Differences in the polarization angles between $^{28}$SiO $v$=0 $J$=1-0 and $v$=0 $J$=2-1 lines are also shown in Figure \ref{fig-chdiff}. 

\clearpage
\subsection{$^{28}$SiO $v$=0 $J$=1-0}

Figure B1 presents the moment maps and integrated emission maps of the $^{28}$SiO $v=0 J=1-0$ line and the 43 GHz continuum emission. Figures B2-B6 show channel maps of the $^{28}$SiO $v=0 J=1-0$ line. 

\begin{figure}[th]
\begin{center}
\includegraphics[width=7.5cm]{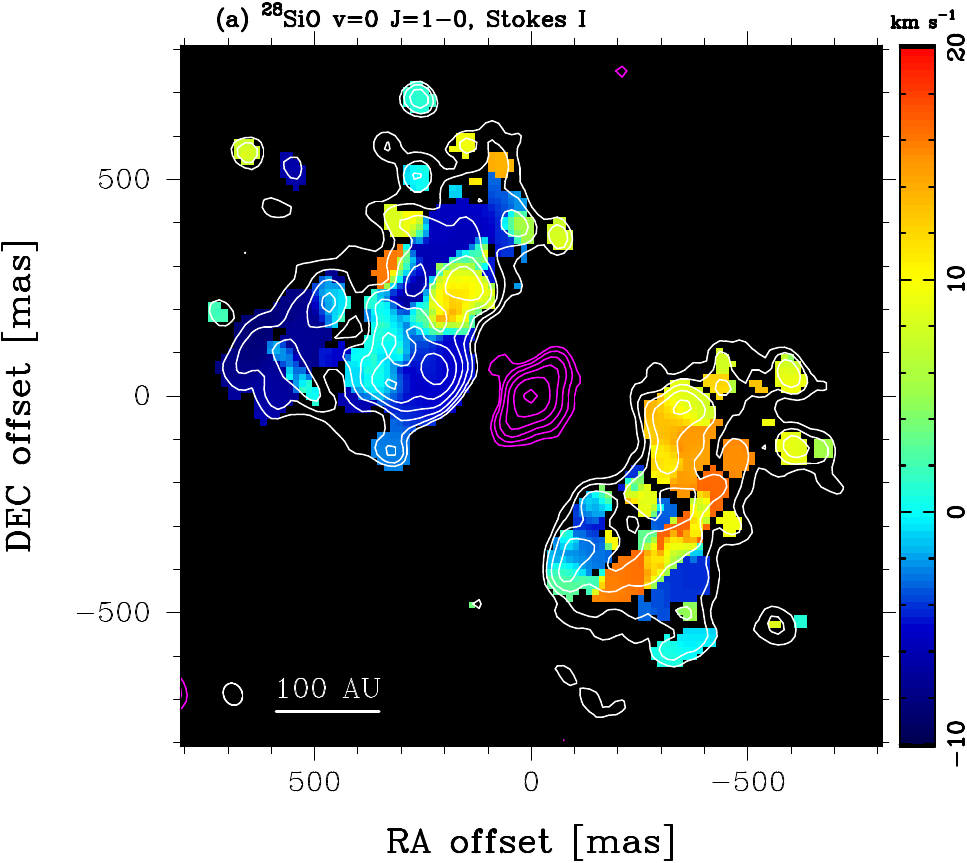}
\includegraphics[width=7.5cm]{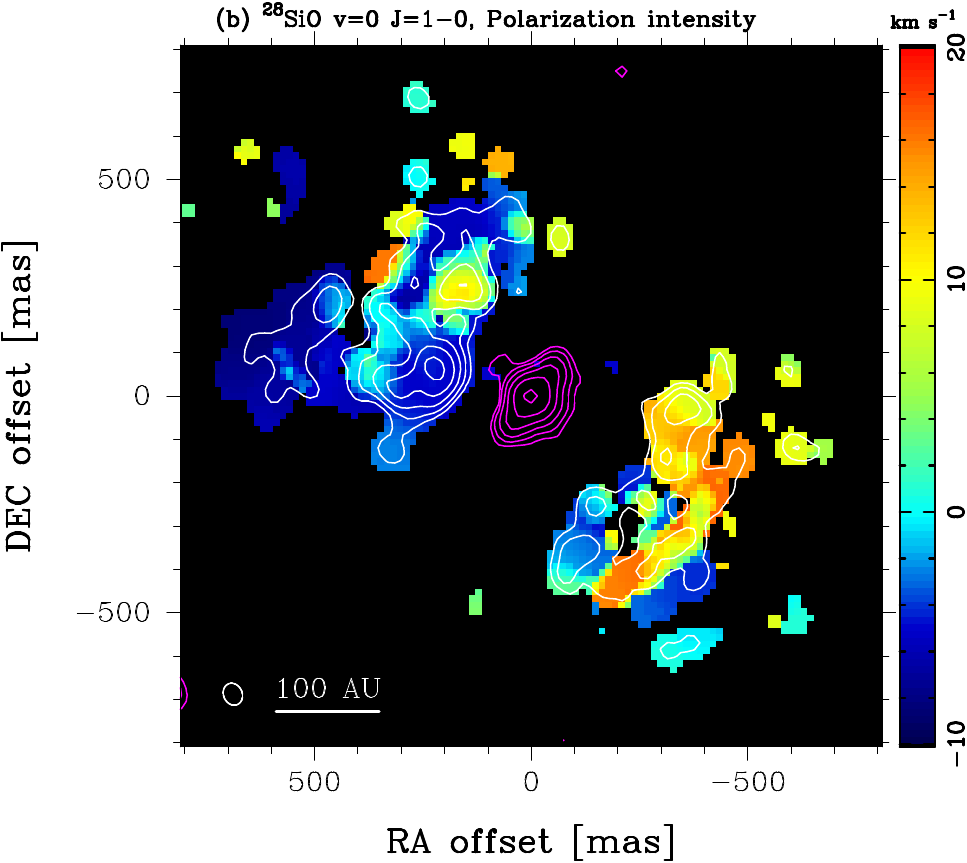}
\includegraphics[width=7.5cm]{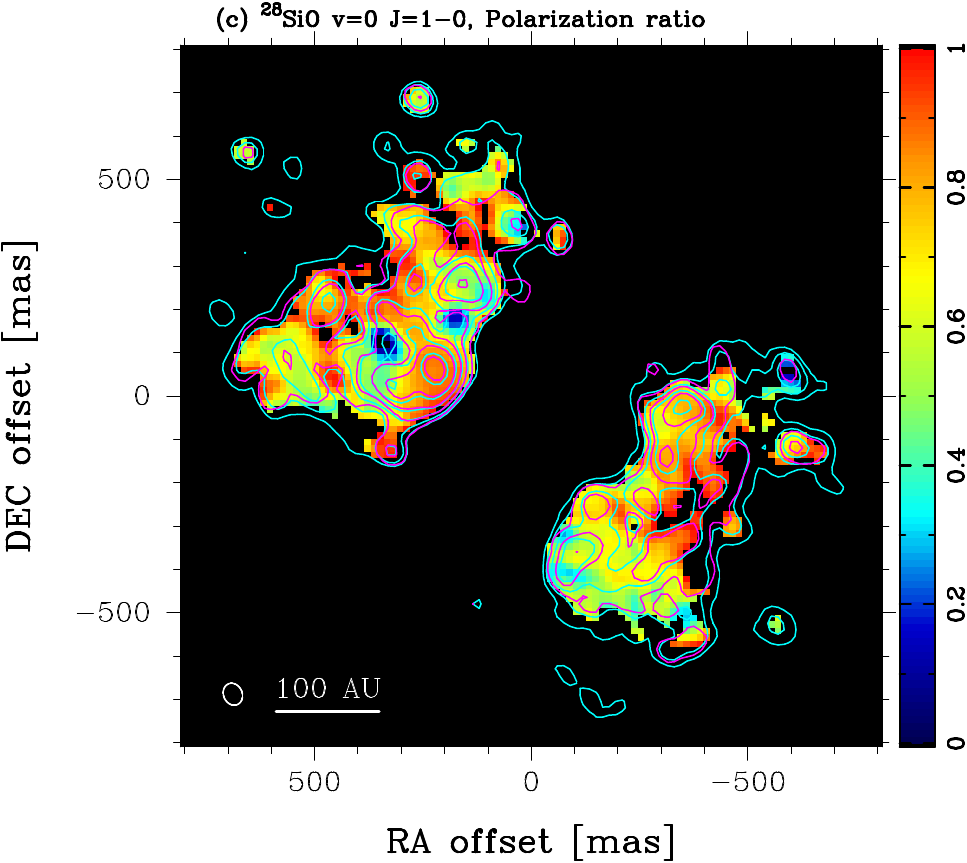}
\includegraphics[width=7.5cm]{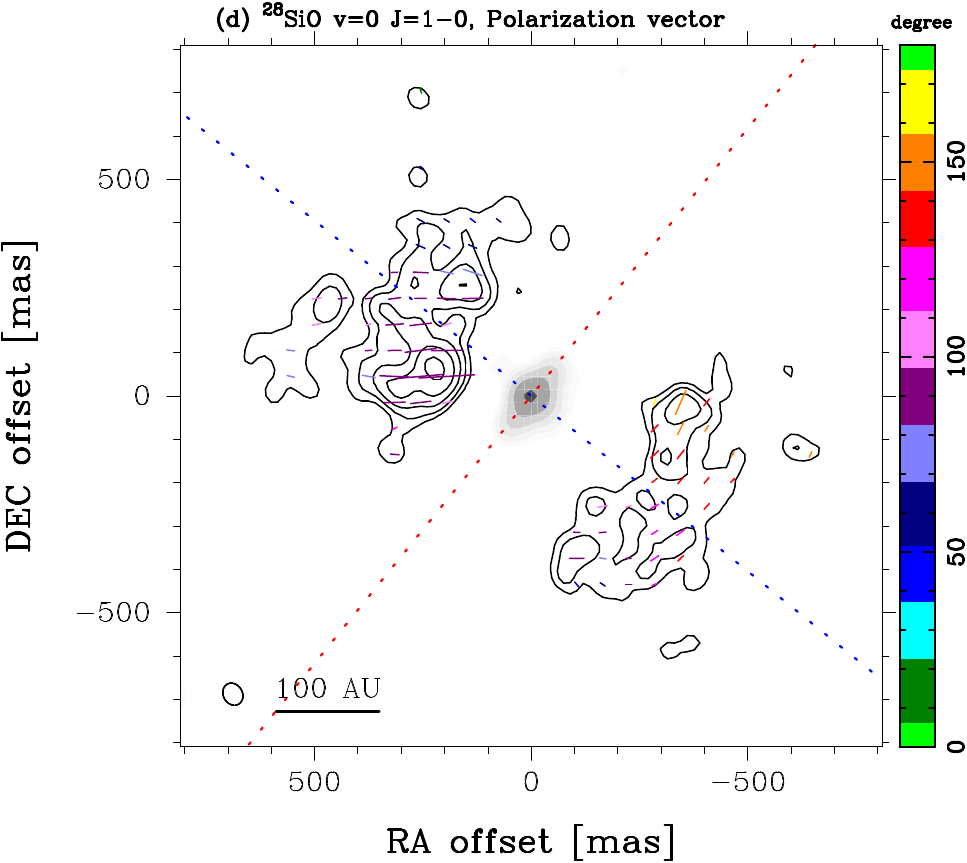}
\caption{
Maps of $^{28}$SiO $v$=0 $J$=1-0 line and the 43~GHz continuum emission. 
The beam size is indicated at the bottom-left corner of each panel. 
Moment maps are produced by using the velocity range from -10 to 20~km~s$^{-1}$. 
(a) Moment 0 (white contour) and Moment 1 (color) maps of Stokes I and the 43~GHz continuum (magenta contour). 
The contour  levels are 4, 8, 16, ... $\sigma$ with the rms noise level of 62~mJy~beam$^{-1}$~km~s$^{-1}$ for the Moment 0 and 0.017~mJy~beam$^{-1}$ for the continuum. 
(b) Same as (a) but for the linear polarization. 
The contour  levels are 4, 8, 16, ... $\sigma$ with the rms noise level of 136~mJy~beam$^{-1}$~km~s$^{-1}$ for the Moment 0. 
(c) The linear polarization ratio (color), the Moment 0 of Stokes I (cyan contour) and linear polarization (magenta contour). 
The contour levels are the same as in (a) and (b). 
(d) Polarization vectors (color) superposed on the linear polarization intensity (contour) and the 43~GHz continuum (gray). 
Polarization angles are calculated from Stokes Q and U images with velocity range from -10 to 20~km~s$^{-1}$. 
Color codes represent the polarization angles as shown in the vertical bar at the right of the panel. 
The contour levels are the same as in (a) and (b). 
The error in the polarization angle is smaller than 7~degrees for the linear polarization intensity higher than 4$\sigma$. 
The blue and red dashed lines indicate the outflow axis (51~degrees) and disk midplane (141~degrees), respectively \citep{Plambeck2016}. 
}
\end{center}
\end{figure}

\begin{figure}[th]
\begin{center}
\includegraphics[width=13cm]{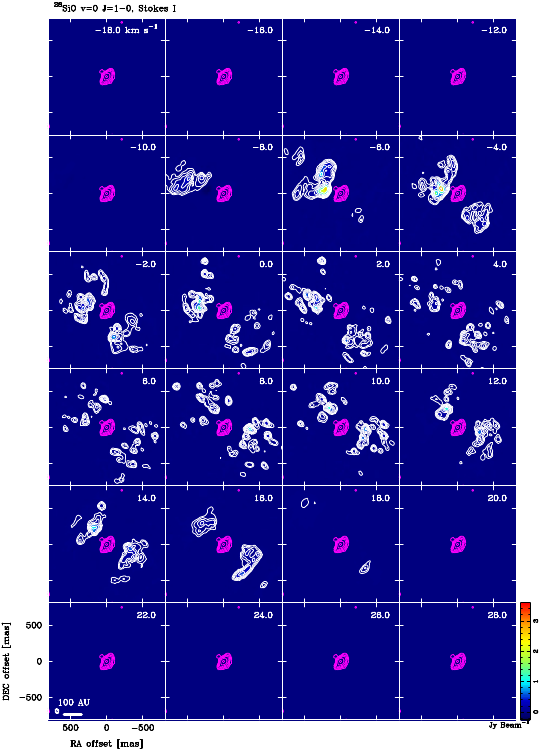}
\caption{
Stokes I channel map of the $^{28}$SiO $v$=0 $J$=1-0 line (color and white contours).
The contour levels are 4, 8, 16, ... $\sigma$ with the rms noise level of 6.6~mJy~beam$^{-1}$. 
Magenta contours show the 43~GHz continuum emission. 
Radial velocity ($v_{lsr}$) is indicated at the top-right corner of each panel. 
The beam sizes are plotted at the bottom-left corner of the bottm-left panel. 
}
\end{center}
\end{figure}

\begin{figure}[th]
\begin{center}
\includegraphics[width=13cm]{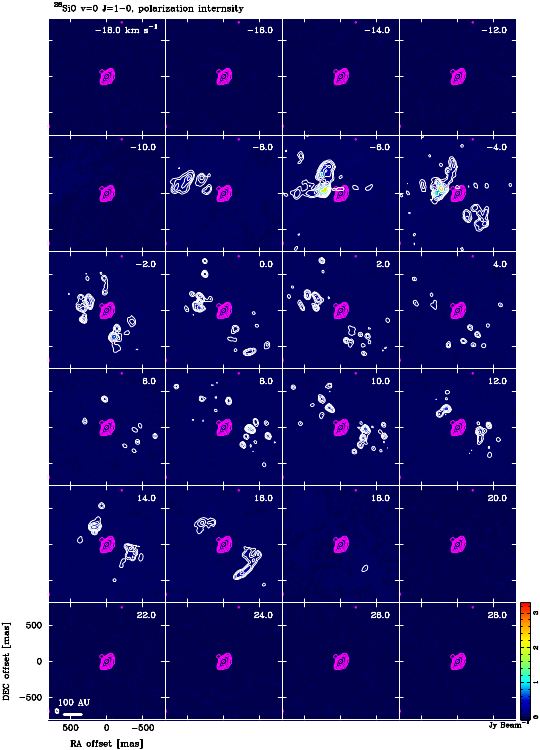}
\caption{
Linear polarization intensity channel map of the $^{28}$SiO $v$=0 $J$=1-0 line (color and white contours).
The contour levels are 4, 8, 16, ... $\sigma$ with the rms noise level of 11.9~mJy~beam$^{-1}$. 
Magenta contours show the 43~GHz continuum emission. 
Radial velocity ($v_{lsr}$) is indicated at the top-right corner of each panel. 
The beam sizes are plotted at the bottom-left corner of the bottm-left panel. 
}
\end{center}
\end{figure}

\begin{figure}[th]
\begin{center}
\includegraphics[width=13cm]{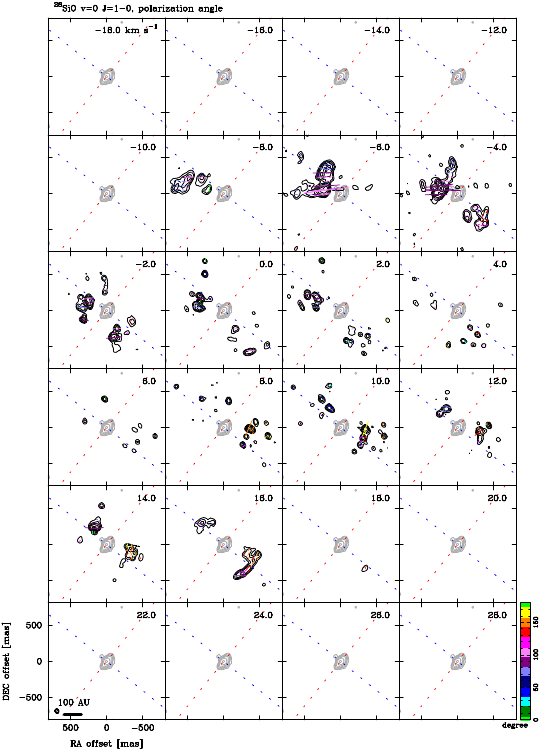}
\caption{
Channel map of the linear polarization intensity (black contours) and vectors (color lines) of the $^{28}$SiO $v$=0 $J$=1-0 line. 
Color codes represent the polarization angles as shown in the vertical bar at the right of the bottom-right panel. 
The error in the polarization angle is smaller than 7~degrees for the linear polarization intensity higher than 4$\sigma$. 
Gray contours show the 43~GHz continuum emission. 
Radial velocity ($v_{lsr}$) is indicated at the top-right corner of each panel. 
The beam sizes are plotted at the bottom-left corner of the bottm-left panel. 
The blue and red dashed lines indicate the outflow axis (51~degrees) and disk midplane (141~degrees), respectively \citep{Plambeck2016}. 
}
\end{center}
\end{figure}

\begin{figure}[th]
\begin{center}
\includegraphics[width=13cm]{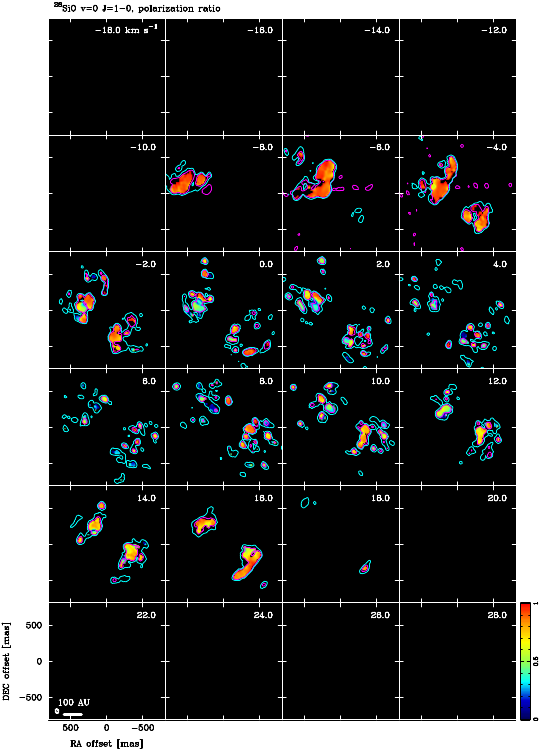}
\caption{
Channel map of the linear polarization ratio (color), Stokes I (cyan contour), and linear polarization intensity (magenta contours) of the $^{28}$SiO $v$=0 $J$=1-0 line. 
The contours show only the 4$\sigma$ levels to outline the distributions of the total and linearly polarized emission. 
Radial velocity ($v_{lsr}$) is indicated at the top-right corner of each panel. 
The beam sizes are plotted at the bottom-left corner of the bottm-left panel. 
}
\end{center}
\end{figure}

\begin{figure}[th]
\begin{center}
\includegraphics[width=13cm]{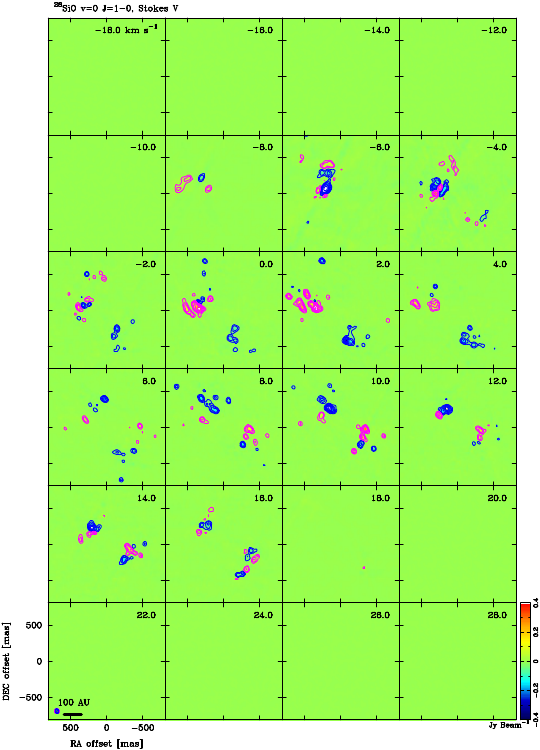}
\caption{Stokes V of the $^{28}$SiO $v$=0 $J$=1-0 line. 
The contour  levels are $\pm$4, $\pm$8, $\pm$16, $\pm$32, $\pm$64, and $\pm$128 times the rms noise levels of 4.6~mJy~beam$^{-1}$. 
Magenta and blue solid lines show the positive and negative levels, respectively. }
\end{center}
\end{figure}

\clearpage
\subsection{$^{28}$SiO $v$=1 $J$=1-0}

Figure B7 presents the moment maps and integrated emission maps of the $^{28}$SiO $v=1 J=1-0$ line and the 43 GHz continuum emission. Figures B8-B11 show channel maps of the $^{28}$SiO $v=1 J=1-0$ line. 

\begin{figure}[th]
\begin{center}
\includegraphics[width=7.5cm]{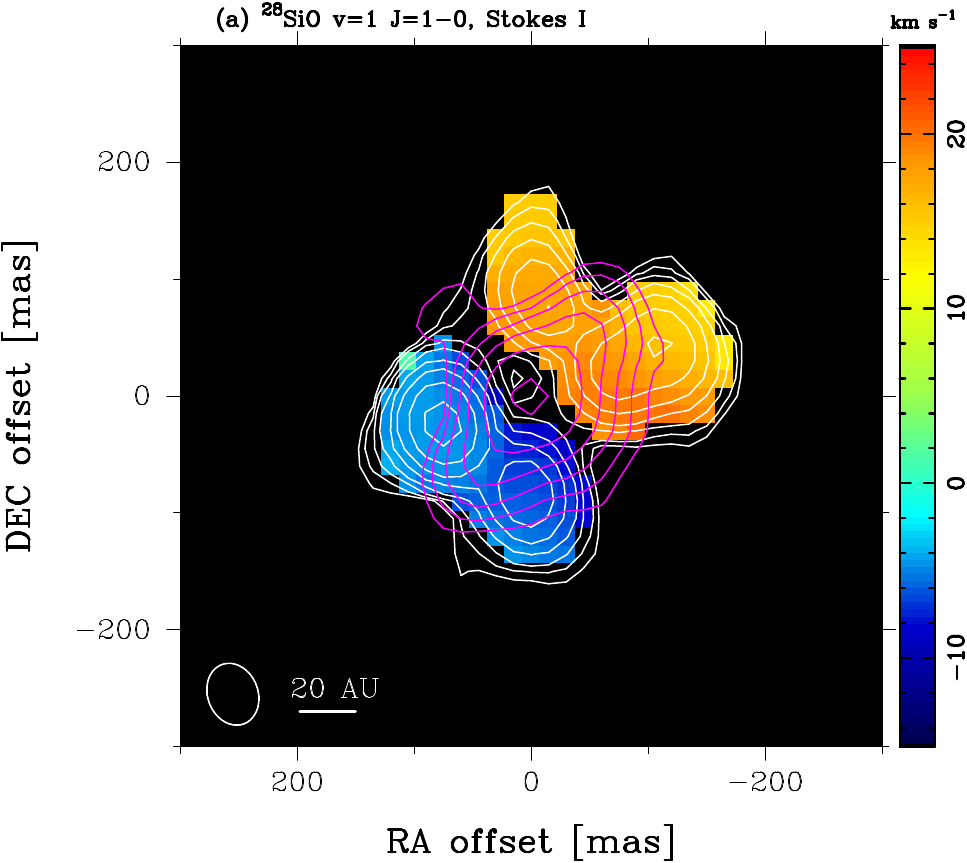}
\includegraphics[width=7.5cm]{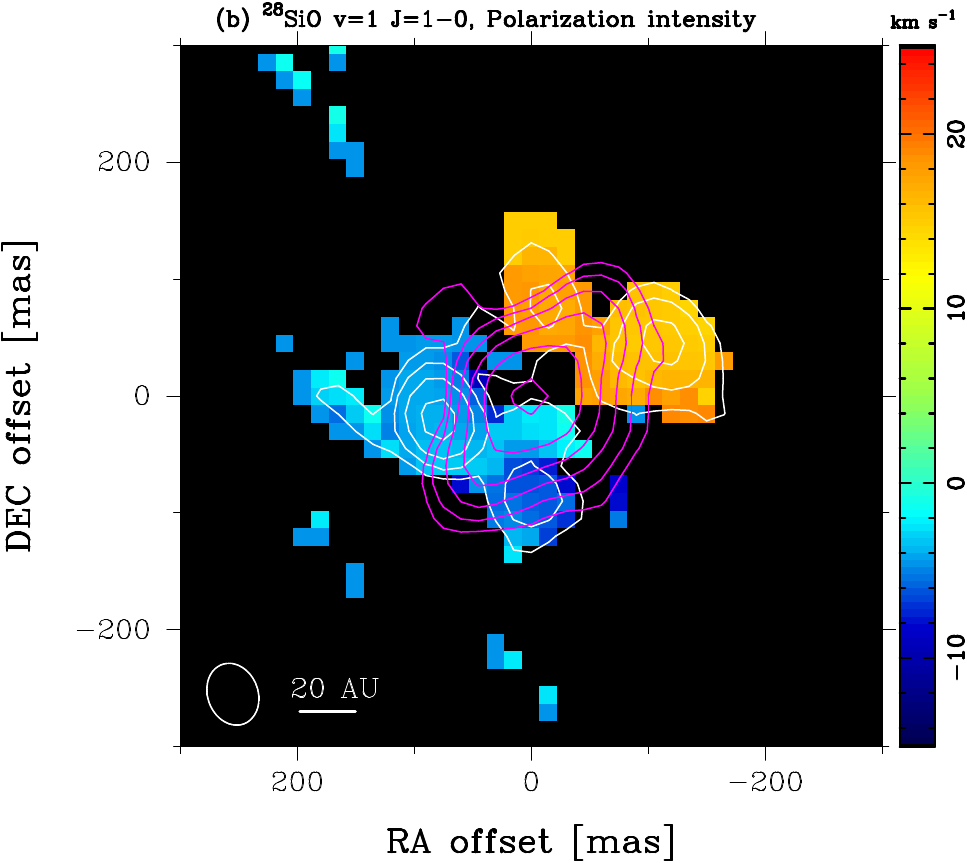}
\includegraphics[width=7.5cm]{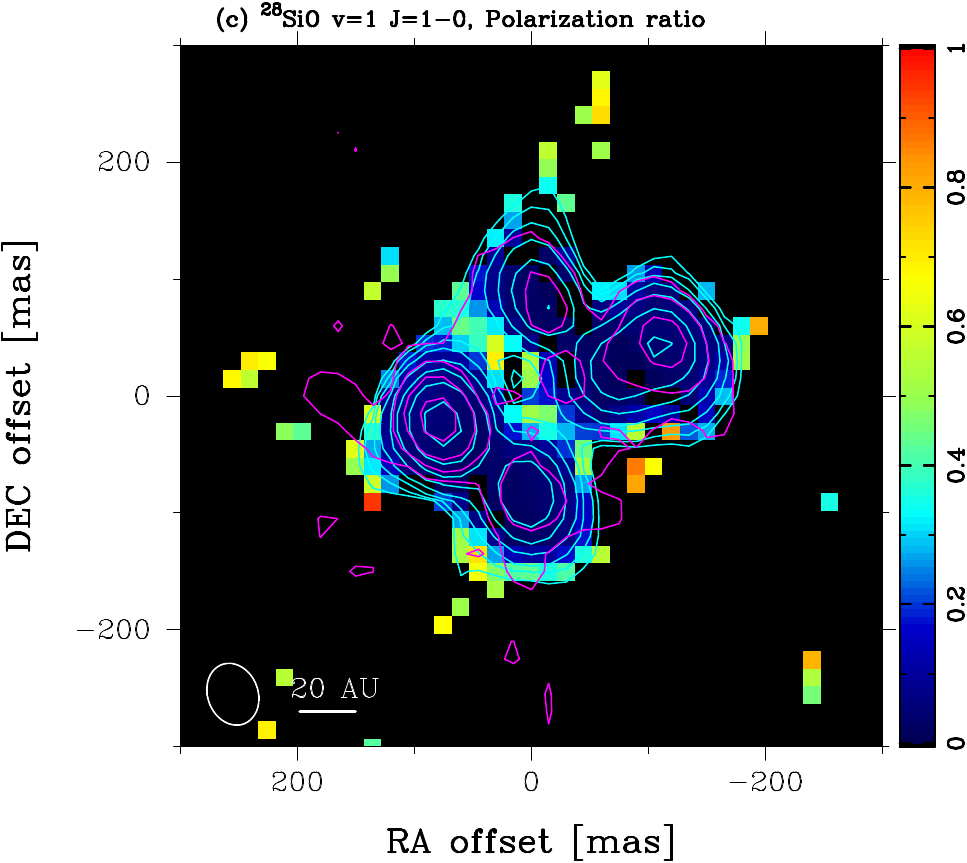}
\includegraphics[width=7.5cm]{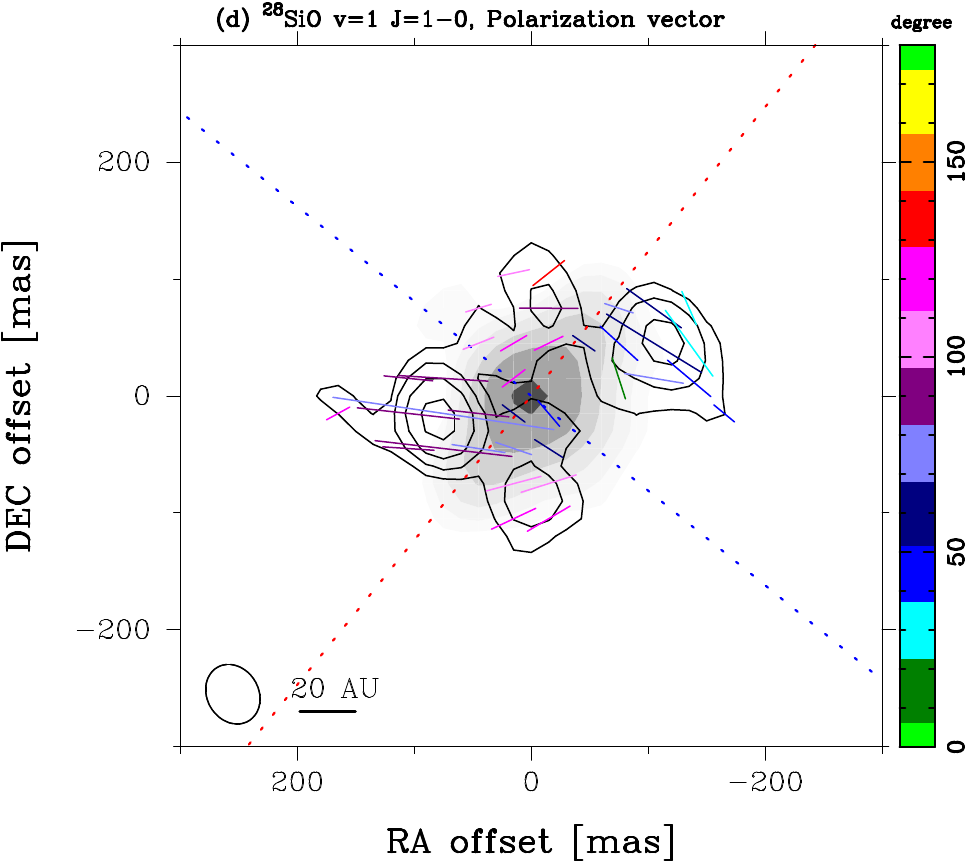}
\caption{
Maps of $^{28}$SiO $v$=1 $J$=1-0 line and the 43~GHz continuum emission. 
The beam size is indicated at the bottom-left corner of each panel. 
Moment maps are produced by using the velocity range from -15 to 25~km~s$^{-1}$. 
(a) Moment 0 (white contour) and Moment 1 (color) maps of Stokes I and the 43~GHz continuum (magenta contour). 
The contour  levels are 4, 8, 16, ... $\sigma$ with the rms noise level of 3.1~Jy~beam$^{-1}$~km~s$^{-1}$ for the Moment 0 and 0.017~mJy~beam$^{-1}$ for the continuum. 
(b) Same as (a) but for the linear polarization. 
The contour  levels are 4, 8, 16, ... $\sigma$ with the rms noise level of 3.0~Jy~beam$^{-1}$~km~s$^{-1}$ for the Moment 0. 
(c) The linear polarization ratio (color), the Moment 0 of Stokes I (cyan contour) and linear polarization (magenta contour). 
The contour levels are the same as in (a) and (b). 
(d) Polarization vectors (color) superposed on the linear polarization intensity (contour) and the 43~GHz continuum (gray). 
Polarization angles are calculated from Stokes Q and U images with velocity range from -10 to 20~km~s$^{-1}$. 
Color codes represent the polarization angles as shown in the vertical bar at the right of the panel. 
The contour levels are the same as in (a) and (b). 
The error in the polarization angle is smaller than 7~degrees for the linear polarization intensity higher than 4$\sigma$. 
The blue and red dashed lines indicate the outflow axis (51~degrees) and disk midplane (141~degrees), respectively \citep{Plambeck2016}. 
}
\end{center}
\end{figure}

\begin{figure}[th]
\begin{center}
\includegraphics[width=13cm]{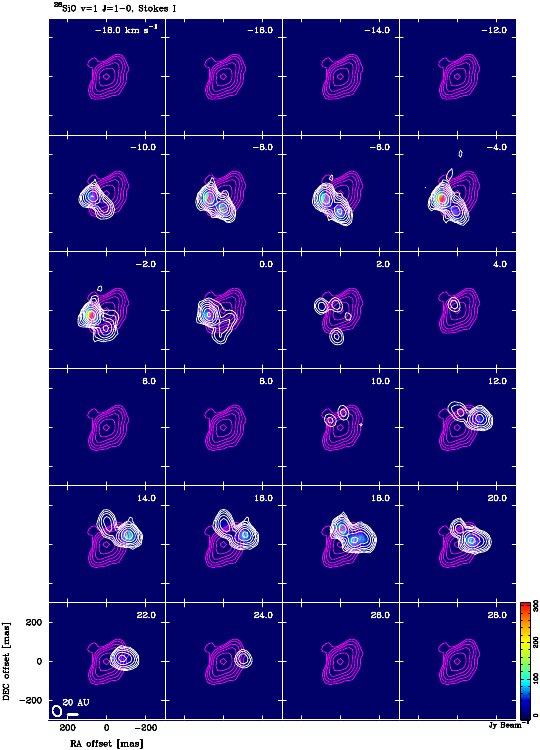}
\caption{
Stokes I channel map of the $^{28}$SiO $v$=1 $J$=1-0 line (color and white contours).
The contour levels are 4, 8, 16, ... $\sigma$ with the rms noise level of 0.34~Jy~beam$^{-1}$. 
Magenta contours show the 43~GHz continuum emission. 
Radial velocity ($v_{lsr}$) is indicated at the top-right corner of each panel. 
The beam sizes are plotted at the bottom-left corner of the bottm-left panel. 
}
\end{center}
\end{figure}

\begin{figure}[th]
\begin{center}
\includegraphics[width=13cm]{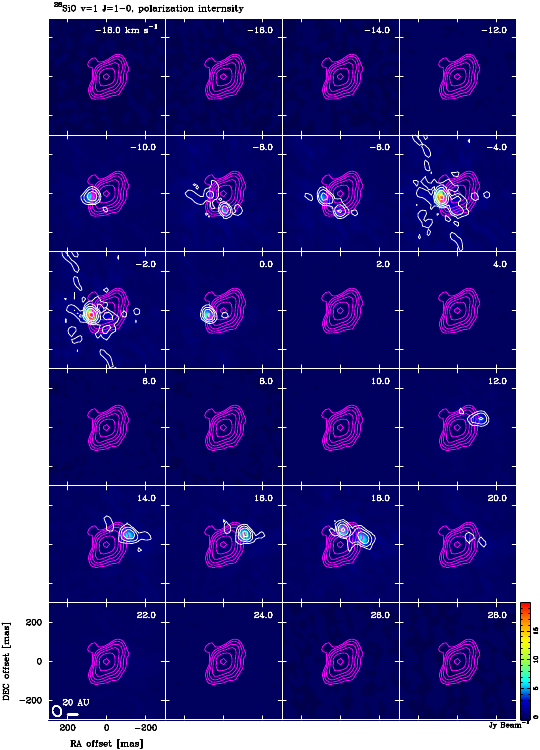}
\caption{
Linear polarization intensity channel map of the $^{28}$SiO $v$=1 $J$=1-0 line (color and white contours).
The contour levels are 4, 8, 16, ... $\sigma$ with the rms noise level of 0.25~Jy~beam$^{-1}$. 
Magenta contours show the 43~GHz continuum emission. 
Radial velocity ($v_{lsr}$) is indicated at the top-right corner of each panel. 
The beam sizes are plotted at the bottom-left corner of the bottm-left panel. 
}
\end{center}
\end{figure}

\begin{figure}[th]
\begin{center}
\includegraphics[width=13cm]{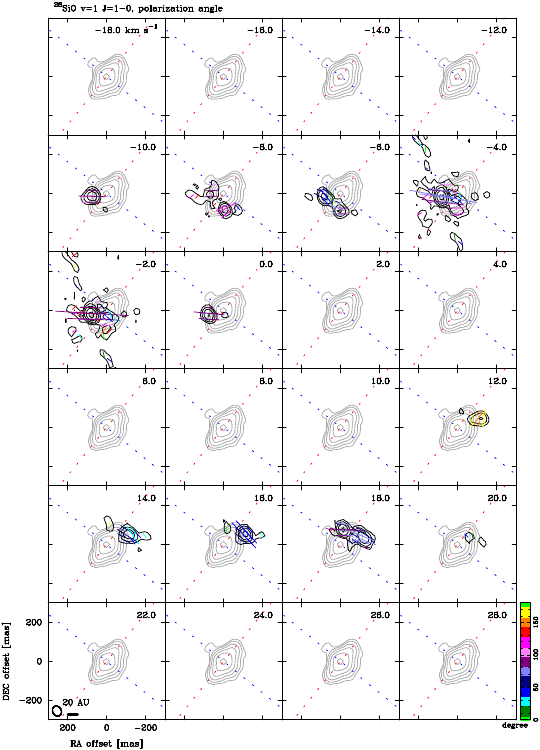}
\caption{
Channel map of the linear polarization intensity (black contours) and vectors (color lines) of the $^{28}$SiO $v$=1 $J$=1-0 line. 
Color codes represent the polarization angles as shown in the vertical bar at the right of the bottom-right panel. 
The error in the polarization angle is smaller than 7~degrees for the linear polarization intensity higher than 4$\sigma$. 
Gray contours show the 43~GHz continuum emission. 
Radial velocity ($v_{lsr}$) is indicated at the top-right corner of each panel. 
The beam sizes are plotted at the bottom-left corner of the bottm-left panel. 
The blue and red dashed lines indicate the outflow axis (51~degrees) and disk midplane (141~degrees), respectively \citep{Plambeck2016}. 
}
\end{center}
\end{figure}

\begin{figure}[th]
\begin{center}
\includegraphics[width=13cm]{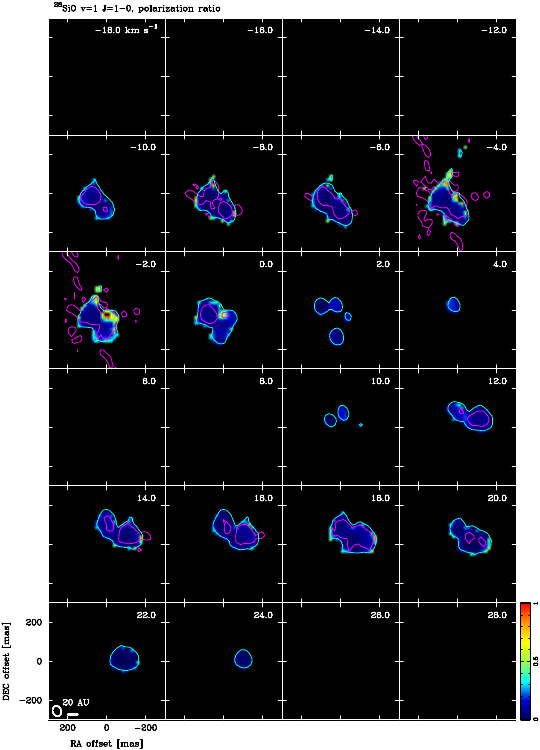}
\caption{
Channel map of the linear polarization ratio (color), Stokes I (cyan contour), and linear polarization intensity (magenta contours) of the $^{28}$SiO $v$=1 $J$=1-0 line. 
The contours show only the 4$\sigma$ levels to outline the distributions of the total and linearly polarized emission. 
Radial velocity ($v_{lsr}$) is indicated at the top-right corner of each panel. 
The beam sizes are plotted at the bottom-left corner of the bottm-left panel. 
}
\end{center}
\end{figure}

\clearpage
\subsection{$^{28}$SiO $v$=2 $J$=1-0}

Figure B12 presents the moment maps and integrated emission maps of the $^{28}$SiO $v=2 J=1-0$ line and the 43 GHz continuum emission. Figures B13-B16 show channel maps of the $^{28}$SiO $v=2 J=1-0$ line. 

\begin{figure}[th]
\begin{center}
\includegraphics[width=7.5cm]{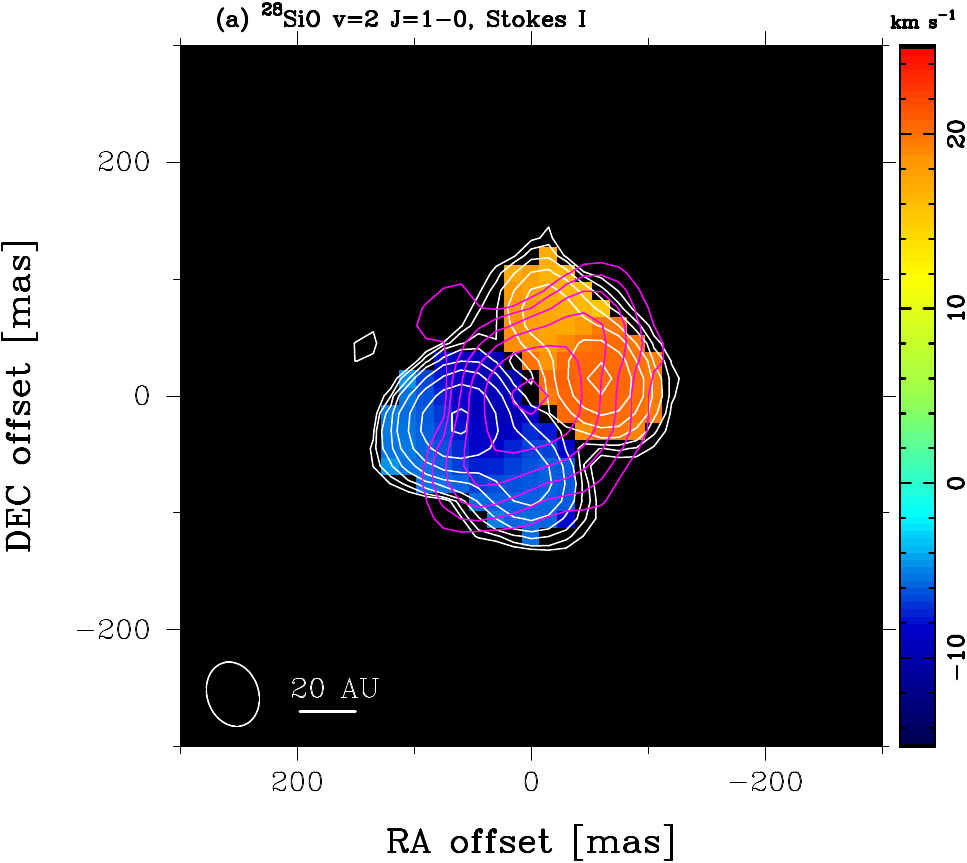}
\includegraphics[width=7.5cm]{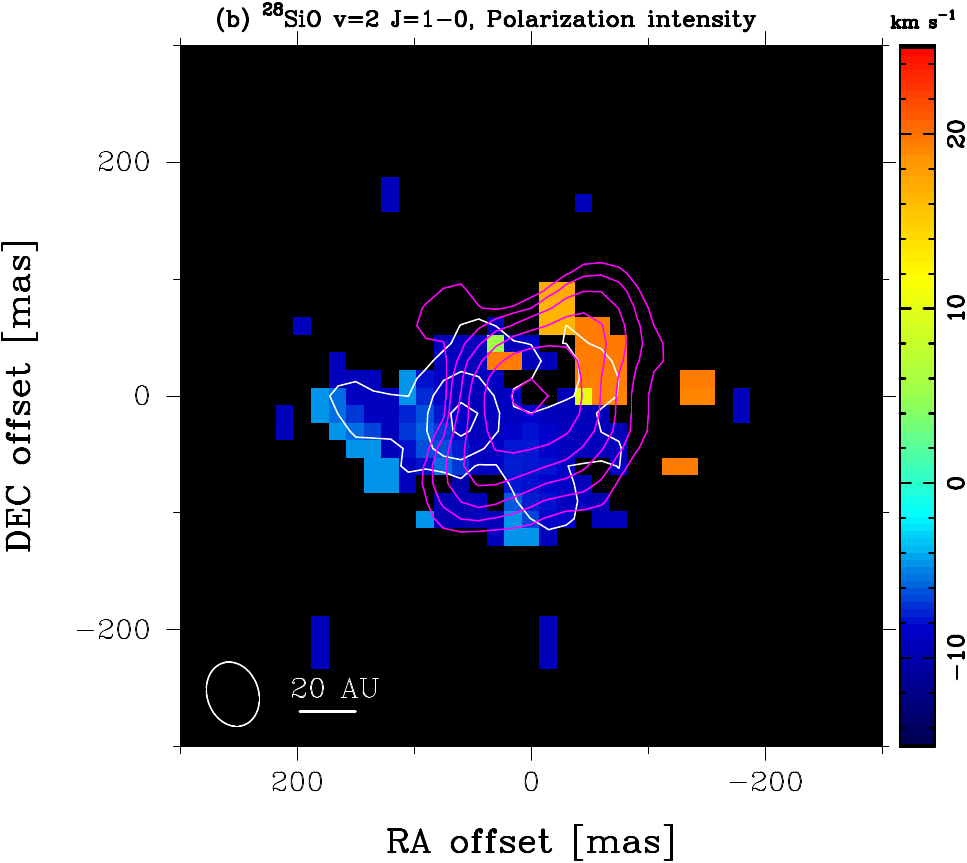}
\includegraphics[width=7.5cm]{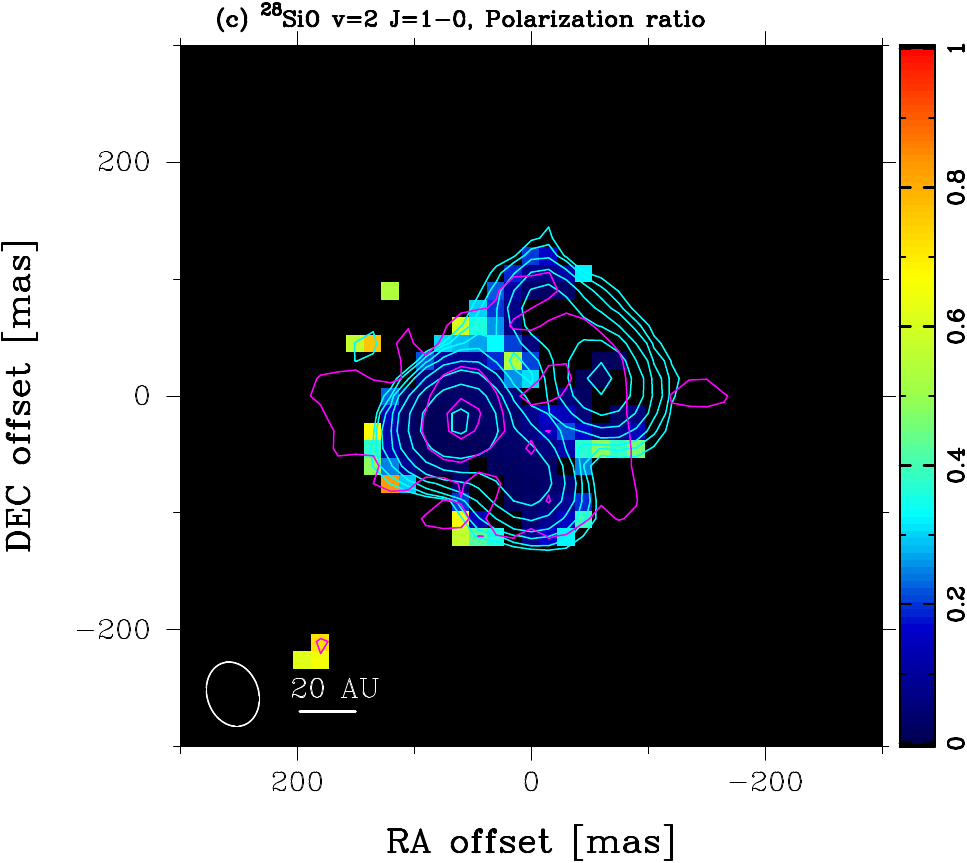}
\includegraphics[width=7.5cm]{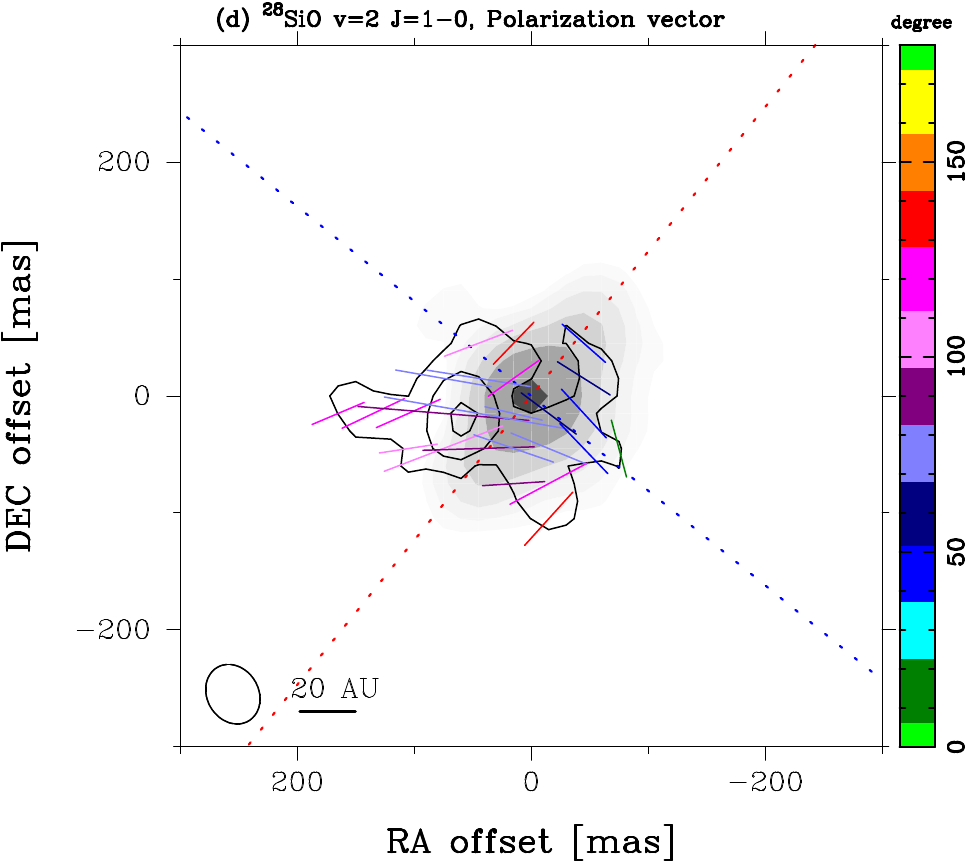}
\caption{
Maps of $^{28}$SiO $v$=2 $J$=1-0 line and the 43~GHz continuum emission. 
The beam size is indicated at the bottom-left corner of each panel. 
Moment maps are produced by using the velocity range from -15 to 25~km~s$^{-1}$. 
(a) Moment 0 (white contour) and Moment 1 (color) maps of Stokes I and the 43~GHz continuum (magenta contour). 
The contour  levels are 4, 8, 16, ... $\sigma$ with the rms noise level of 2.1~Jy~beam$^{-1}$~km~s$^{-1}$ for the Moment 0 and 0.017~mJy~beam$^{-1}$ for the continuum. 
(b) Same as (a) but for the linear polarization. 
The contour  levels are 4, 8, 16, ... $\sigma$ with the rms noise level of 1.96~Jy~beam$^{-1}$~km~s$^{-1}$ for the Moment 0. 
(c) The linear polarization ratio (color), the Moment 0 of Stokes I (cyan contour) and linear polarization (magenta contour). 
The contour levels are the same as in (a) and (b). 
(d) Polarization vectors (color) superposed on the linear polarization intensity (contour) and the 43~GHz continuum (gray). 
Polarization angles are calculated from Stokes Q and U images with velocity range from -10 to 20~km~s$^{-1}$. 
Color codes represent the polarization angles as shown in the vertical bar at the right of the panel. 
The contour levels are the same as in (a) and (b). 
The error in the polarization angle is smaller than 7~degrees for the linear polarization intensity higher than 4$\sigma$. 
The blue and red dashed lines indicate the outflow axis (51~degrees) and disk midplane (141~degrees), respectively \citep{Plambeck2016}. 
}
\end{center}
\end{figure}

\begin{figure}[th]
\begin{center}
\includegraphics[width=13cm]{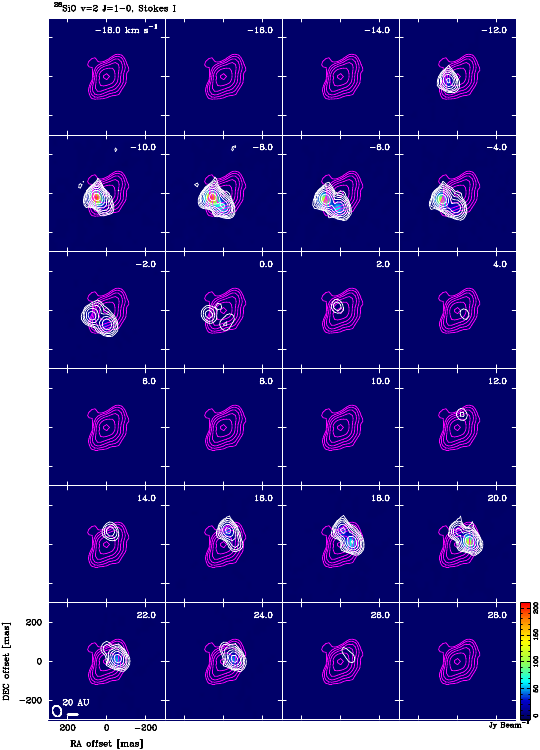}
\caption{
Stokes I channel map of the $^{28}$SiO $v$=2 $J$=1-0 line (color and white contours).
The contour levels are 4, 8, 16, ... $\sigma$ with the rms noise level of 0.28~Jy~beam$^{-1}$. 
Magenta contours show the 43~GHz continuum emission. 
Radial velocity ($v_{lsr}$) is indicated at the top-right corner of each panel. 
The beam sizes are plotted at the bottom-left corner of the bottm-left panel. 
}
\end{center}
\end{figure}

\begin{figure}[th]
\begin{center}
\includegraphics[width=13cm]{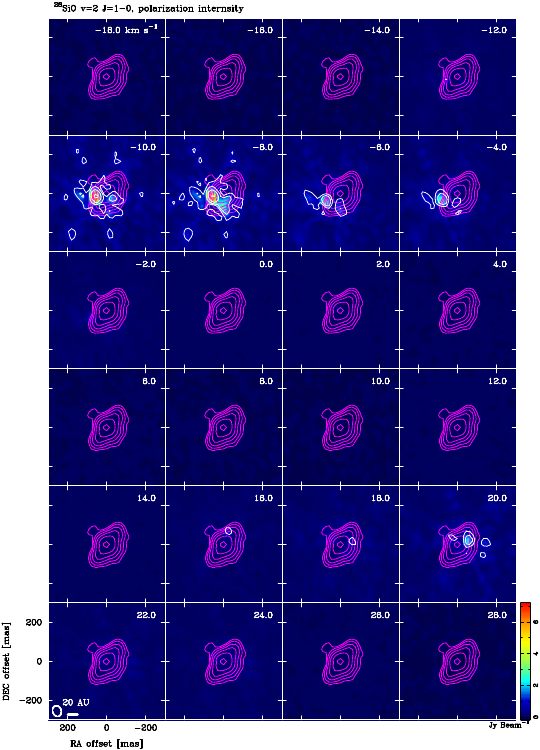}
\caption{
Linear polarization intensity channel map of the $^{28}$SiO $v$=2 $J$=1-0 line (color and white contours).
The contour levels are 4, 8, 16, ... $\sigma$ with the rms noise level of 0.196~Jy~beam$^{-1}$. 
Magenta contours show the 43~GHz continuum emission. 
Radial velocity ($v_{lsr}$) is indicated at the top-right corner of each panel. 
The beam sizes are plotted at the bottom-left corner of the bottm-left panel. 
}
\end{center}
\end{figure}

\begin{figure}[th]
\begin{center}
\includegraphics[width=13cm]{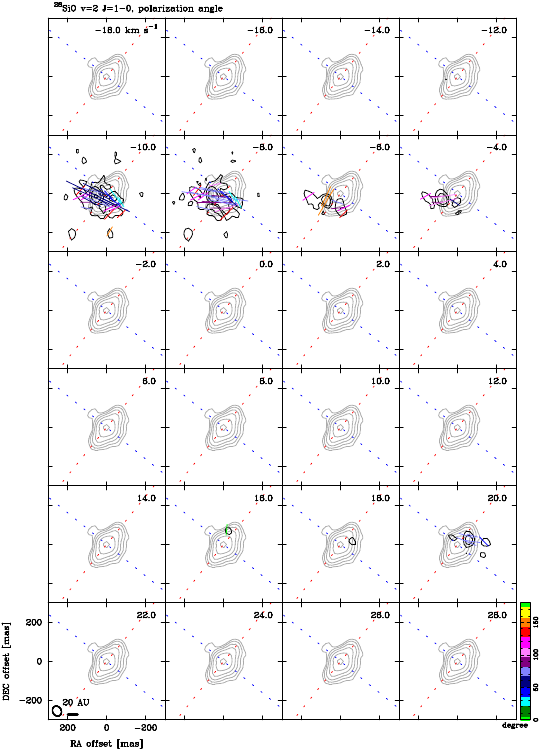}
\caption{
Channel map of the linear polarization intensity (black contours) and vectors (color lines) of the $^{28}$SiO $v$=2 $J$=1-0 line. 
Color codes represent the polarization angles as shown in the vertical bar at the right of the bottom-right panel. 
The error in the polarization angle is smaller than 7~degrees for the linear polarization intensity higher than 4$\sigma$. 
Gray contours show the 43~GHz continuum emission. 
Radial velocity ($v_{lsr}$) is indicated at the top-right corner of each panel. 
The beam sizes are plotted at the bottom-left corner of the bottm-left panel. 
The blue and red dashed lines indicate the outflow axis (51~degrees) and disk midplane (141~degrees), respectively \citep{Plambeck2016}. 
}
\end{center}
\end{figure}

\begin{figure}[th]
\begin{center}
\includegraphics[width=13cm]{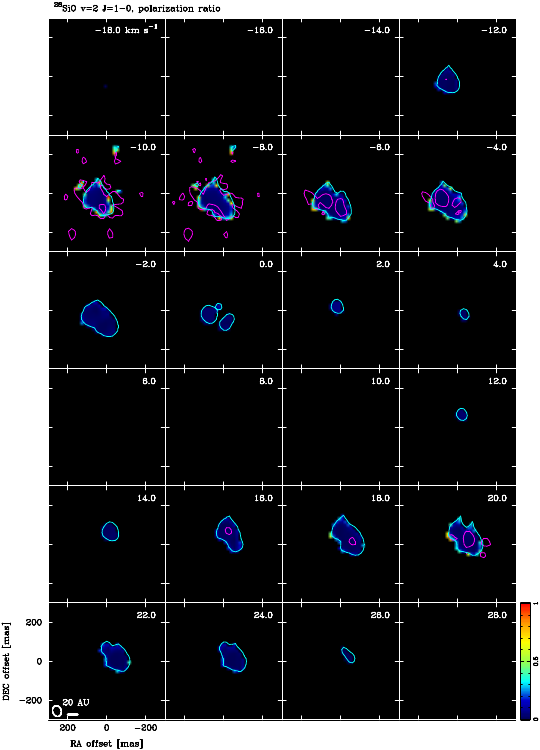}
\caption{
Channel map of the linear polarization ratio (color), Stokes I (cyan contour), and linear polarization intensity (magenta contours) of the $^{28}$SiO $v$=2 $J$=1-0 line. 
The contours show only the 4$\sigma$ levels to outline the distributions of the total and linear polarized emission. 
Radial velocity ($v_{lsr}$) is indicated at the top-right corner of each panel. 
The beam sizes are plotted at the bottom-left corner of the bottm-left panel. 
}
\end{center}
\end{figure}

\clearpage
\subsection{$^{29}$SiO $v$=0, and $^{30}$SiO $v$=0 $J$=1-0}

Figure B17 presents the moment maps of the $^{29}$SiO and $^{30}$SiO $v=0 J=1-0$ lines and the 43 GHz continuum emission. Figures B18 and B19 show channel maps of the $^{29}$SiO and $^{30}$SiO $v=0 J=1-0$ lines, respectively. 

\begin{figure}[th]
\begin{center}
\includegraphics[width=7.5cm]{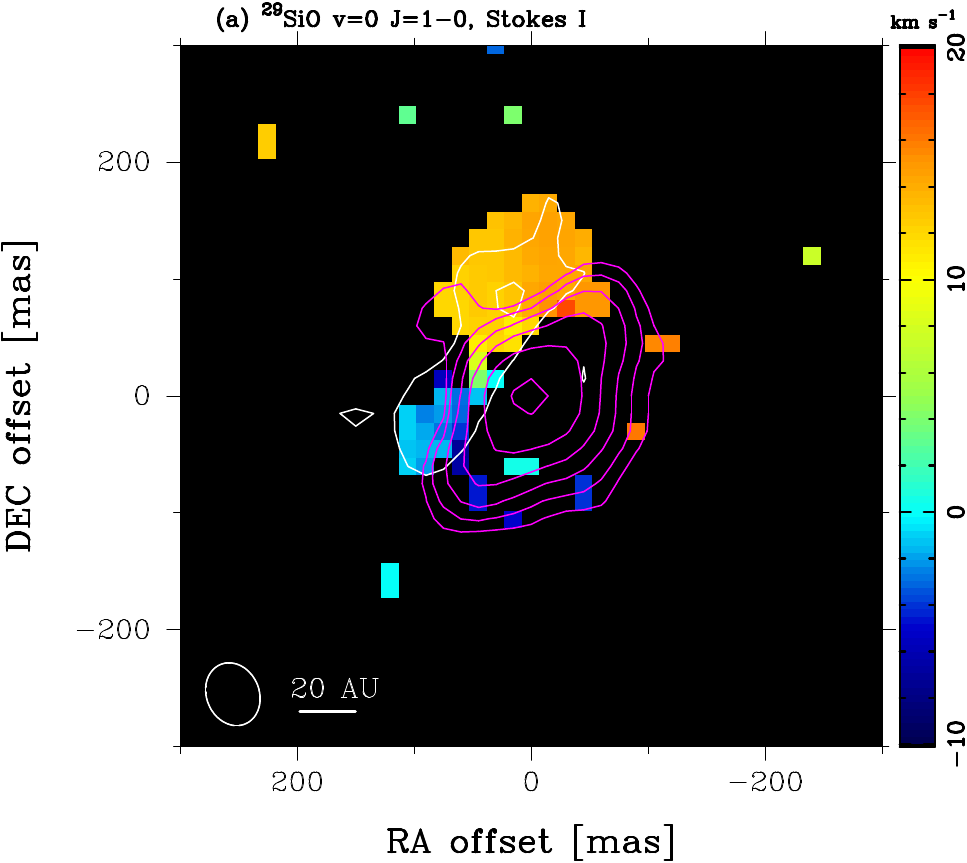}
\includegraphics[width=7.5cm]{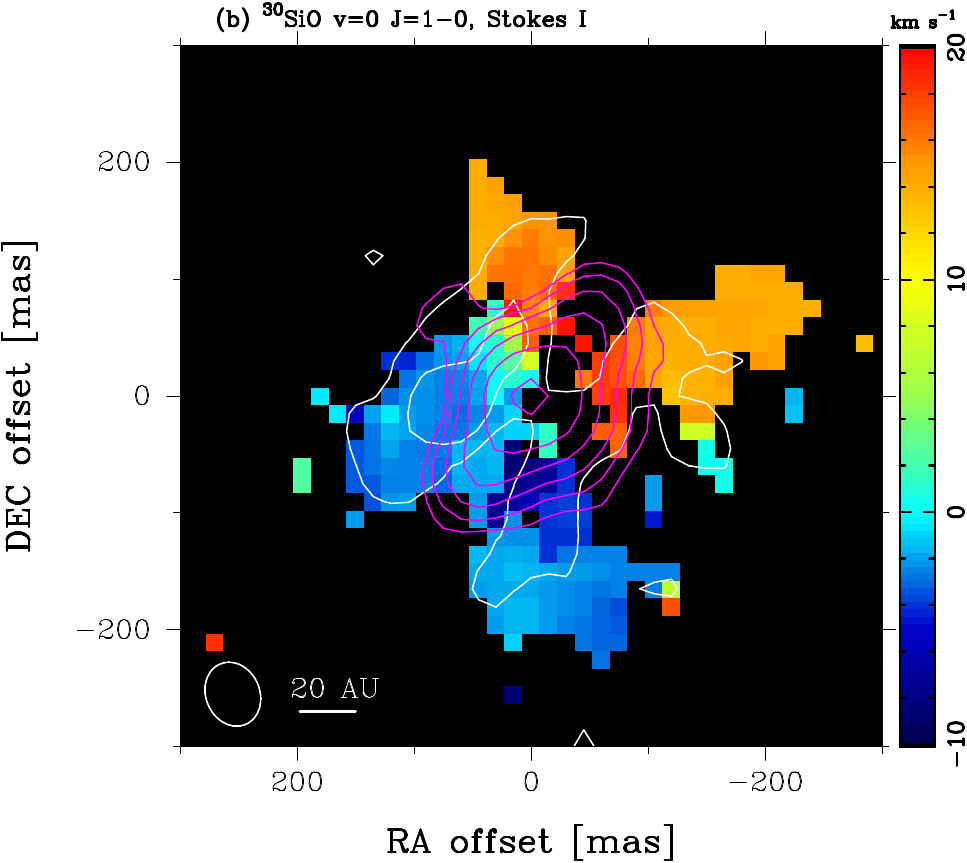}
\caption{
Moment 0 (white contour) and Moment 1 (color) maps of Stokes I and the 43~GHz continuum (magenta contour). 
The beam size is indicated at the bottom-left corner of each panel. 
Moment maps are produced by using the velocity range from -10 to 20~km~s$^{-1}$ for both lines. 
(a) The $^{29}$SiO $v$=0 $J$=1-0 line. 
The contour  levels are 4, 8, 16, ... $\sigma$ with the rms noise level of 9.5~mJy~beam$^{-1}$~km~s$^{-1}$ for the Moment 0 and 0.017~mJy~beam$^{-1}$ for the continuum. 
(b) Same as (a) but $^{30}$SiO $v$=0 $J$=1-0 line. 
The contour  levels are 4, 8, 16, ... $\sigma$ with the rms noise level of 9.3~mJy~beam$^{-1}$~km~s$^{-1}$ for the Moment 0. 
}
\end{center}
\end{figure}

\begin{figure}[th]
\begin{center}
\includegraphics[width=13cm]{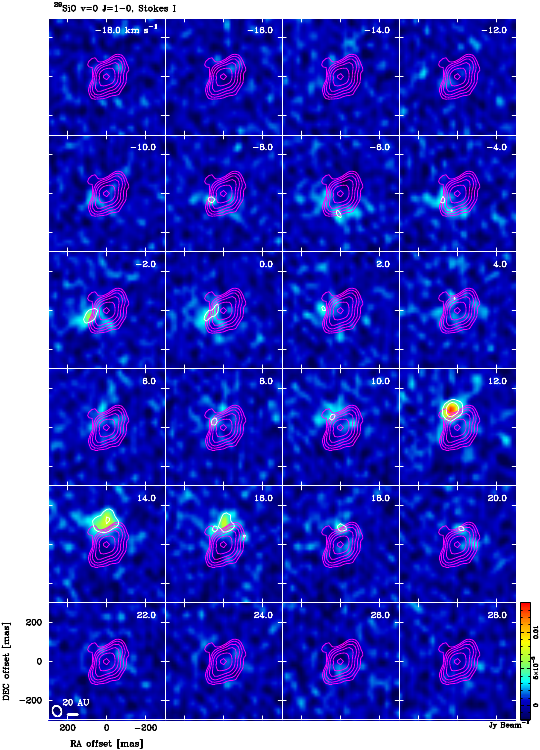}
\caption{
Stokes I channel map of the $^{29}$SiO $v$=0 $J$=1-0 line (color and white contours).
The contour levels are 4, 8, 16, ... $\sigma$ with the rms noise level of 0.91~mJy~beam$^{-1}$. 
Magenta contours show the 43~GHz continuum emission. 
Radial velocity ($v_{lsr}$) is indicated at the top-right corner of each panel. 
The beam sizes are plotted at the bottom-left corner of the bottm-left panel. 
}
\end{center}
\end{figure}

\begin{figure}[th]
\begin{center}
\includegraphics[width=13cm]{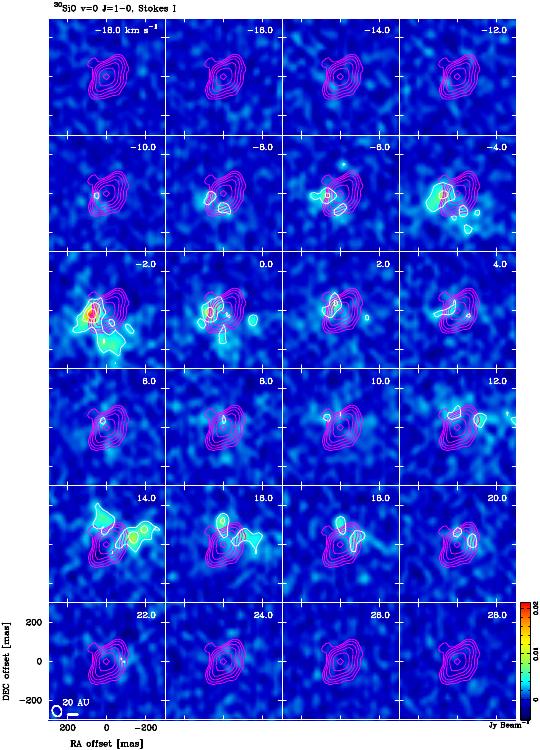}
\caption{
Stokes I channel map of the $^{30}$SiO $v$=0 $J$=1-0 line (color and white contours).
The contour levels are 4, 8, 16, ... $\sigma$ with the rms noise level of 0.84~mJy~beam$^{-1}$. 
Magenta contours show the 43~GHz continuum emission. 
Radial velocity ($v_{lsr}$) is indicated at the top-right corner of each panel. 
The beam sizes are plotted at the bottom-left corner of the bottm-left panel. 
}
\end{center}
\end{figure}

\clearpage
\subsection{$^{28}$SiO $v$=0 $J$=2-1}

Figure B20 presents the moment maps and integrated emission maps of the $^{28}$SiO $v=0 J=2-1$ line and the 96 GHz continuum emission. Figures B21-B25 show channel maps of the $^{28}$SiO $v=0 J=2-1$ line. 

\begin{figure}[th]
\begin{center}
\includegraphics[width=7.5cm]{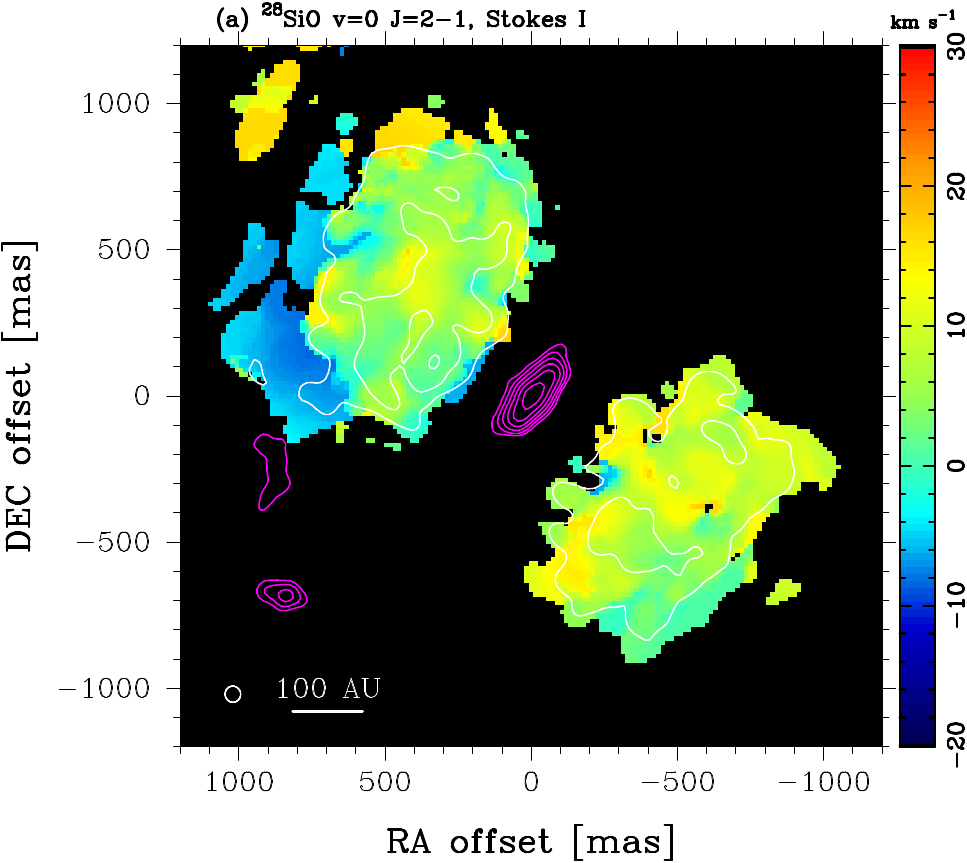}
\includegraphics[width=7.5cm]{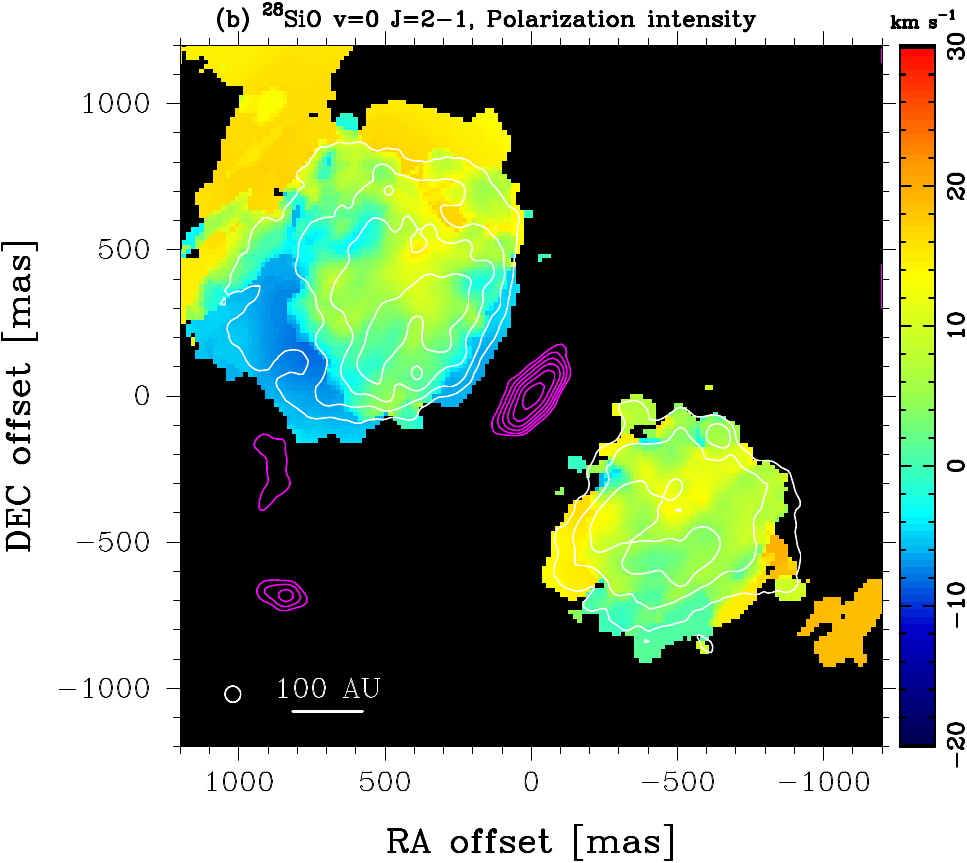}
\includegraphics[width=7.5cm]{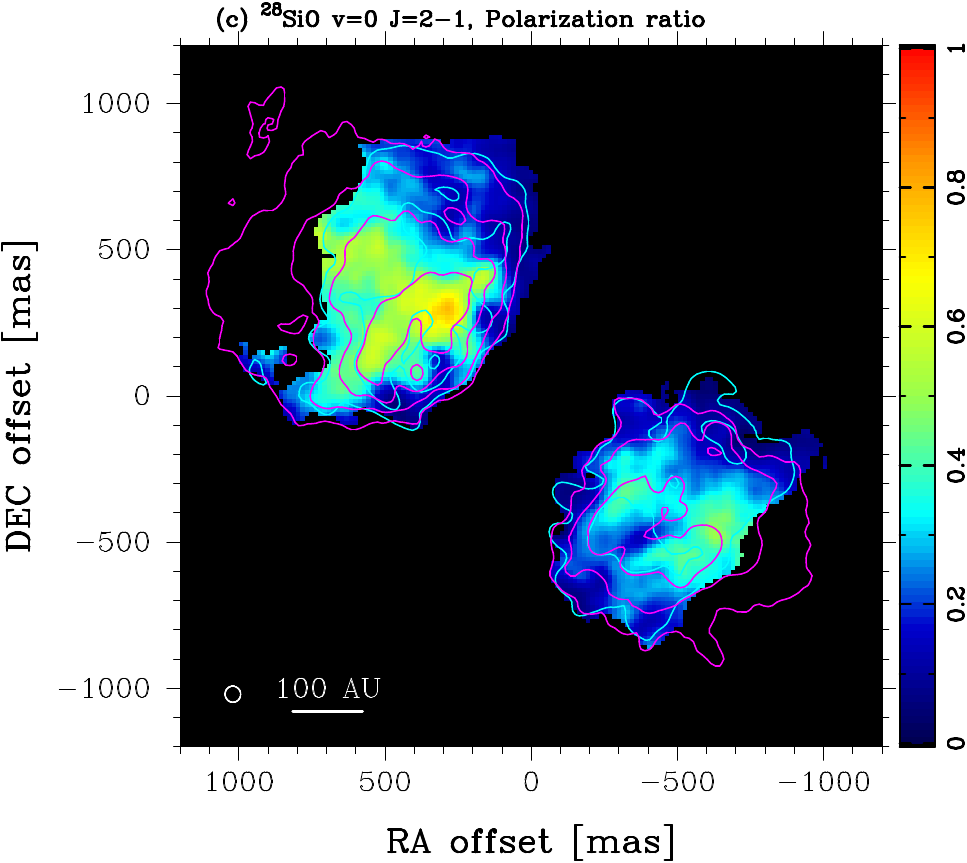}
\includegraphics[width=7.5cm]{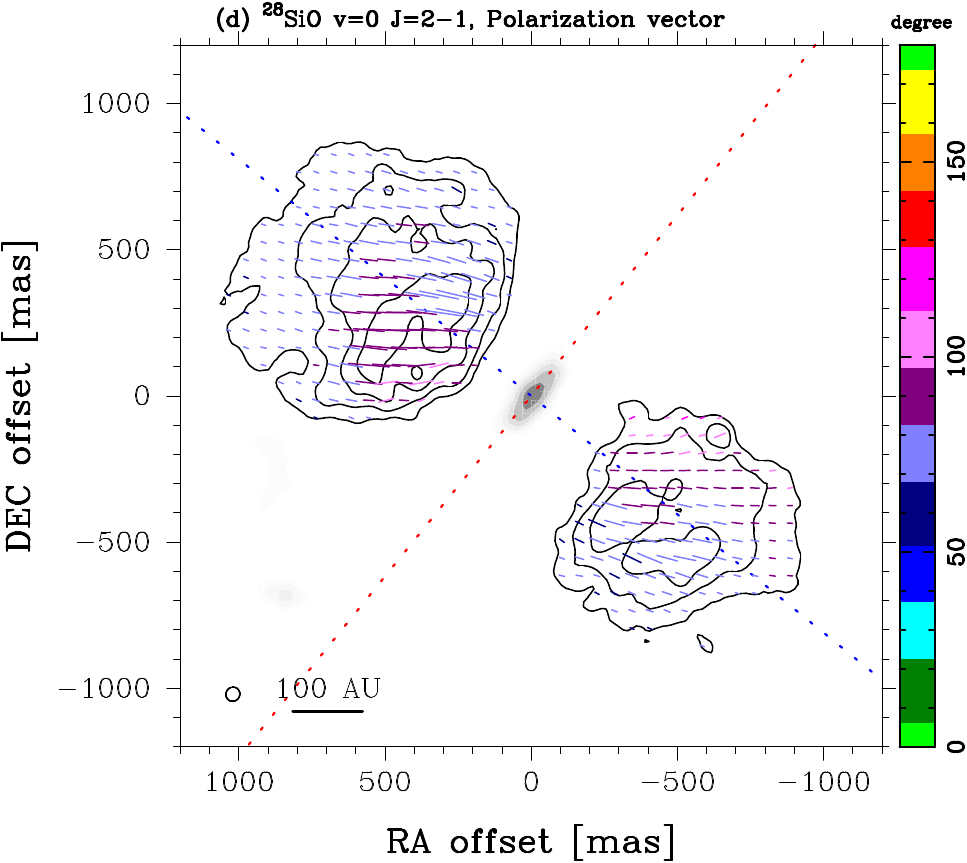}
\caption{
Maps of $^{28}$SiO $v$=0 $J$=2-1 line and the 96~GHz continuum emission. 
The beam size is indicated at the bottom-left corner of each panel. 
Moment maps are produced by using the velocity range from -20 to 30~km~s$^{-1}$. 
(a) Moment 0 (white contour) and Moment 1 (color) maps of Stokes I and the 96~GHz continuum (magenta contour). 
The contour  levels are 4, 8, 16, ... $\sigma$ with the rms noise level of 182~mJy~beam$^{-1}$~km~s$^{-1}$ for the Moment 0 and 0.070~mJy~beam$^{-1}$ for the continuum. 
(b) Same as (a) but for the linear polarization. 
The contour  levels are 4, 8, 16, ... $\sigma$ with the rms noise level of 27~mJy~beam$^{-1}$~km~s$^{-1}$ for the Moment 0. 
(c) The linear polarization ratio (color), the Moment 0 of Stokes I (cyan contour) and linear polarization (magenta contour). 
The contour levels are the same as in (a) and (b). 
(d) Polarization vectors (color) superposed on the linear polarization intensity (contour) and the 96~GHz continuum (gray). 
Polarization angles are calculated from Stokes Q and U images with velocity range from -10 to 20~km~s$^{-1}$. 
Color codes represent the polarization angles as shown in the vertical bar at the right of the panel. 
The contour levels are the same as in (a) and (b). 
The error in the polarization angle is smaller than 7~degrees for the linear polarization intensity higher than 4$\sigma$. 
The blue and red dashed lines indicate the outflow axis (51~degrees) and disk midplane (141~degrees), respectively \citep{Plambeck2016}. 
}
\end{center}
\end{figure}

\begin{figure}[th]
\begin{center}
\includegraphics[width=13cm]{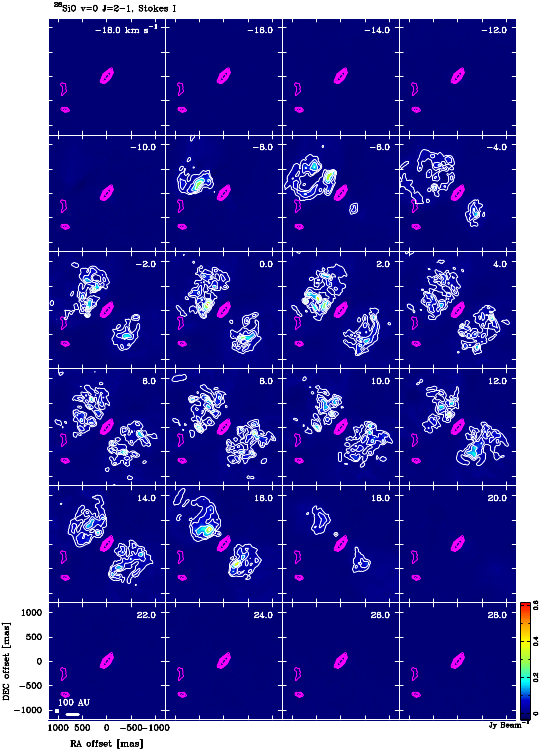}
\caption{
Stokes I channel map of the $^{28}$SiO $v$=0 $J$=2-1 line (color and white contours).
The contour levels are 4, 8, 16, ... $\sigma$ with the rms noise level of 6.8~mJy~beam$^{-1}$. 
Magenta contours show the 96~GHz continuum emission. 
Radial velocity ($v_{lsr}$) is indicated at the top-right corner of each panel. 
The beam sizes are plotted at the bottom-left corner of the bottm-left panel. 
}
\end{center}
\end{figure}

\begin{figure}[th]
\begin{center}
\includegraphics[width=13cm]{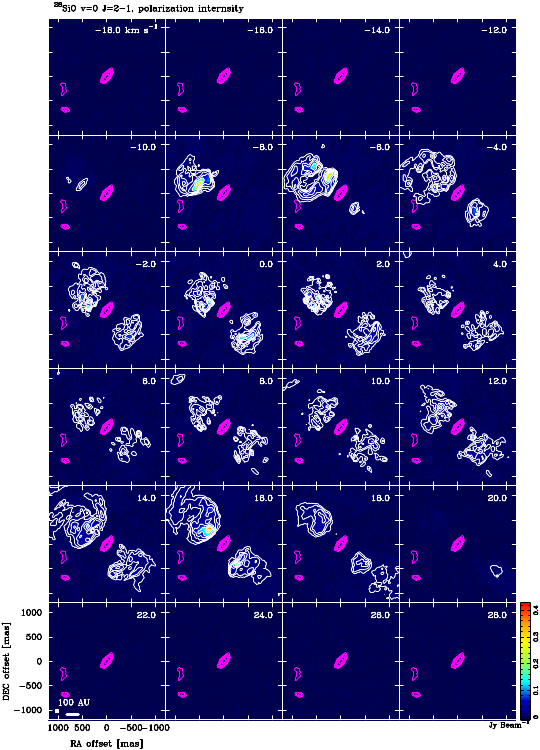}
\caption{
Polarization intensity channel map of the $^{28}$SiO $v$=0 $J$=2-1 line (color and white contours).
The contour levels are 4, 8, 16, ... $\sigma$ with the rms noise level of 1.55~mJy~beam$^{-1}$. 
Magenta contours show the 96~GHz continuum emission. 
Radial velocity ($v_{lsr}$) is indicated at the top-right corner of each panel. 
The beam sizes are plotted at the bottom-left corner of the bottm-left panel. 
}
\end{center}
\end{figure}

\begin{figure}[th]
\begin{center}
\includegraphics[width=13cm]{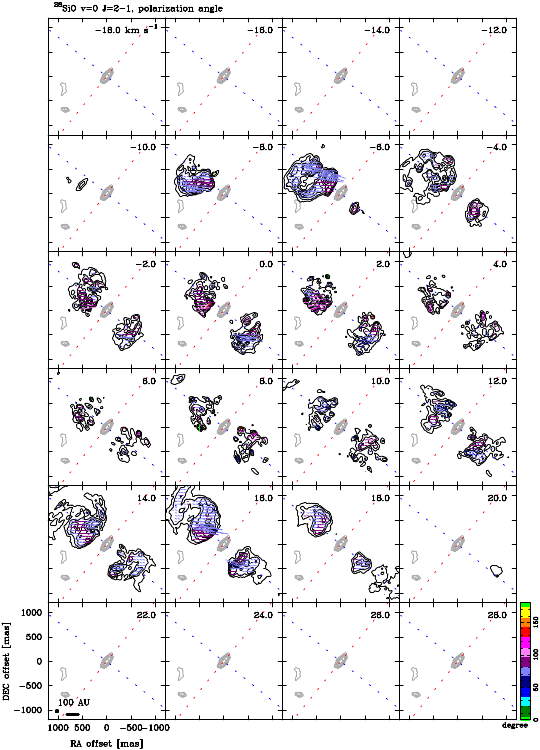}
\caption{
Channel map of the linear polarization intensity (black contours) and vectors (color lines) of the $^{28}$SiO $v$=0 $J$=2-1 line. 
Color codes represent the polarization angles as shown in the vertical bar at the right of the bottom-right panel. 
The error in the polarization angle is smaller than 7~degrees for the linear polarization intensity higher than 4$\sigma$. 
Gray contours show the 96~GHz continuum emission. 
Radial velocity ($v_{lsr}$) is indicated at the top-right corner of each panel. 
The beam sizes are plotted at the bottom-left corner of the bottm-left panel. 
The blue and red dashed lines indicate the outflow axis (51~degrees) and disk midplane (141~degrees), respectively \citep{Plambeck2016}. 
}
\end{center}
\end{figure}

\begin{figure}[th]
\begin{center}
\includegraphics[width=13cm]{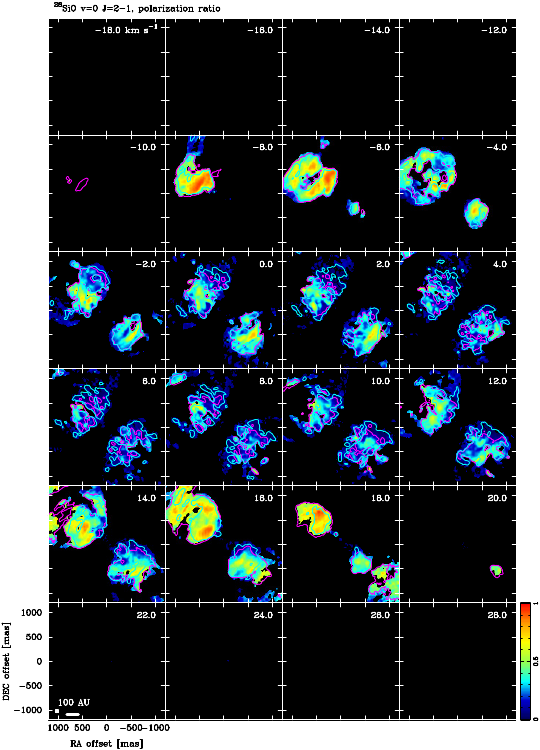}
\caption{
Channel map of the linear polarization ratio (color), Stokes I (cyan contour), and linear polarization intensity (magenta contours) of the $^{28}$SiO $v$=0 $J$=2-1 line. 
The contours show only the 4$\sigma$ levels to outline the distributions of the total and linearly polarized emission. 
Radial velocity ($v_{lsr}$) is indicated at the top-right corner of each panel. 
The beam sizes are plotted at the bottom-left corner of the bottm-left panel. 
}
\end{center}
\end{figure}

\begin{figure}[th]
\begin{center}
\includegraphics[width=13cm]{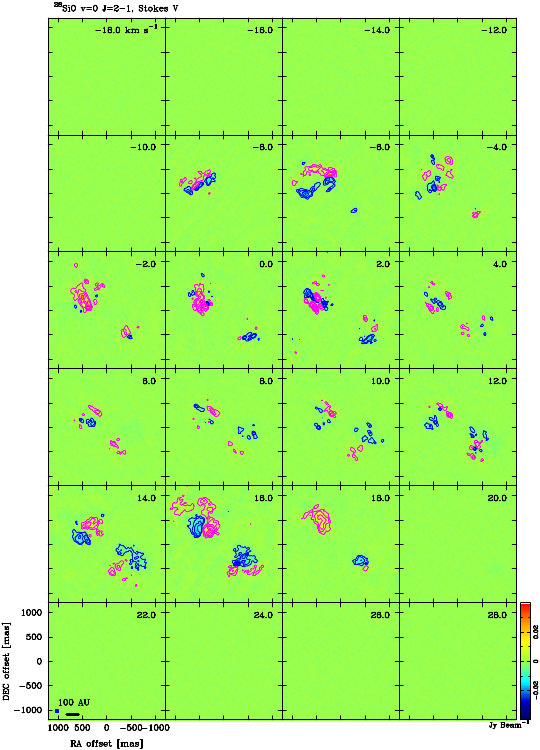}
\caption{Stokes V of the $^{28}$SiO $v$=0 $J$=2-1 line. 
The contour  levels are $\pm$4, $\pm$8, $\pm$16, and $\pm$32 times the rms noise levels of 0.70~mJy~beam$^{-1}$. 
Magenta and blue solid lines show the positive and negative levels, respectively. }
\end{center}
\end{figure}

\clearpage
\subsection{$^{28}$SiO $v$=1 $J$=2-1}

Figure B26 presents the moment maps and integrated emission maps of the $^{28}$SiO $v=1 J=2-1$ line and the 96 GHz continuum emission. Figures B27-B30 show channel maps of the $^{28}$SiO $v=1 J=2-1$ line. 

\begin{figure}[th]
\begin{center}
\includegraphics[width=7.5cm]{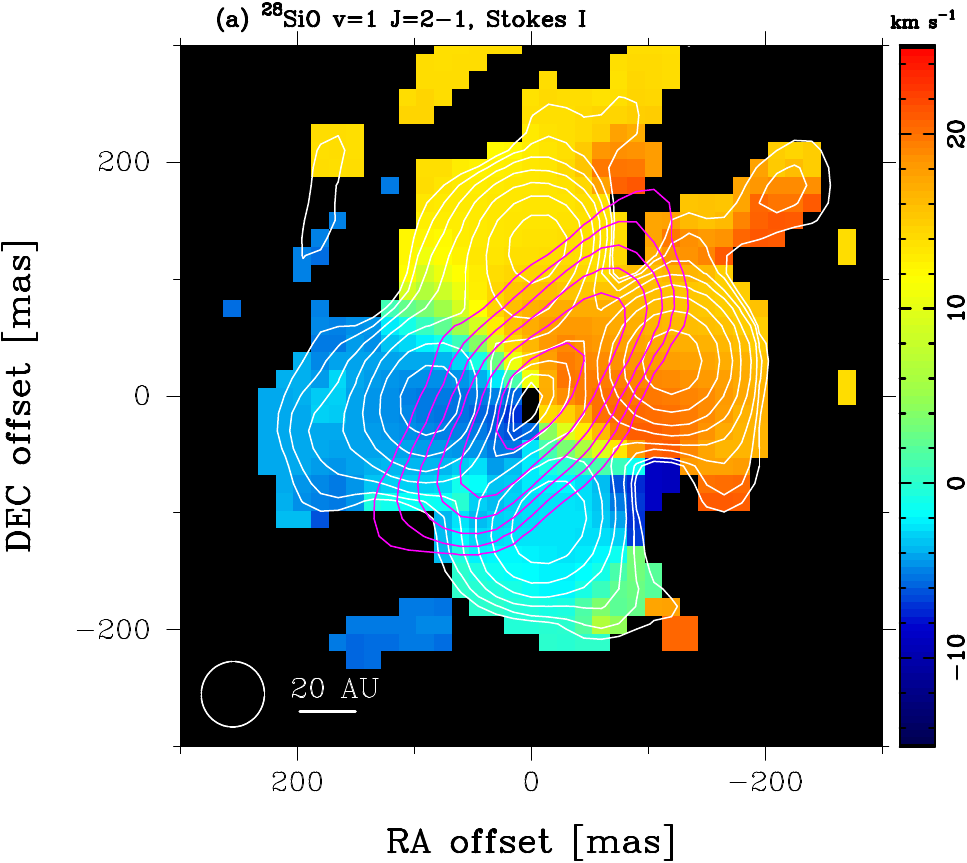}
\includegraphics[width=7.5cm]{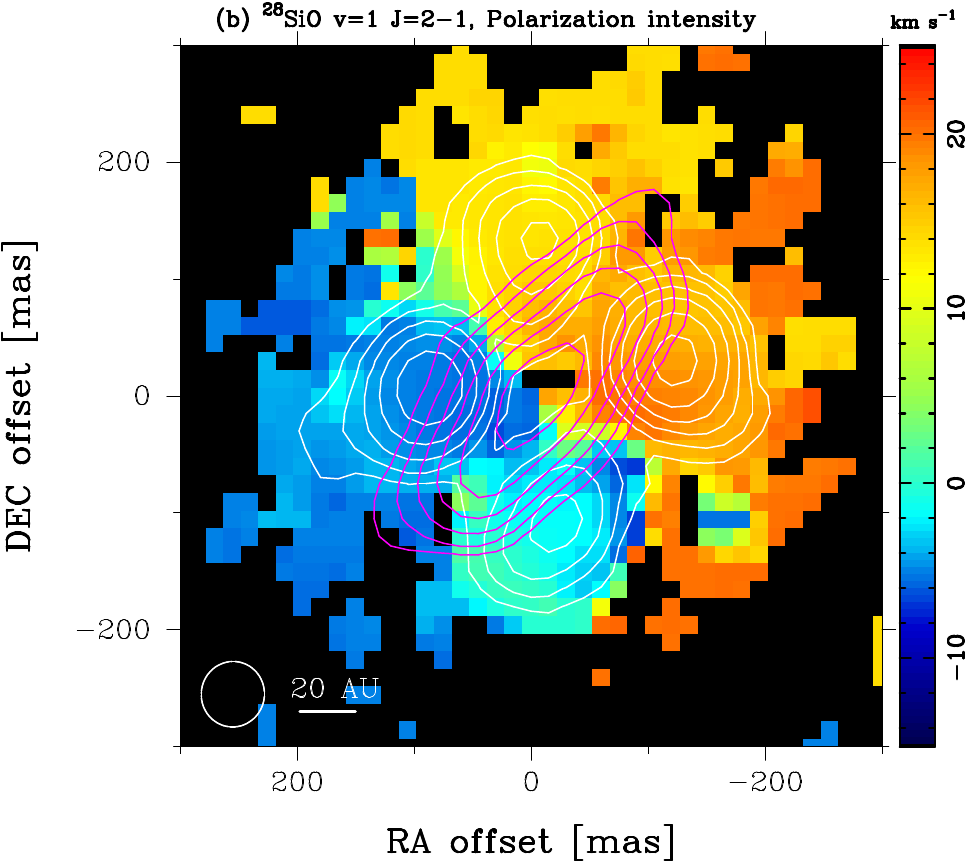}
\includegraphics[width=7.5cm]{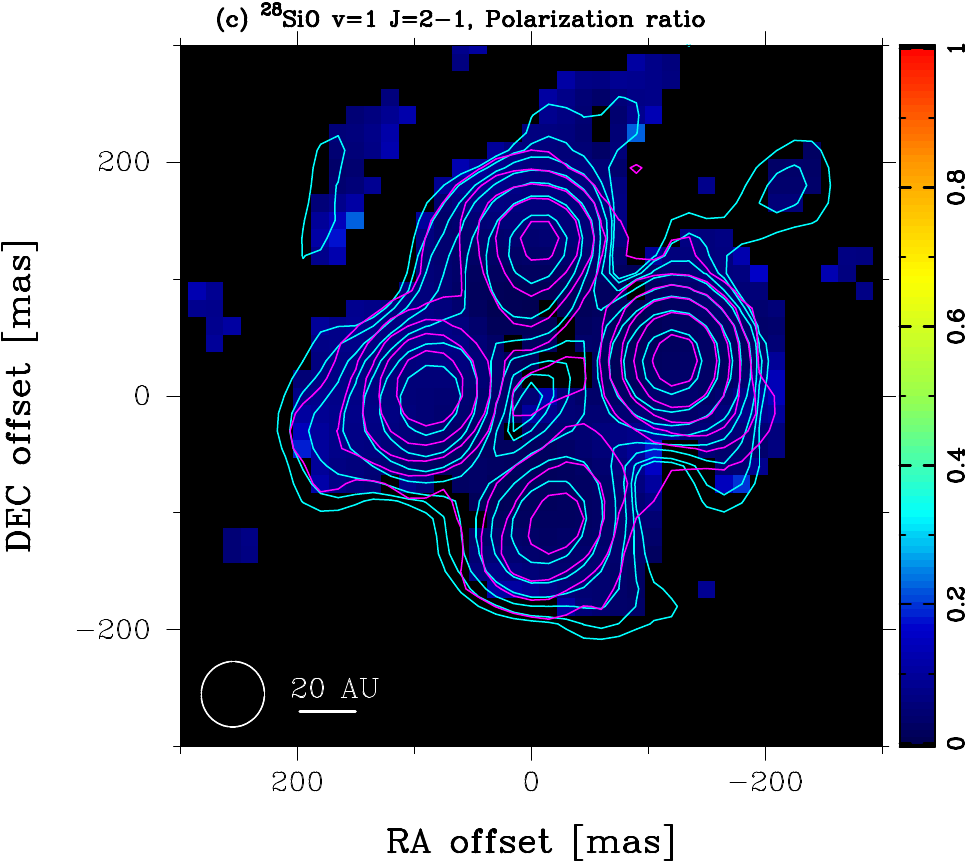}
\includegraphics[width=7.5cm]{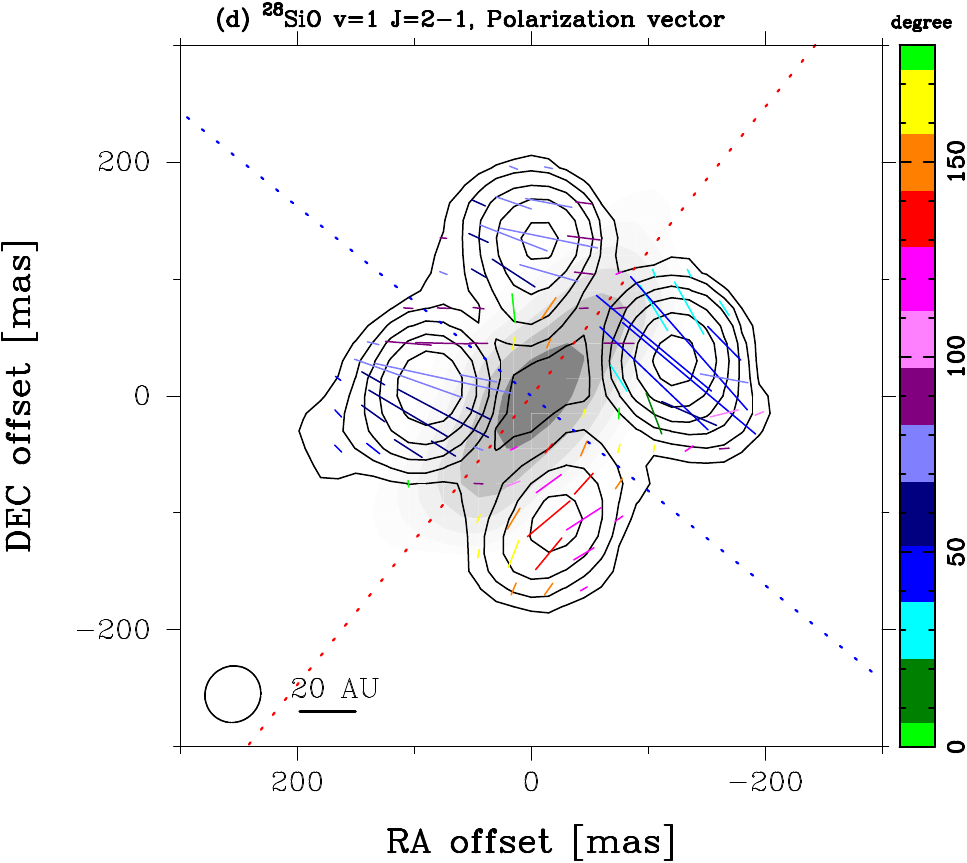}
\caption{
Maps of $^{28}$SiO $v$=1 $J$=2-1 line and the 96~GHz continuum emission. 
The beam size is indicated at the bottom-left corner of each panel. 
Moment maps are produced by using the velocity range from -15 to 25~km~s$^{-1}$. 
(a) Moment 0 (white contour) and Moment 1 (color) maps of Stokes I and the 96~GHz continuum (magenta contour). 
The contour  levels are 4, 8, 16, ... $\sigma$ with the rms noise level of 3.0~Jy~beam$^{-1}$~km~s$^{-1}$ for the Moment 0 and 0.070~mJy~beam$^{-1}$ for the continuum. 
(b) Same as (a) but for the linear polarization. 
The contour  levels are 4, 8, 16, ... $\sigma$ with the rms noise level of 0.92~Jy~beam$^{-1}$~km~s$^{-1}$ for the Moment 0. 
(c) The linear polarization ratio (color), the Moment 0 of Stokes I (cyan contour) and linear polarization (magenta contour). 
The contour levels are the same as in (a) and (b). 
(d) Polarization vectors (color) superposed on the linear polarization intensity (contour) and the 96~GHz continuum (gray). 
Polarization angles are calculated from Stokes Q and U images with velocity range from -10 to 20~km~s$^{-1}$. 
Color codes represent the polarization angles as shown in the vertical bar at the right of the panel. 
The contour levels are the same as in (a) and (b). 
The error in the polarization angle is smaller than 7~degrees for the linear polarization intensity higher than 4$\sigma$. 
The blue and red dashed lines indicate the outflow axis (51~degrees) and disk midplane (141~degrees), respectively \citep{Plambeck2016}. 
}
\end{center}
\end{figure}

\begin{figure}[th]
\begin{center}
\includegraphics[width=13cm]{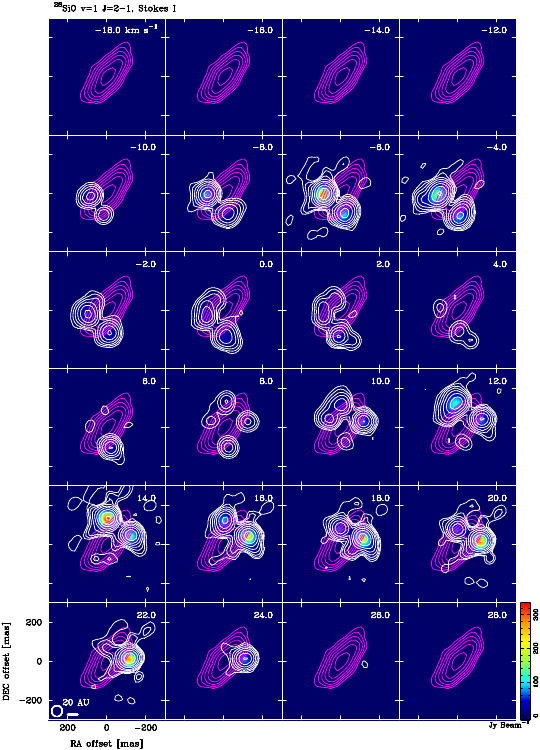}
\caption{
Stokes I channel map of the $^{28}$SiO $v$=1 $J$=2-1 line (color and white contours).
The contour levels are 4, 8, 16, ... $\sigma$ with the rms noise level of 0.28~Jy~beam$^{-1}$. 
Magenta contours show the 96~GHz continuum emission. 
Radial velocity ($v_{lsr}$) is indicated at the top-right corner of each panel. 
The beam sizes are plotted at the bottom-left corner of the bottm-left panel. 
}
\end{center}
\end{figure}

\begin{figure}[th]
\begin{center}
\includegraphics[width=13cm]{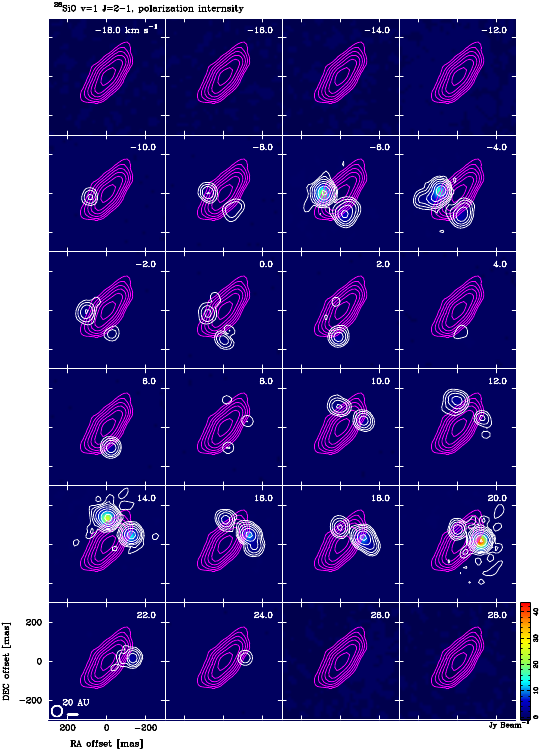}
\caption{
Linear polarization intensity channel map of the $^{28}$SiO $v$=1 $J$=2-1 line (color and white contours).
The contour levels are 4, 8, 16, ... $\sigma$ with the rms noise level of 78~mJy~beam$^{-1}$. 
Magenta contours show the 96~GHz continuum emission. 
Radial velocity ($v_{lsr}$) is indicated at the top-right corner of each panel. 
The beam sizes are plotted at the bottom-left corner of the bottm-left panel. 
}
\end{center}
\end{figure}

\begin{figure}[th]
\begin{center}
\includegraphics[width=13cm]{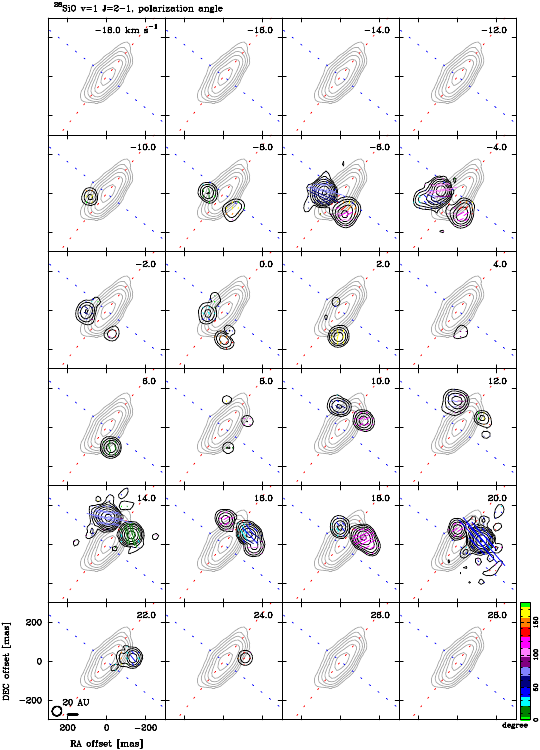}
\caption{
Channel map of the linear polarization intensity (black contours) and vectors (color lines) of the $^{28}$SiO $v$=1 $J$=2-1 line. 
Color codes represent the polarization angles as shown in the vertical bar at the right of the bottom-right panel. 
The error in the polarization angle is smaller than 7~degrees for the linear polarization intensity higher than 4$\sigma$. 
Gray contours show the 96~GHz continuum emission. 
Radial velocity ($v_{lsr}$) is indicated at the top-right corner of each panel. 
The beam sizes are plotted at the bottom-left corner of the bottm-left panel. 
The blue and red dashed lines indicate the outflow axis (51~degrees) and disk midplane (141~degrees), respectively \citep{Plambeck2016}. 
}
\end{center}
\end{figure}

\begin{figure}[th]
\begin{center}
\includegraphics[width=13cm]{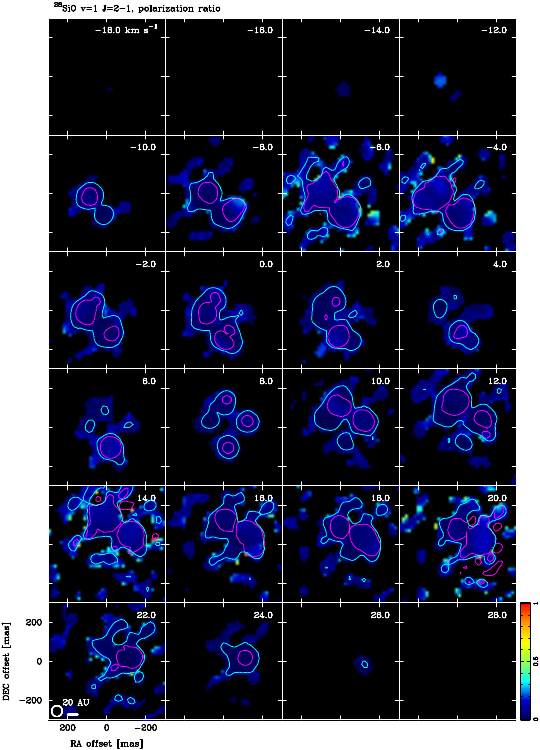}
\caption{
Channel map of the linear polarization ratio (color), Stokes I (cyan contour), and linear polarization intensity (magenta contours) of the $^{28}$SiO $v$=1 $J$=2-1 line. 
The contours show only the 4$\sigma$ levels to outline the distributions of the total and linearly polarized emission. 
Radial velocity ($v_{lsr}$) is indicated at the top-right corner of each panel. 
The beam sizes are plotted at the bottom-left corner of the bottm-left panel. 
}
\end{center}
\end{figure}

\clearpage
\subsection{$^{29}$SiO $v$=0 $J$=2-1}

Figure B31 presents the moment maps and integrated emission maps of the $^{29}$SiO $v=0 J=2-1$ line and the 96 GHz continuum emission. Figures B32-B35 show channel maps of the $^{29}$SiO $v=0 J=2-1$ line. 

\begin{figure}[th]
\begin{center}
\includegraphics[width=7.5cm]{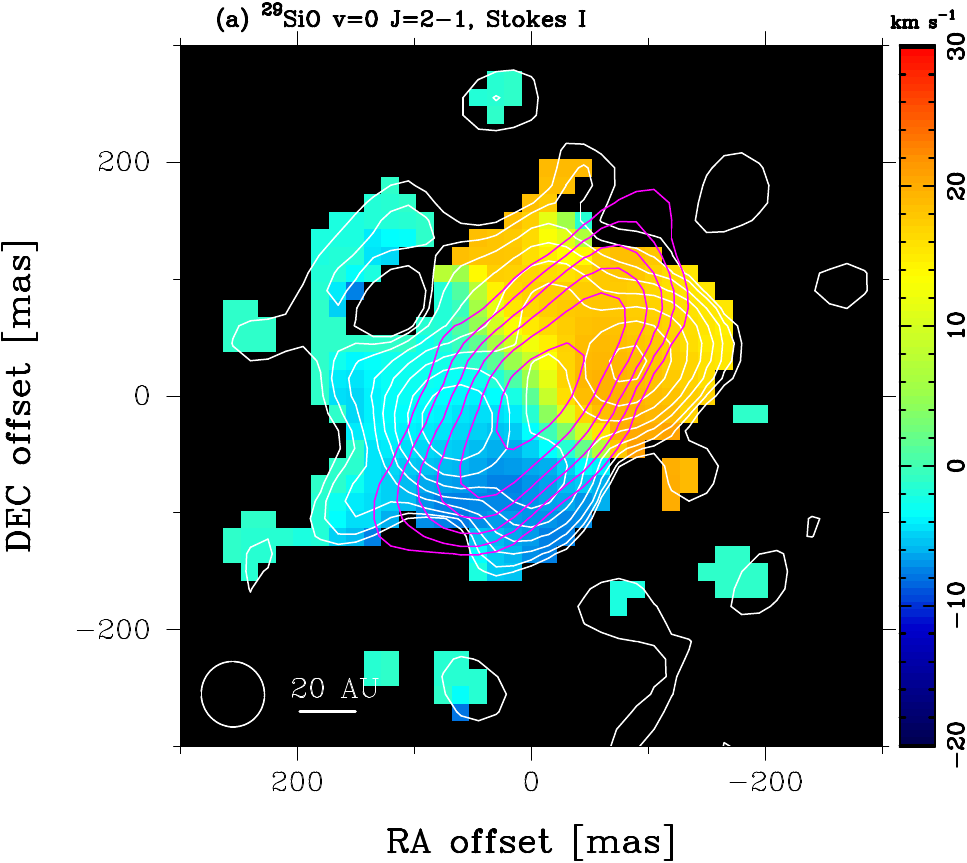}
\includegraphics[width=7.5cm]{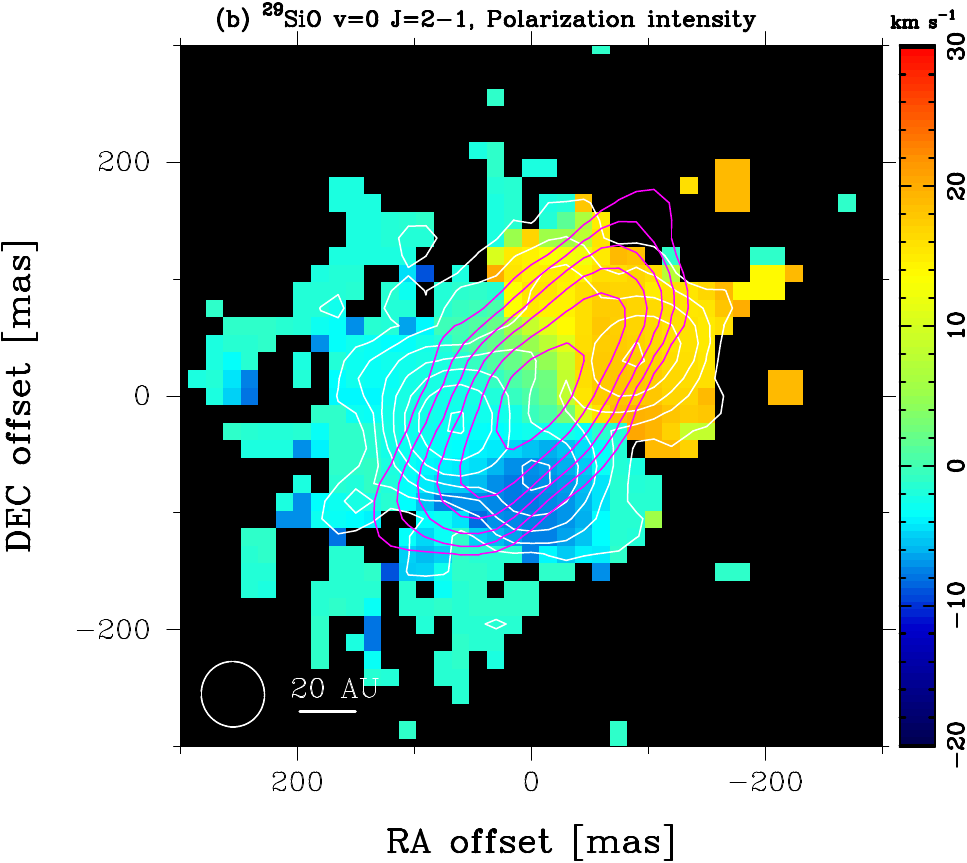}
\includegraphics[width=7.5cm]{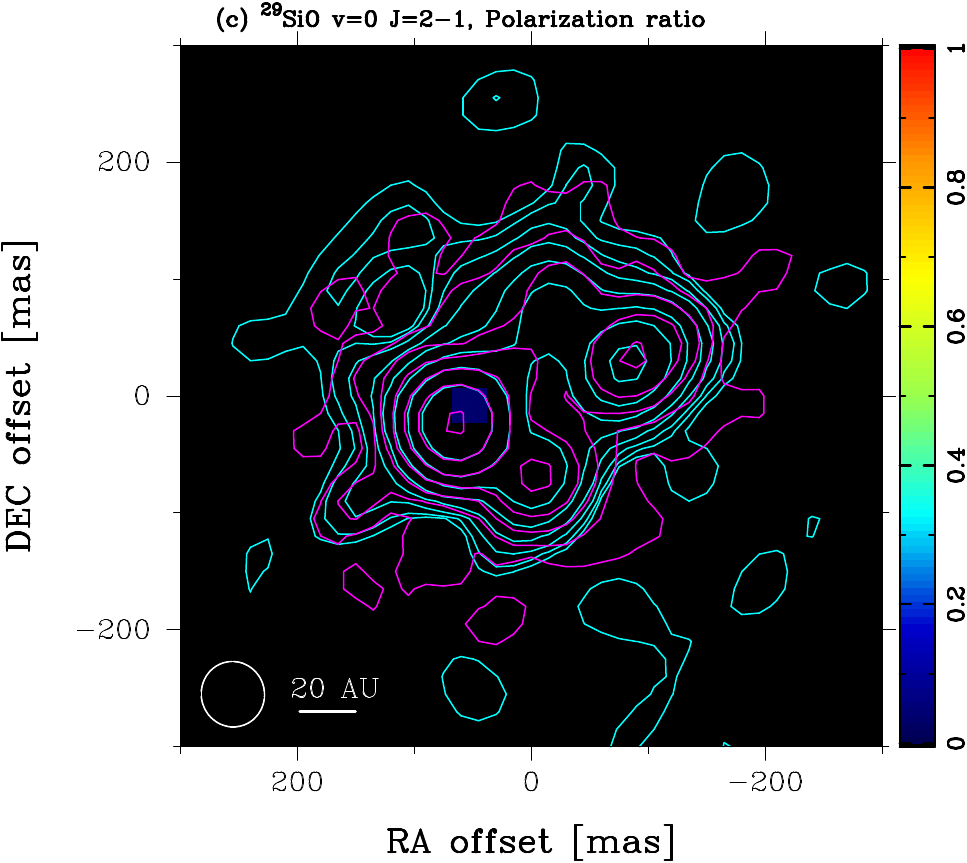}
\includegraphics[width=7.5cm]{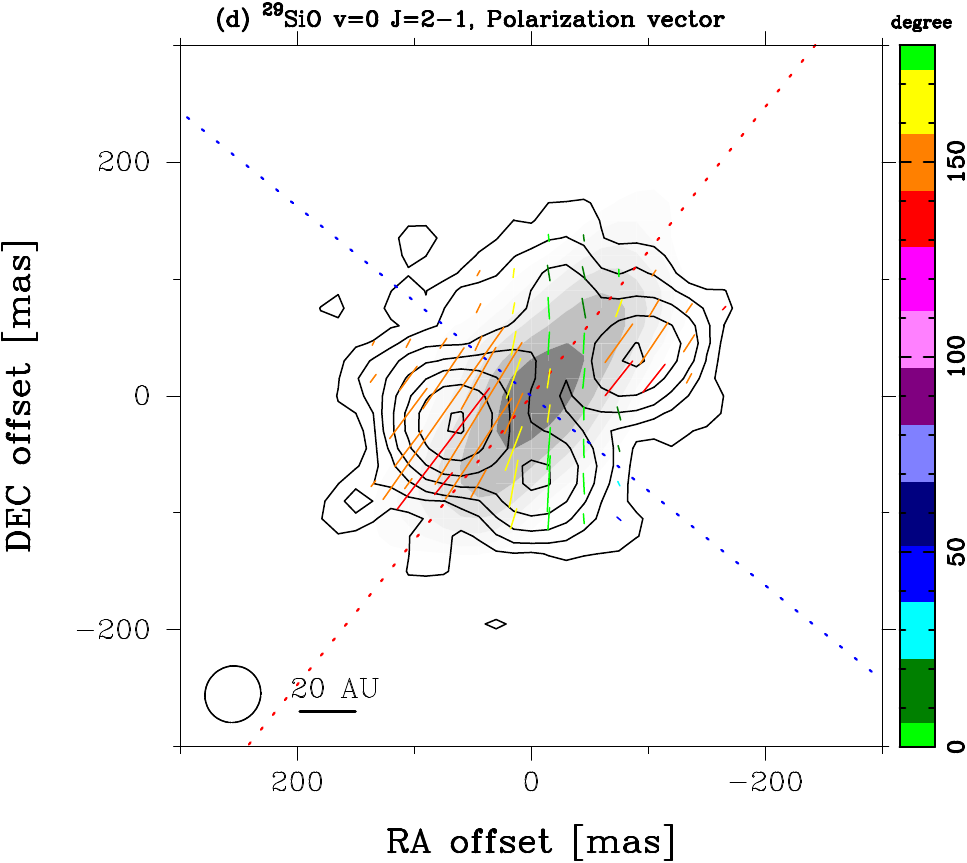}
\caption{
Maps of $^{29}$SiO $v$=0 $J$=2-1 line and the 96~GHz continuum emission. 
The beam size is indicated at the bottom-left corner of each panel. 
Moment maps are produced by using the velocity range from -20 to 30~km~s$^{-1}$. 
(a) Moment 0 (white contour) and Moment 1 (color) maps of Stokes I and the 96~GHz continuum (magenta contour). 
The contour  levels are 4, 8, 16, ... $\sigma$ with the rms noise level of 107~mJy~beam$^{-1}$~km~s$^{-1}$ for the Moment 0 and 0.070~mJy~beam$^{-1}$ for the continuum. 
(b) Same as (a) but for the linear polarization. 
The contour  levels are 4, 8, 16, ... $\sigma$ with the rms noise level of 15.9~mJy~beam$^{-1}$~km~s$^{-1}$ for the Moment 0. 
(c) The linear polarization ratio (color), the Moment 0 of Stokes I (cyan contour) and linear polarization (magenta contour). 
The contour levels are the same as in (a) and (b). 
(d) Polarization vectors (color) superposed on the linear polarization intensity (contour) and the 96~GHz continuum (gray). 
Polarization angles are calculated from Stokes Q and U images with velocity range from -10 to 20~km~s$^{-1}$. 
Color codes represent the polarization angles as shown in the vertical bar at the right of the panel. 
The contour levels are the same as in (a) and (b). 
The error in the polarization angle is smaller than 7~degrees for the linear polarization intensity higher than 4$\sigma$. 
The blue and red dashed lines indicate the outflow axis (51~degrees) and disk midplane (141~degrees), respectively \citep{Plambeck2016}. 
}
\end{center}
\end{figure}

\begin{figure}[th]
\begin{center}
\includegraphics[width=13cm]{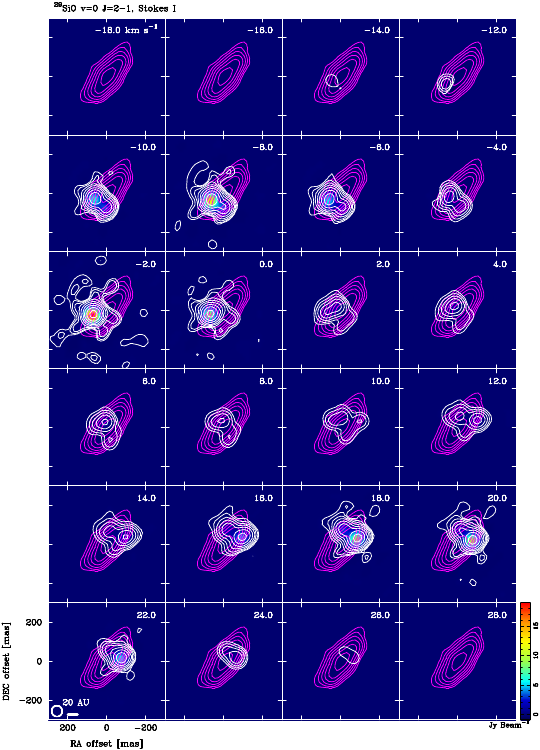}
\caption{
Stokes I channel map of the $^{29}$SiO $v$=0 $J$=2-1 line (color and white contours).
The contour levels are 4, 8, 16, ... $\sigma$ with the rms noise level of 13.8~mJy~beam$^{-1}$. 
Magenta contours show the 96~GHz continuum emission. 
Radial velocity ($v_{lsr}$) is indicated at the top-right corner of each panel. 
The beam sizes are plotted at the bottom-left corner of the bottm-left panel. 
}
\end{center}
\end{figure}

\begin{figure}[th]
\begin{center}
\includegraphics[width=13cm]{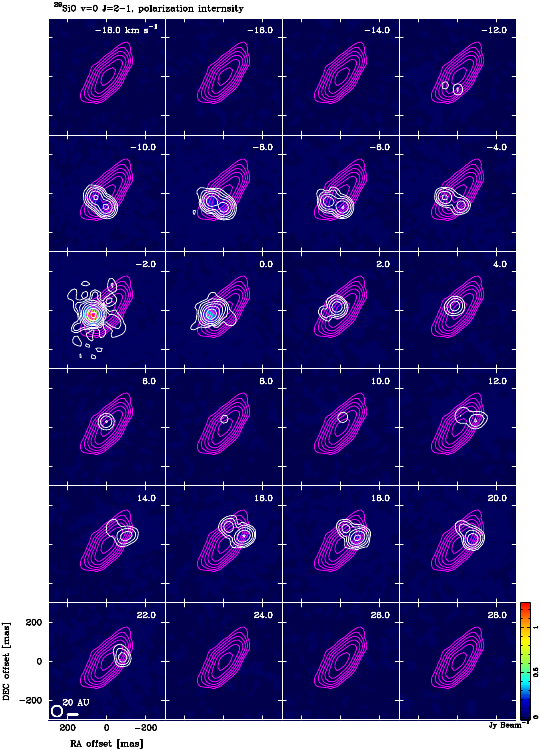}
\caption{
Linear polarization intensity channel map of the $^{29}$SiO $v$=0 $J$=2-1 line (color and white contours).
The contour levels are 4, 8, 16, ... $\sigma$ with the rms noise level of 2.1~mJy~beam$^{-1}$. 
Magenta contours show the 96~GHz continuum emission. 
Radial velocity ($v_{lsr}$) is indicated at the top-right corner of each panel. 
The beam sizes are plotted at the bottom-left corner of the bottm-left panel. 
}
\end{center}
\end{figure}

\begin{figure}[th]
\begin{center}
\includegraphics[width=13cm]{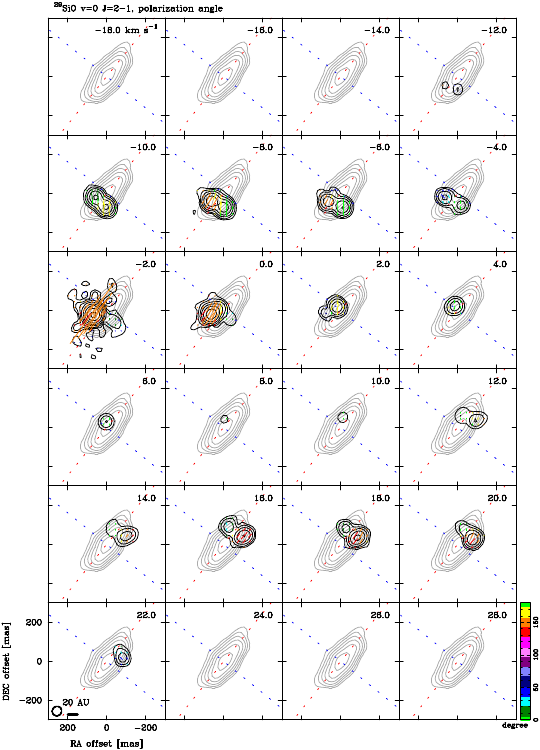}
\caption{
Channel map of the linear polarization intensity (black contours) and vectors (color lines) of the $^{29}$SiO $v$=0 $J$=2-1 line. 
Color codes represent the polarization angles as shown in the vertical bar at the right of the bottom-right panel. 
The error in the polarization angle is smaller than 7~degrees for the linear polarization intensity higher than 4$\sigma$. 
Gray contours show the 96~GHz continuum emission. 
Radial velocity ($v_{lsr}$) is indicated at the top-right corner of each panel. 
The beam sizes are plotted at the bottom-left corner of the bottm-left panel. 
The blue and red dashed lines indicate the outflow axis (51~degrees) and disk midplane (141~degrees), respectively \citep{Plambeck2016}. 
}
\end{center}
\end{figure}

\begin{figure}[th]
\begin{center}
\includegraphics[width=13cm]{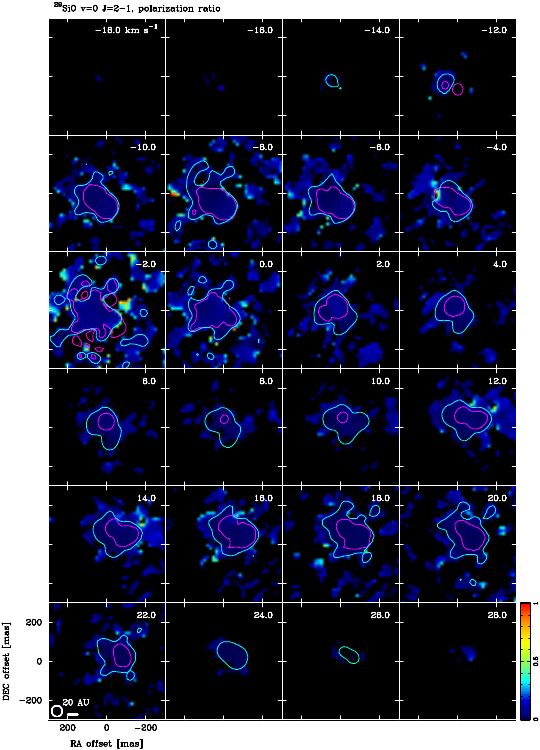}
\caption{
Channel map of the linear polarization ratio (color), Stokes I (cyan contour), and linear polarization intensity (magenta contours) of the $^{29}$SiO $v$=0 $J$=2-1 line. 
The contours show only the 4$\sigma$ levels to outline the distributions of the total and linearly polarized emission. 
Radial velocity ($v_{lsr}$) is indicated at the top-right corner of each panel. 
The beam sizes are plotted at the bottom-left corner of the bottm-left panel. 
}
\end{center}
\end{figure}

\clearpage
\subsection{$^{28}$SiO $v$=2, $^{29}$SiO $v$=1, and $^{30}$SiO $v$=0 $J$=2-1}

Figure B36 presents the moment maps of the $^{28}$SiO $v=2$, $^{29}$SiO $v=1$, and $^{30}$SiO $v=0 J=2-1$ lines and the 96 GHz continuum emission. Figures B37, B38, and B39 show channel maps of the $^{28}$SiO $v=2$, $^{29}$SiO $v=1$, and $^{30}$SiO $v=0 J=2-1$ lines, respectively. 

\begin{figure}[th]
\begin{center}
\includegraphics[width=7.5cm]{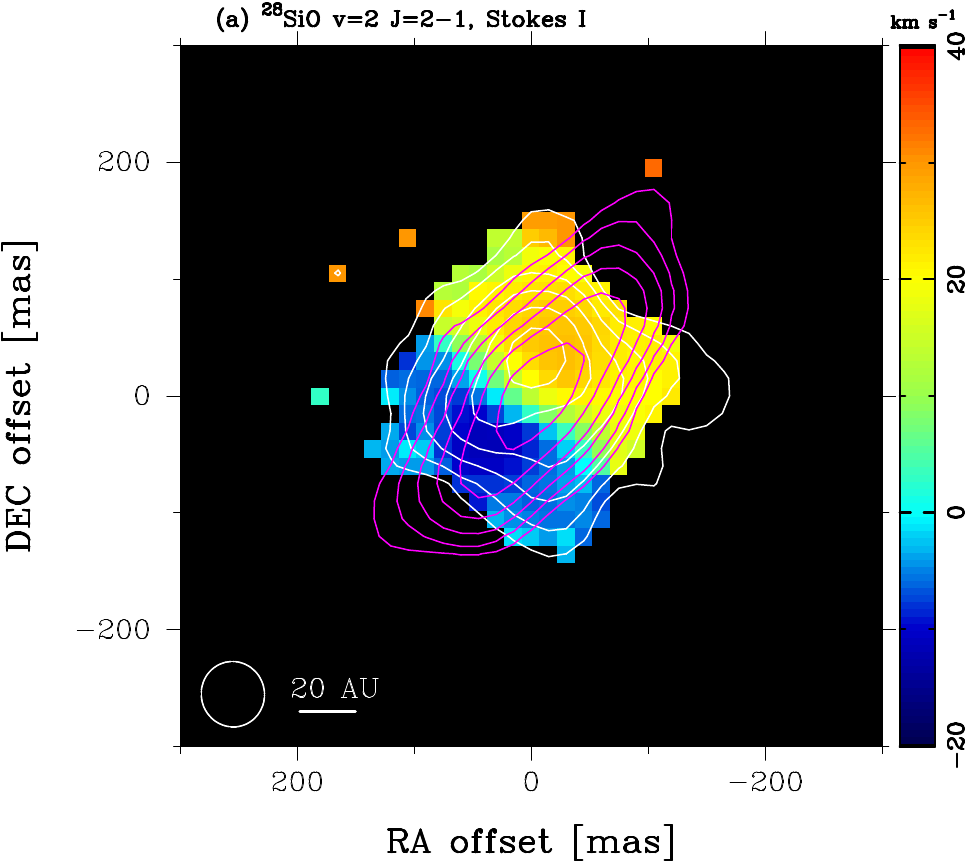}
\includegraphics[width=7.5cm]{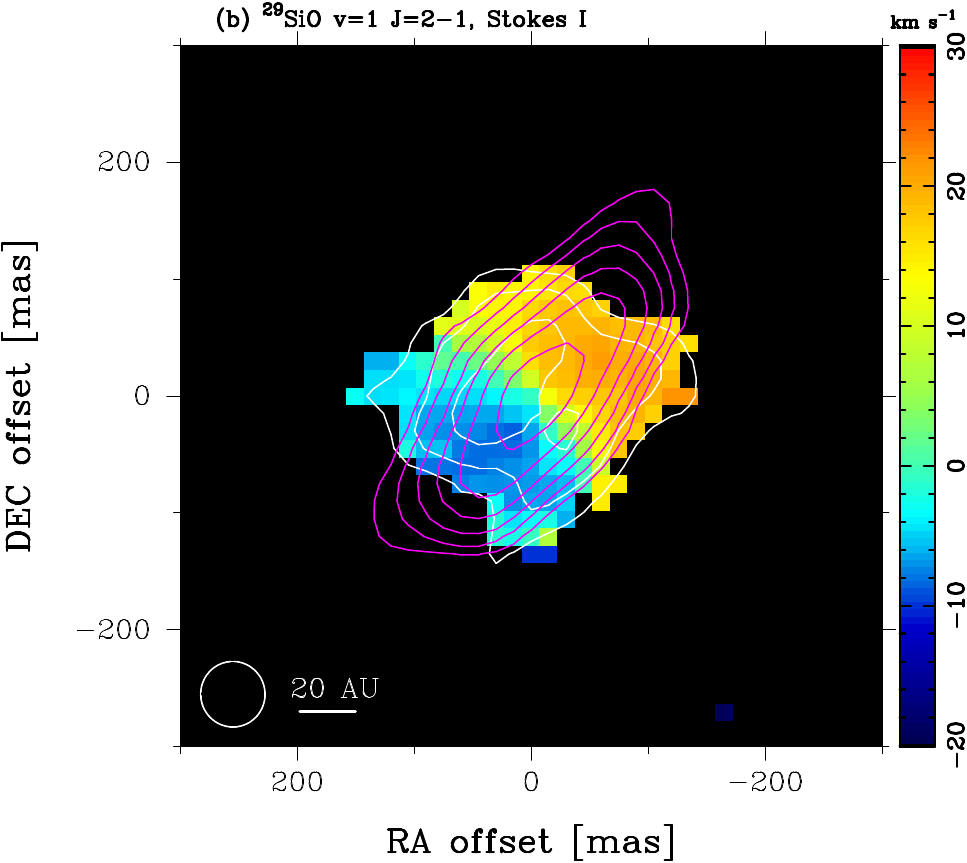}
\includegraphics[width=7.5cm]{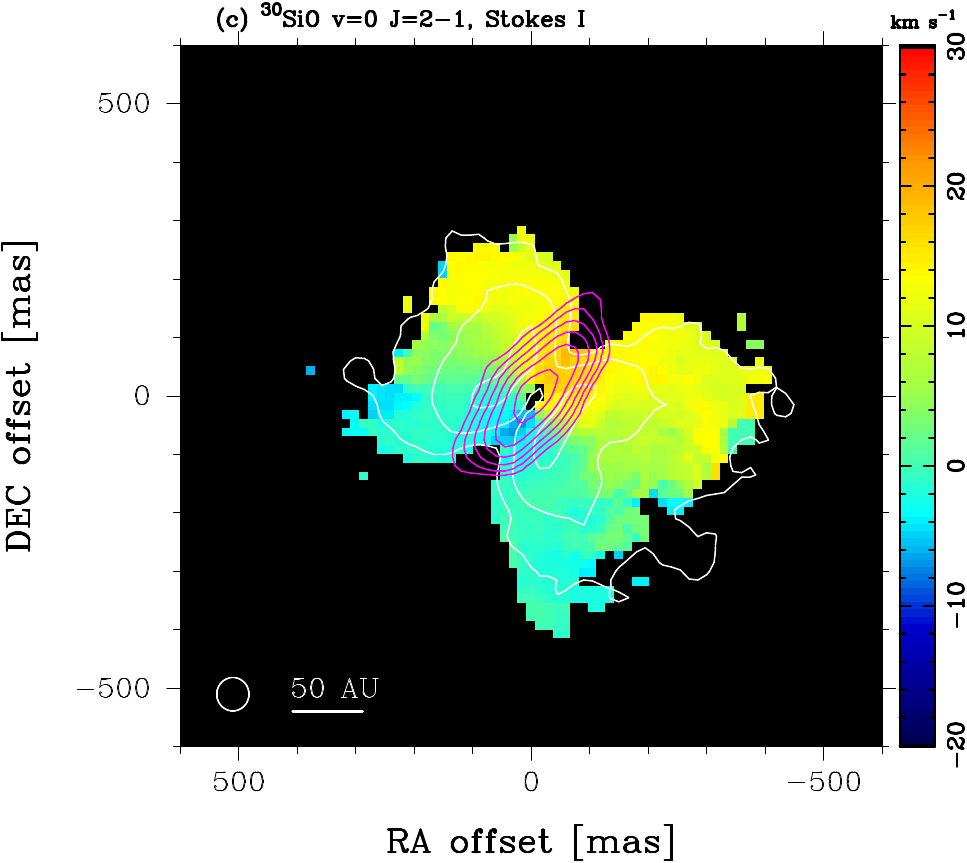}
\caption{
Moment 0 (white contour) and Moment 1 (color) maps of Stokes I and the 96~GHz continuum (magenta contour). 
The beam size is indicated at the bottom-left corner of each panel. 
(a) The $^{28}$SiO $v$=2 $J$=2-1 line. 
Moment maps are produced by using the velocity range from -20 to 40~km~s$^{-1}$. 
The contour  levels are 4, 8, 16, ... $\sigma$ with the rms noise level of 7.5~mJy~beam$^{-1}$~km~s$^{-1}$ for the Moment 0 and 0.070~mJy~beam$^{-1}$ for the continuum. 
(b) Same as (a) but $^{29}$SiO $v$=1 $J$=2-1 line. 
Moment maps are produced by using the velocity range from -20 to 30~km~s$^{-1}$. 
The contour  levels are 4, 8, 16, ... $\sigma$ with the rms noise level of 6.4~mJy~beam$^{-1}$~km~s$^{-1}$ for the Moment 0. 
(c) Same as (a) but $^{30}$SiO $v$=0 $J$=2-1 line. 
Moment maps are produced by using the velocity range from -20 to 30~km~s$^{-1}$. 
The contour  levels are 4, 8, 16, ... $\sigma$ with the rms noise level of 23~mJy~beam$^{-1}$~km~s$^{-1}$ for the Moment 0. 
}
\end{center}
\end{figure}

\begin{figure}[th]
\begin{center}
\includegraphics[width=13cm]{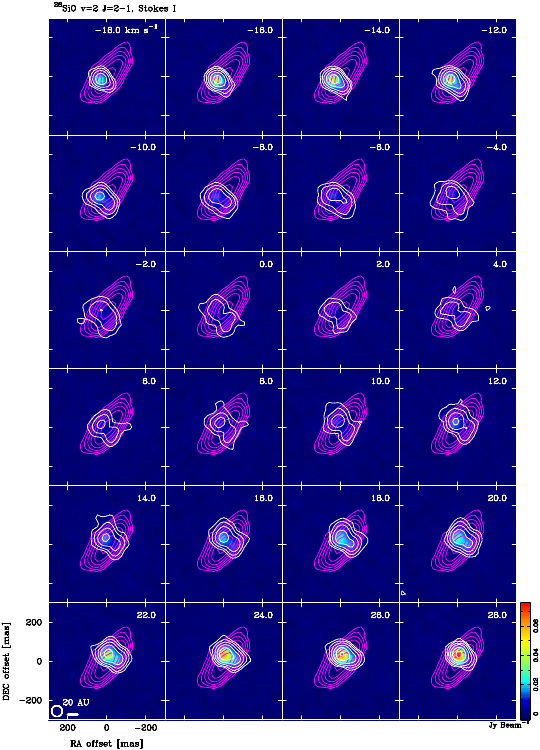}
\caption{
Stokes I channel map of the $^{28}$SiO $v$=2 $J$=2-1 line (color and white contours).
The contour levels are 4, 8, 16, ... $\sigma$ with the rms noise level of 0.48~mJy~beam$^{-1}$. 
Magenta contours show the 96~GHz continuum emission. 
Radial velocity ($v_{lsr}$) is indicated at the top-right corner of each panel. 
The beam sizes are plotted at the bottom-left corner of the bottm-left panel. 
}
\end{center}
\end{figure}

\begin{figure}[th]
\begin{center}
\includegraphics[width=13cm]{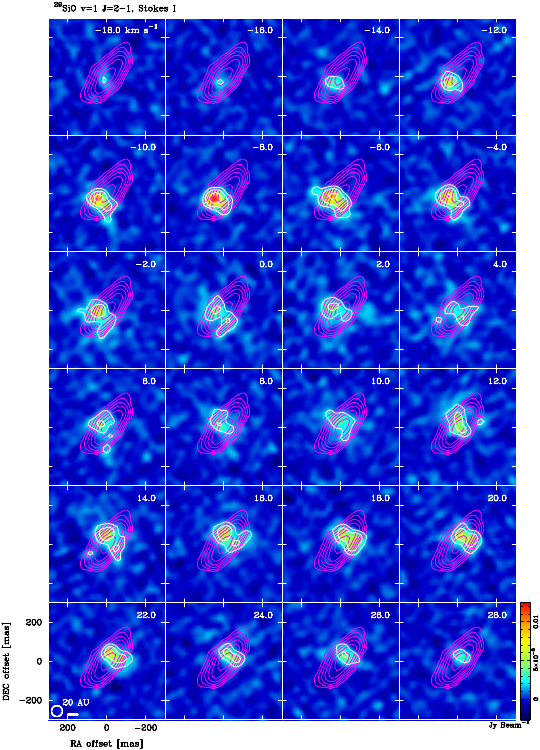}
\caption{
Stokes I channel map of the $^{29}$SiO $v$=1 $J$=2-1 line (color and white contours).
The contour levels are 4, 8, 16, ... $\sigma$ with the rms noise level of 0.49~mJy~beam$^{-1}$. 
Magenta contours show the 96~GHz continuum emission. 
Radial velocity ($v_{lsr}$) is indicated at the top-right corner of each panel. 
The beam sizes are plotted at the bottom-left corner of the bottm-left panel. 
}
\end{center}
\end{figure}

\begin{figure}[th]
\begin{center}
\includegraphics[width=13cm]{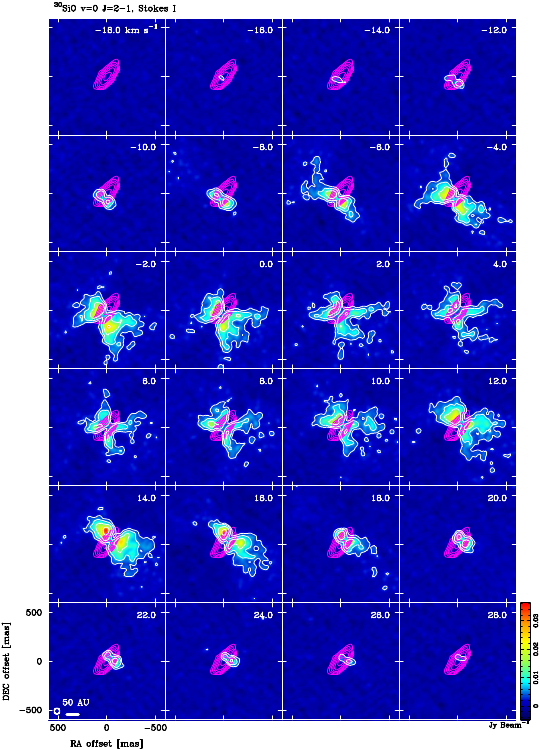}
\caption{
Stokes I channel map of the $^{30}$SiO $v$=0 $J$=2-1 line (color and white contours).
The contour levels are 4, 8, 16, ... $\sigma$ with the rms noise level of 0.83~mJy~beam$^{-1}$. 
Magenta contours show the 96~GHz continuum emission. 
Radial velocity ($v_{lsr}$) is indicated at the top-right corner of each panel. 
The beam sizes are plotted at the bottom-left corner of the bottm-left panel. 
}
\end{center}
\end{figure}

\clearpage
\subsection{Difference between $^{28}$SiO $v$=0 $J$=1-0 and $v$=0 $J$=2-1}

Figure B40 shows the polarization angle difference between $^{28}$SiO $v=0 J=1-0$ and $J=2-1$ lines.

\begin{figure}[th]
\begin{center}
\includegraphics[width=13cm]{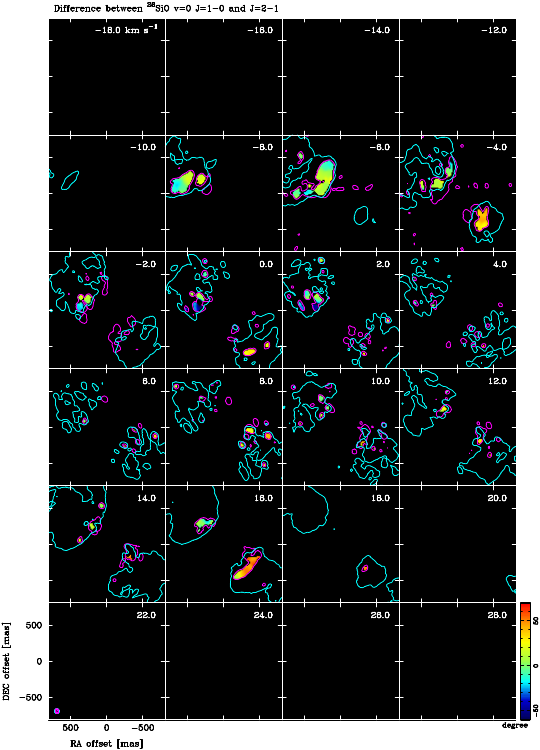}
\caption{Difference in polarization angle (PA) between $^{28}$SiO $v$=0 $J$=1-0 and $J$=2-1, PA($J$=1-0)$-$PA($J$=2-1).
The contour  shows the 4$\sigma$ levels of linear polarization intensities of $^{28}$SiO $v$=0 $J$=1-0 (47~mJy~beam$^{-1}$) and 2-1 (6.2~mJy~beam$^{-1}$) lines. 
The polarization angle rotation of 1~degree corresponds to $RM$ of 460~rad~m$^{-2}$.
}
\label{fig-chdiff}
\end{center}
\end{figure}

\end{document}